\documentclass[10 pt]{article}

\usepackage[a4paper, top=2cm, bottom=2cm]{geometry}  
\usepackage[english]{babel}         
\usepackage{graphicx}               
\usepackage{float}                  
\usepackage{amsmath}
\usepackage{mathrsfs}
\usepackage{amsfonts}

\usepackage{csvsimple}

\usepackage{tabularx}
\usepackage{subcaption}

\numberwithin{equation}{section}
\graphicspath{ {images/} }          

\nocite{*}

\title{Bethe ansatz solutions and hidden $sl(2)$ algebraic structure for a class of quasi-exactly solvable systems}
\author{Siyu Li, Ian Marquette and Yao-Zhong Zhang}

\begin{document}

\maketitle

\begin{abstract}
 \noindent The construction of analytic solutions for quasi-exactly solvable systems is an interesting problem. We revisit a class of models for which the odd solutions were largely missed previously in the literature: the anharmonic oscillator, the singular anharmonic oscillator, the generalized quantum isotonic oscillator, non-polynomially deformed oscillator, and the Schr\"odinger system from the kink stability analysis of $\phi^6$-type field theory. We present a systematic and unified treatment for the odd and even sectors of these models. We find generic closed-form expressions for constraints to the allowed model parameters for quasi-exact solvability, the corresponding energies and wavefunctions. We also make progress in the analysis of solutions to the Bethe ansatz equations in the spaces of model parameters and provide insight into the curves/surfaces of the allowed parameters in the parameter spaces.  Most previous analyses in this aspect were on a case-by-case basis and restricted to the first excited states. We present analysis of the solutions (i.e. roots) of the Bethe ansatz equations for higher excited states (up to levels $n$=30 or 50). The shapes of the root distributions change drastically across different regions of model parameters, illustrating phenomena analogous to phase transition in context of integrable models. Furthermore, we also  obtain the $sl(2)$ algebraization for the class of models in their respective even and odd sectors in a unified way. 
 \end{abstract}

\section {Introduction}

Quasi-exactly solvable (QES) Hamiltonians were discovered in the 80ties \cite{Tur1988, Tur1994, Gon1993,Ush1994}. Over time they have become an important class of quantum systems and attracted much international attention (see e.g. the review paper \cite{turbiner2016} and references therein). QES models are models for which a finite number of energy levels and the corresponding wave functions can be obtained via analytical or algebraic methods.  For special model parameters, the wave functions of these systems are connected with polynomials, among which the most well known are the so-called Bender-Dunne  polynomials \cite{Ben1996}.  In the seminar work of Turbiner \cite{Tur1988,Tur1994}, connection of many QES models to Lie algebra $sl(2)$ was established via a process called algebraization. The work on classification by Olver, Gomez-Ulate and Kamran allowed one to obtain QES systems with with rational, trigonometric, hyperbolic and even elliptic potentials \cite{Gon1993}. From a mathematical perspective QES systems have their origin in the work on Lam\'e polynomials and are often connected with the Heun differential equation and its various confluent versions.

Among various approaches of solving QES systems is the (functional) Bethe ansatz method \cite{Sas2009,Zha2012}. This method allows one to find the constraints of model parameters and construct closed-form solutions (energies and wave functions) of the model in terms of a set of algebraic equations (the so-called Bethe ansatz equations) \cite{Sas2009,Zha2012,Agb2012,agboola2014,quesne2018,Que2018,quesne2017}. It converts the problem of solving differential equations into the one of solving sets of algebraic Bethe ansatz equations (BAEs). 
Such a system of algebraic equations are usually non-linear which are in general hard to solve analytically. For higher-level excited wave functions, numerical techniques are usually needed to solve the BAEs. In \cite{Agb2012}, transformations were applied to recast the problem in alternative form but they were limited to the even sector. 

In this paper we revisit an interesting class of QES models partially solved previously in the literature and provide a unified treatment for exact solutions  in their even and odd sectors. As far as we know, such unified derivation of closed-form solutions in the even and odd sectors of the models had not been achieved previously. In particular, our approach enables us to obtain new higher-level analytic expressions for wave functions of the Schr\"odinger system from the kink stability analysis of $\phi^6$-type field theory. We also aim to make progress in solving the BAEs related to higher level excited states. We propose a method that allows us to gain insight into solutions of the BAEs via curves/surfaces in the spaces of model parameters. These curves/surfaces are connected with possible solutions of the BAEs. We present analysis of the spaces of model parameters for which the BAEs admit solutions. We also provide plots of the roots of the polynomials associated with the wave functions up to levels $n=30$ or 50 in the complex plane. We observe that the shapes of the root distributions show phenomena analogous to the phase transitions in the pairing models. We provide a complete description for the quasi-exact solvability of the models by means of analytical and numerical techniques. To our knowledge, our study on root distributions in the complex plane for polynomials associated with quasi-exactly solvable systems is new. 

Algebraization is a powerful way for proving the quasi-exact solvability of a quantum system. (We remark, however, that there exist examples of 1D quantum systems which do not have the usual $sl(2)$ algebraization \cite{Tur1988,Tur1994}.) For the class of QES models considered in this paper, we obtain their $sl(2)$ algebraization in the even and odd sectors in a unified way. As far as we know, the hidden $sl(2)$ algebra symmetry for these models had not been realized previously.

This paper is organized as follows. In Section 2, we provide a unified derivation of Bethe ansatz solution and $sl(2)$ algebraization in the odd and even sectors for the sextic anharmonic oscillator. In section 3, We study the singular anharmonic oscillator and construct its exact solutions via Bethe ansatz method. We prove that this model has no odd polynomial solutions.  In Sections 4 and 5, a detailed analysis of the generalized quantum isotonic oscillator and non-polynomially modified oscillator are respectively performed. Section 6 is devoted to the study of the quasi-exact solvability of the Schr\"odinger equation from kink stability analysis of $\phi^6$ field theory. In section 7, we present the conclusions of the paper. In the appendix, we review some results in \cite{Zha2012,zhang2016} which will be used in this paper.

\section{Anharmonic oscillator: Sextic potential}

Anharmonic oscillator with sextic potential has been widely studied in the literature (see e.g. the review paper \cite{turbiner2016} and references therein). Solutions to this model consist of even and odd sectors \cite{Ush1994}. For special model parameters, solutions in even sector are given by the so-called Bender-Dunne polynomials \cite{Ben1996,Kra1997}. Solutions of this model were also studied in \cite{Atre2003}.

In this section, we reexamine this model by using the Bethe ansatz method \cite{Zha2012}. We will present the derivation of closed-form solutions and $sl(2)$ algebraization  in odd and even sectors of this model in a unified way. This unified $sl(2)$ algebraization in the even and odd sectors seems a new result. We also examine the parameter space for which the BAEs have solutions and find the root distributions by novel numerical techniques for $n=30$.  

Consider the Hamiltonian of the model with sextic potential \cite{Ush1994}
\begin{equation}\label{Sixtic Hamiltonian}
    {H}=-\frac{d^2}{dx^2}+\alpha x^2+\beta x^4+\gamma x^6,
\end{equation}
where $\alpha, \beta$ and $\gamma>0$ are real model parameters. The time-independent Schr\"odinger equation is $H\psi(x)=E\psi(x)$. Let
\begin{equation}\label{ground state}
\psi_0(x)= \exp\left(-\frac{\beta}{4\sqrt{\gamma}}x^2-\frac{\sqrt{\gamma}}{4}x^4\right). 
\end{equation}
It can be checked that a similarity or gauge transformation  $\psi_0(x)^{-1}H\psi_0(x)$ eliminates the terms involving $x^4$ and $x^6$ in the transformed Hamiltonian. So we assume that the wave function $\psi(x)$ has the factorizable form, $\psi(x)=\psi_0(x) \phi(x)$. That is, we set
\begin{equation}\label{Sixtic wavefunction}
    \psi(x)=\exp\left(-\frac{\beta}{4\sqrt{\gamma}}x^2-\frac{\sqrt{\gamma}}{4}x^4\right)\, \phi(x).
\end{equation}
Substituting this into the Schr\"odinger equation yields the ODE for $\phi(x)$,
\begin{equation}\label{ODE x}
    \phi''(x)-\frac{1}{\sqrt{\gamma} }(\beta x+2\gamma x^3)\phi'(x)
        - \left\{\frac{\beta}{2\sqrt{\gamma}}- E  +  \left[-\frac{\beta^2}{4\gamma}+  \alpha+3 \sqrt{\gamma} \right]x^2  \right\} \phi(x)=0.
\end{equation}

To get solutions in the odd and even sectors 
in unified way, we set $\phi(x)$ to the form
\begin{equation}
    \phi(x)=x^p\, y(x),\quad p=0, 1,
\end{equation}
where $p=0,1$ correspond to even and odd sectors, respectively. Under this replacement, the ODE (\ref{ODE x}) becomes 
\begin{equation}\label{Sixtic ODE x}
    \begin{split}
    y''(x)&-\left(\frac{\beta}{\sqrt{\gamma}} x-\frac{2p}{x}+2\sqrt{\gamma} x^3\right)y'(x)\\
        &- \left\{\frac{(2p+1)\beta}{2\sqrt{\gamma}}- E  +  \left[-\frac{\beta^2}{4\gamma}+  \alpha+(3+2p) \sqrt{\gamma} \right] x^2 \right\} y(x)=0,
    \end{split}
\end{equation}
Making the variable change $z=x^2$, we obtain
\begin{equation}\label{Sixtic ODE z}
    \begin{split}
        4 z y''(z) & -\left[4\sqrt{\gamma} z^2+ \frac{2\beta}{\sqrt{\gamma}} z-2(2p+1) \right] y'(z)\\
        &- \left\{\frac{(2p+1)\beta}{2\sqrt{\gamma}}- E  +  \left[-\frac{\beta^2}{4\gamma}+  \alpha+(3+2p) \sqrt{\gamma} \right] z \right\}y(z)=0.
    \end{split}
\end{equation}

The ODE (\ref{Sixtic ODE z}) is quasi-exactly solvable provided that the potential parameters satisfy certain constraints, and exact solutions are given by the degree $n$ polynomials,
\begin{equation}\label{polynomial soln Sextic}
    y(z)=\prod_{i-1}^n(z-z_i),\quad y(z)\equiv 1 ~ {\rm for} ~ n=0.
\end{equation}
This is seen as follows. Applying the Bethe ansatz method \cite{Zha2012}, (\ref{c2}), (\ref{general Bethe ansatz}) and (\ref{c1}), we get the energies
\begin{equation}\label{Sixtic En}
        E_n^{(p)}=(4n+2p+1)\frac{\beta}{2 \sqrt{\gamma}}+4 \sqrt{\gamma} \sum_{i=1}^n z_i,\qquad n=0,1,2,\ldots,
\end{equation}
and the constraint for the model parameters
\begin{equation}\label{Sixtic parameter constrain n}
       \alpha= \frac{\beta^2}{4\gamma}-(3+4n+2p)\sqrt{\gamma},
\end{equation}
where the roots $z_1, z_2,\dots z_n$ are determined by the BAEs
\begin{equation}\label{Sixtic roots of Bethe ansatz n}
   \sum_{i=1}^n \frac{2}{z_i-z_j}-\frac{2\gamma z_i^2+\beta z_i-(1+2p)\sqrt{\gamma}}{2 \sqrt{\gamma} z_i}=0,\quad i=1,2,\dots,n.
\end{equation}
It follows that the closed-form expressions for wave functions in the even and odd sectors are
 \begin{equation}\label{Sixtic wavefunction n}
       \psi^{(p)}_n(x)= \exp\left[-\frac{\beta}{4\sqrt{\gamma}}x^2-\frac{\sqrt{\gamma}}{4}x^4 \right] \,x^p\,\prod_{i=1}^n (x^2-z_i),\qquad p=0,1.
\end{equation}
Similar expressions corresponding to (\ref{Sixtic En})-(\ref{Sixtic wavefunction n}) for the model were presented in \cite{Ush1994}.

The quasi-exact solvability of (\ref{Sixtic ODE z}) can also be seen from the fact that (\ref{Sixtic ODE z}) admits a $sl(2)$ algebraization if the model parameters satisfy the constraint (\ref{Sixtic parameter constrain n}).  To show this, 
we write (\ref{Sixtic ODE z}) in the form, $\mathcal{H}y(z)=Ey(z)$, where
\begin{equation}\label{sixtic algebraic linear differential operator}
\begin{split}
    \mathcal{H}=4z \frac{d^2}{dz^2}-\left(4\sqrt{\gamma}z^2+\frac{2 \beta}{\sqrt{\gamma}}z-2-4p\right)\frac{d}{dz}-\left[-\frac{\beta^2}{4\gamma}+  \alpha+(3+2p) \sqrt{\gamma} \right] z  - \frac{\beta(1+2p)}{2\sqrt{\gamma}}
\end{split}
\end{equation}
is related to Hamiltonian (\ref{Sixtic Hamiltonian}) via the gauge transformations and the variable change. Applying theorem $A2$ (\ref{general algebraization}), the transformed Hamiltonian $\mathcal{H}$ allows for  $sl(2)$ algebraization if 

\begin{equation}
  - \left[ -\frac{\beta^2}{4\gamma}+  \alpha+(3+2p) \sqrt{\gamma}\right] \equiv c_1=-n\left[(n-1)a_3+b_2\right]=4n\sqrt{\gamma}, 
\end{equation}


which is nothing but the constraint (\ref{Sixtic parameter constrain n}) (obtained by the Bethe ansatz method). Indeed, for the model parameters satisfying this constraint, the Hamiltonian $\mathcal{H}$ is an element of the enveloping algebra of Lie-algebra $sl(2)$,

\begin{equation}\label{sixtic algebraic sl(2)}
    \begin{split}
        \mathcal{H}=4 J^0 J^- +4\sqrt{\gamma} J^+ -\frac{2 \beta}{\sqrt{\gamma}} J^0+2(n+2p+1) J^- -\frac{\beta(1+2n+2p)}{2\sqrt{\gamma}},
    \end{split}
\end{equation}
where $J^+, J^-, J^0$ are 1st-order differential operators in (\ref{sl(2) operator}). They
are differential operator realization of the $n + 1$-dimensional representation of the $sl(2)$ algebra.
The expression (\ref{sixtic algebraic sl(2)}) thus gives unified $sl(2)$ algebraization for the even and odd sectors of the model. 

We now study the solutions of the BAEs (\ref{Sixtic roots of Bethe ansatz n}) and present explicit expressions for the first few levels. 

For $n=0$,  we have $y(z)=1$. The constraint in this case is given by $\alpha=\frac{\beta^2}{4\gamma} - (2p+1) \sqrt{\gamma}$. The corresponding energies and wave functions in even and odd sectors are
\begin{equation}
    E^{(p)}_{0}=\frac{(2p+1)\beta}{2\sqrt{\gamma}},\qquad
    \psi^{(p)}_{0}(x)=x^p\,\exp\left[-\frac{\beta}{4\sqrt{\gamma}}x^2-\frac{\sqrt{\gamma}}{4}x^4 \right].
\end{equation}
$\psi_0^{(0)}(x)$ is even and the ground level solution. $\psi_0^{(1)}(x)$ is odd and gives the 1st excited solution.

When $n=1$, the constraint for the model parameters is 
\begin{equation}
     \alpha=\frac{\beta^2}{4\gamma}-(2p+7)\sqrt{\gamma},
\end{equation}
and the Bethe ansatz equation gives the roots 
\begin{equation} 
    z^{(p)}_{1\pm}=\frac{-\beta \pm \sqrt{\beta^2+8(2p+1) \gamma^{3/2}}}{4 \gamma}.
\end{equation}
The associated energies and wave-functions in the even and odd sectors are
\begin{align}
    &\begin{aligned}\label{Sixtic E1 p}
        E^{(p)}_{1\pm}=\frac{(2p+3) \beta \pm 2 \sqrt{\beta^2+8(2p+1) \gamma^{3/2}}}{2\sqrt{\gamma}},
    \end{aligned}\\
    &\begin{aligned}\label{Sixtic wavefunction n1 p}
      \psi^{(p)}_{1\pm}(x)=x^p\,\exp\left[-\frac{\beta}{4\sqrt{\gamma}}x^2-\frac{\sqrt{\gamma}}{4}x^4 \right] \left(x^2-\frac{-\beta \pm \sqrt{\beta^2+8 (2p+1)\gamma^{3/2}}}{4 \gamma} \right).
    \end{aligned}
\end{align}
$\psi_{1\pm}^{(0)}(x)$ and $\psi_{1\pm}^{(1)}(x)$ give the next level even and odd solutions, respectively.

\subsection{Roots and allowed model parameters for $n=2, 3$}

\begin{figure}[ht]
    \centering

    \begin{subfigure}{0.48\textwidth}
        \centering
        \includegraphics[width=\linewidth]{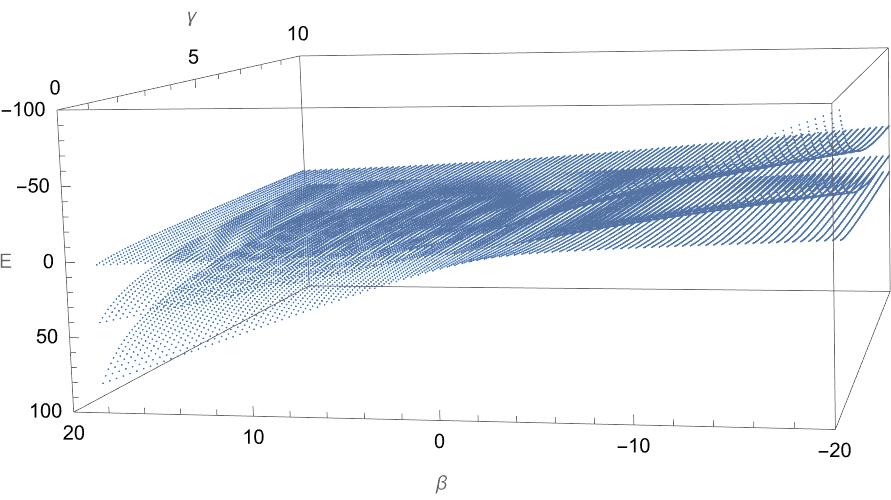}
        \caption{Relationship for $n=2$ in even sector}
        \label{fS n2 p0}
    \end{subfigure}
    \hfill
    \begin{subfigure}{0.48\textwidth}
        \centering
        \includegraphics[width=\linewidth]{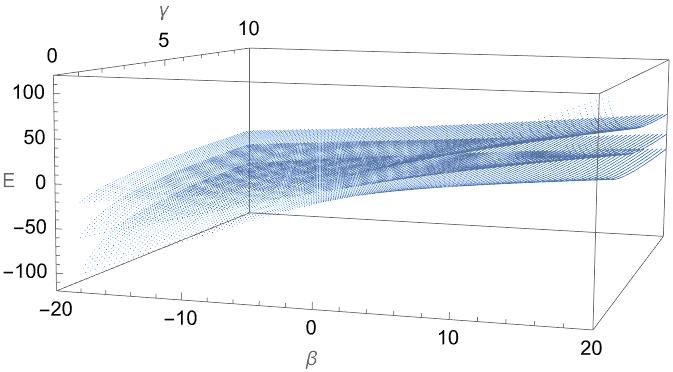}
        \caption{Relationship for $n=2$ in odd sector}
        \label{fS n2 p1}
    \end{subfigure}
    
    \caption{Relationship between the energy and the parameters $\beta$ and $\gamma$ for $n=2$.}
    \label{fS n2}
\end{figure}

For $n=2$, the corresponding energy and parameter constraint are 
\begin{equation}\label{Sixtic E2}
        E_2^{(p)}=(2p+9)\frac{\beta}{2 \sqrt{\gamma}}+4 \sqrt{\gamma} (z_1+z_2),
\end{equation}
\begin{equation}\label{Sixtic parameter constrain 2}
    4 \gamma\left[\alpha+(2p+11)\sqrt{\gamma}\right]-\beta^2=0, \quad p=0,1.
\end{equation}
This model is different from the rest models because the constraint for the model parameters is independent of the roots of the BAEs. As the energies depend on the roots, so in this case we solve the BAEs and the parameter constraint simultaneously via numerical techniques and plot the energies against the allowed values of parameters. The space for the parameters and energies for which the BAEs have solutions is plotted in Fig.\ref{fS n2}. 
There are three different layers in the graphs. This means three different energies for fixed parameters $\beta, \gamma$, corresponding to three different sets of roots of the BAEs for $n=2$. This is in agreement with the prediction of the $sl(2)$ algebraic structure of the system.

\begin{figure}[ht]
    \centering

    \begin{subfigure}{0.48\textwidth}
        \centering
        \includegraphics[width=\linewidth]{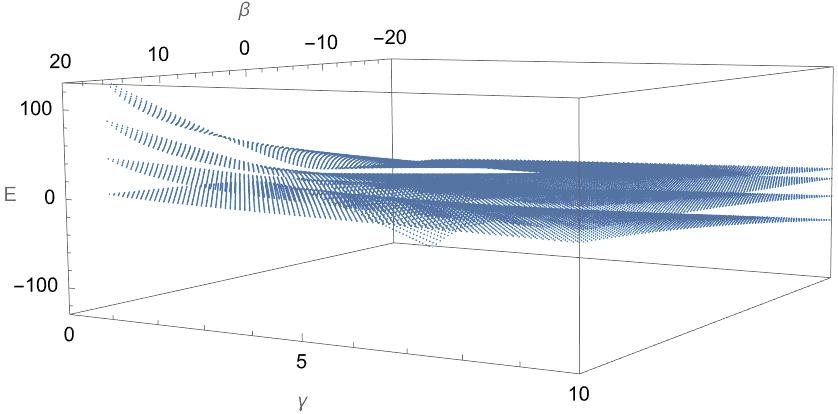}
        \caption{Relationship for $n=3$ in even sector}
        \label{fS n3 p0}
    \end{subfigure}
    \hfill
    \begin{subfigure}{0.48\textwidth}
        \centering
        \includegraphics[width=\linewidth]{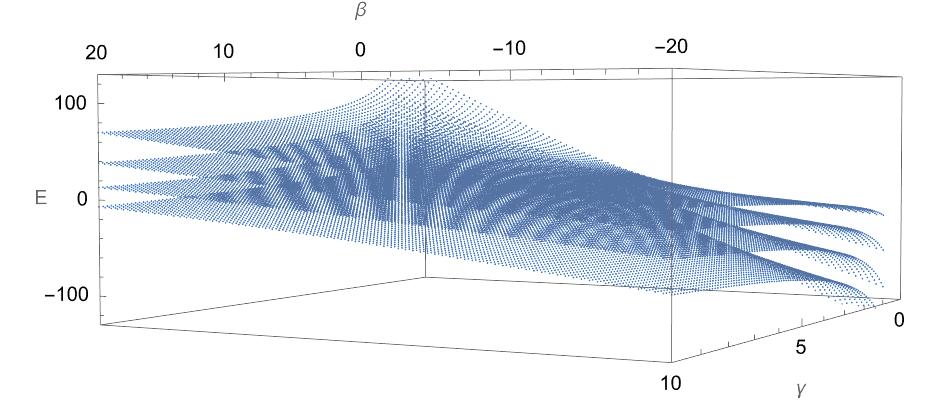}
        \caption{Relationship for $n=3$ in odd sector}
        \label{fS n3 p1}
    \end{subfigure}
    
    \caption{Relationship between the energy and the parameters $\beta$ and $\gamma$ for $n=3$.}
    \label{fS n3}
\end{figure}

For $n=3$, the energy and parameter constraint are  
\begin{equation}\label{Sixtic E3}
        E_2^{(p)}=(2p+11)\frac{\beta}{2 \sqrt{\gamma}}+4 \sqrt{\gamma} (z_1+z_2+z_3),
\end{equation}
\begin{equation}\label{Sixtic parameter constrain 3}
   4\gamma \left[\alpha+(15+2p)\sqrt{\gamma}\right]-\beta^2=0,
\end{equation}
We numerically solve the roots of the BAEs, the constraint of the parameters and energy simultaneously. We obtain the allowed values of parameters for which the BAEs have solutions. Similar to the $n=2$ case, in Fig.\ref{fS n3} we plot the energy (which depends on the roots of the BAEs) against the allowed values of the parameters $\beta, \gamma$ for $n=3$. 
There are 4 layers of energies with fixed parameters $\beta, \gamma$, corresponding to 4 different sets of roots of the BAEs, as expected from the $sl(2)$ algebraic structure of the system for $n=3$.

\subsection{Root distributions for higher level wave functions}

In this subsection we present the distributions and properties of the roots of the BAEs for fixed values of model parameters which satisfy the constraint. 
We use an approach similar to the one in \cite{marquette2012}. We write the solution $y(z)$ of the ODE (\ref{Sixtic ODE z}) in the form 
\begin{equation}\label{S root polynomial of psi(z)}
    y(z)=\sum_{k=0}^n a_k z^k,
\end{equation}
and solve the coefficients $a_k$. We then find the roots of $y(z)$. These roots correspond to the roots of the BAEs. Due to the parameter constrain (\ref{Sixtic parameter constrain n}) is not related to the roots of the BAEs, we focus on the energy expression, which is the function of the sum of the roots of the BAEs. The Energy expression (\ref{Sixtic En}) equation contains 2 model parameters and the energy $E$. We fix two ($\beta, \gamma$) of them and set $a_n=1$. Substituting (\ref{S root polynomial of psi(z)}) into the (\ref{Sixtic ODE z}), the ODE becomes a bilinear $(n+1) \times (n+1)$ matrix equations for the coefficients $a_0, \ldots, a_{n-1}$ and the model energy $E$. Solving this matrix equation gives us all values of the coefficients and the model parameter. Substituting them into (\ref{S root polynomial of psi(z)}) and setting $y(z)$ equal to 0, we obtain the roots of $y(z)$, which correspond to the roots of the BAEs. 

\begin{figure}[ht]
    \centering

    \begin{subfigure}{0.24\textwidth}
        \centering
        \includegraphics[width=\linewidth]{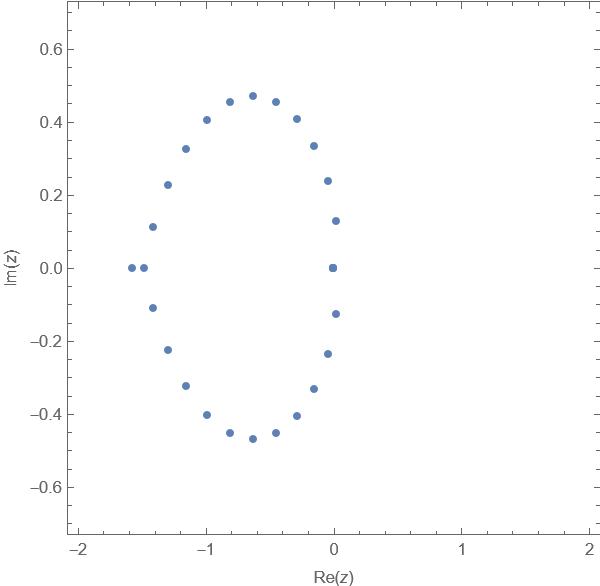}
        \caption{$E=-3815.8946$}
        \label{fS E=-3815.8946 p1}
    \end{subfigure}
     \hfill
    \begin{subfigure}{0.24\textwidth}
        \centering
        \includegraphics[width=\linewidth]{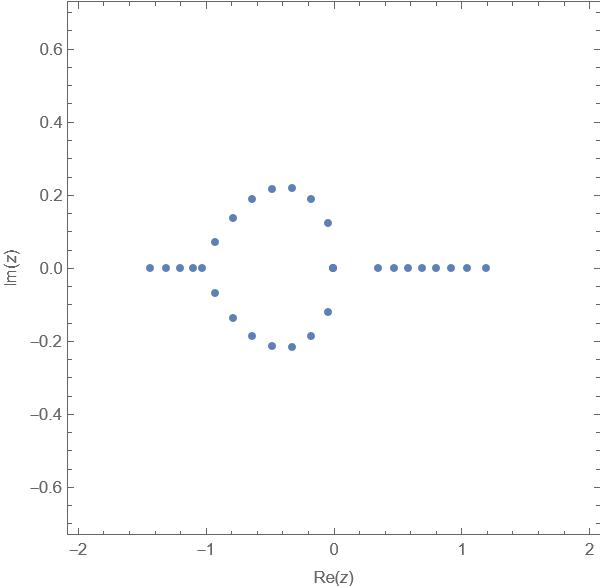}
        \caption{$E=-1468.8522$}
        \label{fS E=-1468.8522 p1}
    \end{subfigure}
     \hfill
    \begin{subfigure}{0.24\textwidth}
        \centering
        \includegraphics[width=\linewidth]{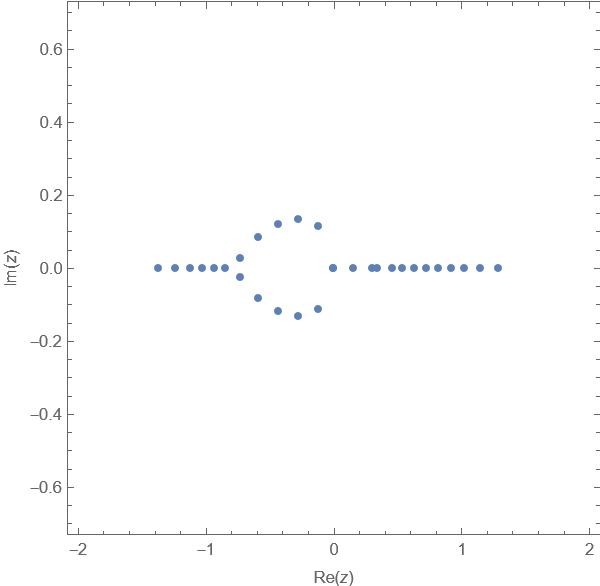}
        \caption{$E=-538.1682$}
        \label{fS E=-538.1682p1}
    \end{subfigure}
    \hfill
    \begin{subfigure}{0.24\textwidth}
        \centering
        \includegraphics[width=\linewidth]{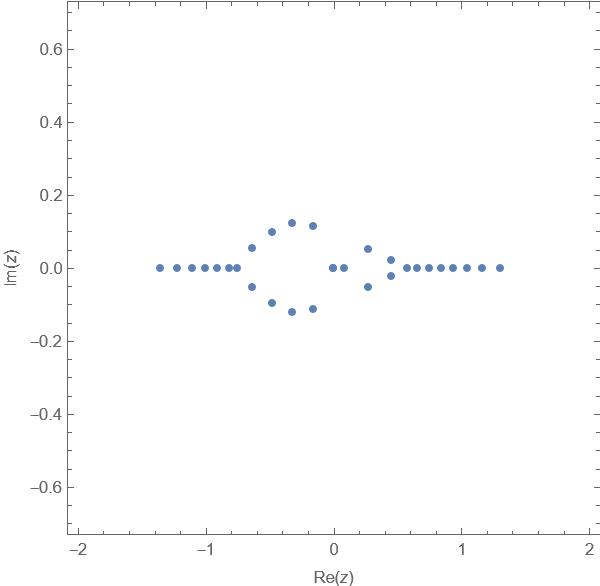}
        \caption{$E=-340.3596$}
        \label{fS E=-340.3596 p1}
    \end{subfigure}
    \hfill
    \begin{subfigure}{0.24\textwidth}
        \centering
        \includegraphics[width=\linewidth]{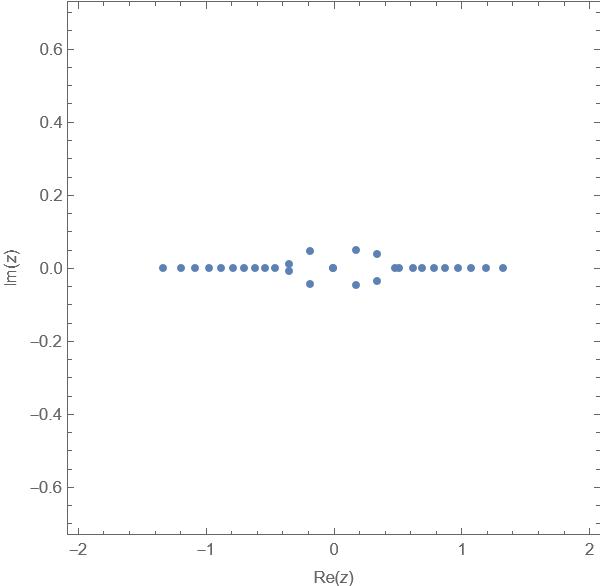}
        \caption{$E=0.3454$}
        \label{fS E=0.3454 p1}
    \end{subfigure}
     \hfill
    \begin{subfigure}{0.24\textwidth}
        \centering
        \includegraphics[width=\linewidth]{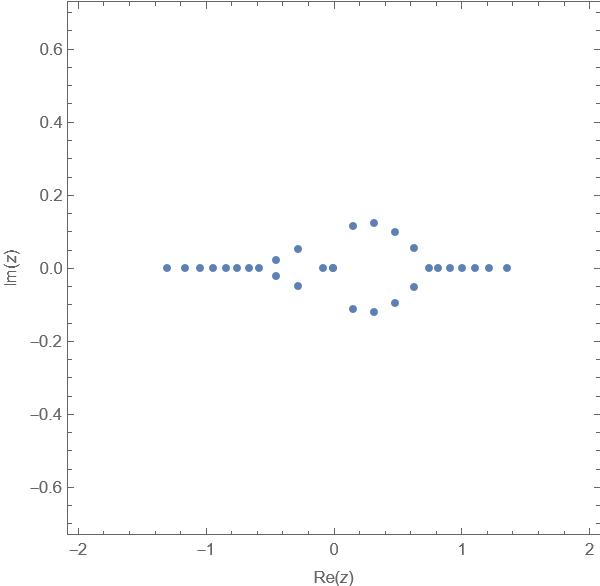  }
        \caption{$E=341.1904$}
        \label{fS E=341.1904 p1}
    \end{subfigure}
     \hfill
    \begin{subfigure}{0.24\textwidth}
        \centering
        \includegraphics[width=\linewidth]{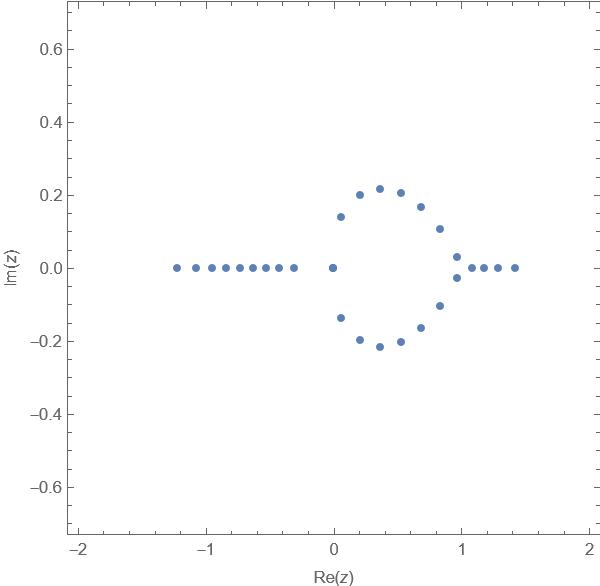  }
        \caption{$E=1219.2647$}
        \label{fS E=1219.2647 p1}
    \end{subfigure}
    \hfill
    \begin{subfigure}{0.24\textwidth}
        \centering
        \includegraphics[width=\linewidth]{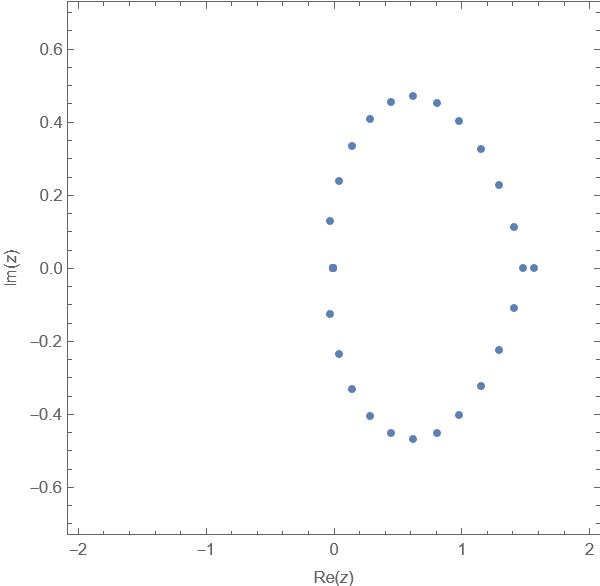  }
        \caption{$E=3817.3949$}
        \label{fS E=3817.3949 p1}
    \end{subfigure}
    \caption{Involution of root distribution with energy for odd solutions.}
    \label{fS E p=1}
\end{figure}

The root distributions with different energies and fixed parameters $\beta=1, \gamma=3000$ are shown in Fig.\ref{fS E p=1}. The order of the graphs is from the lowest energy to the highest energy. All the roots are circularly distributed on the left of the complex plain, with the lowest energy $E=-3815.8946$. With the increase of energy, circle becomes smaller and a line appears on the real axis. For $E=-340.3596$, the line starts to shape like a circle. When energy becomes a positive value $E=0.3454$, we have a circle with line tails which are symmetric about the imaginary axis. Then, the right circle becomes bigger and the left circle becomes a line. For energy $E=3817.3949$, the shape of the roots is opposite to that of the starting one. The root distribution graphs are also symmetric about the one for $E=0.3454$, indicating that root distributions in negative energies are opposite to those in positive energies. It seems that some kind of critical points appear in  Fig.\ref{fS E=-1468.8522 p1} and Fig.\ref{fS E=1219.2647 p1}, because of the appearance and disappearance of the line. 

\begin{figure}[ht]
    \centering

    \begin{subfigure}{0.31\textwidth}
        \centering
        \includegraphics[width=\linewidth]{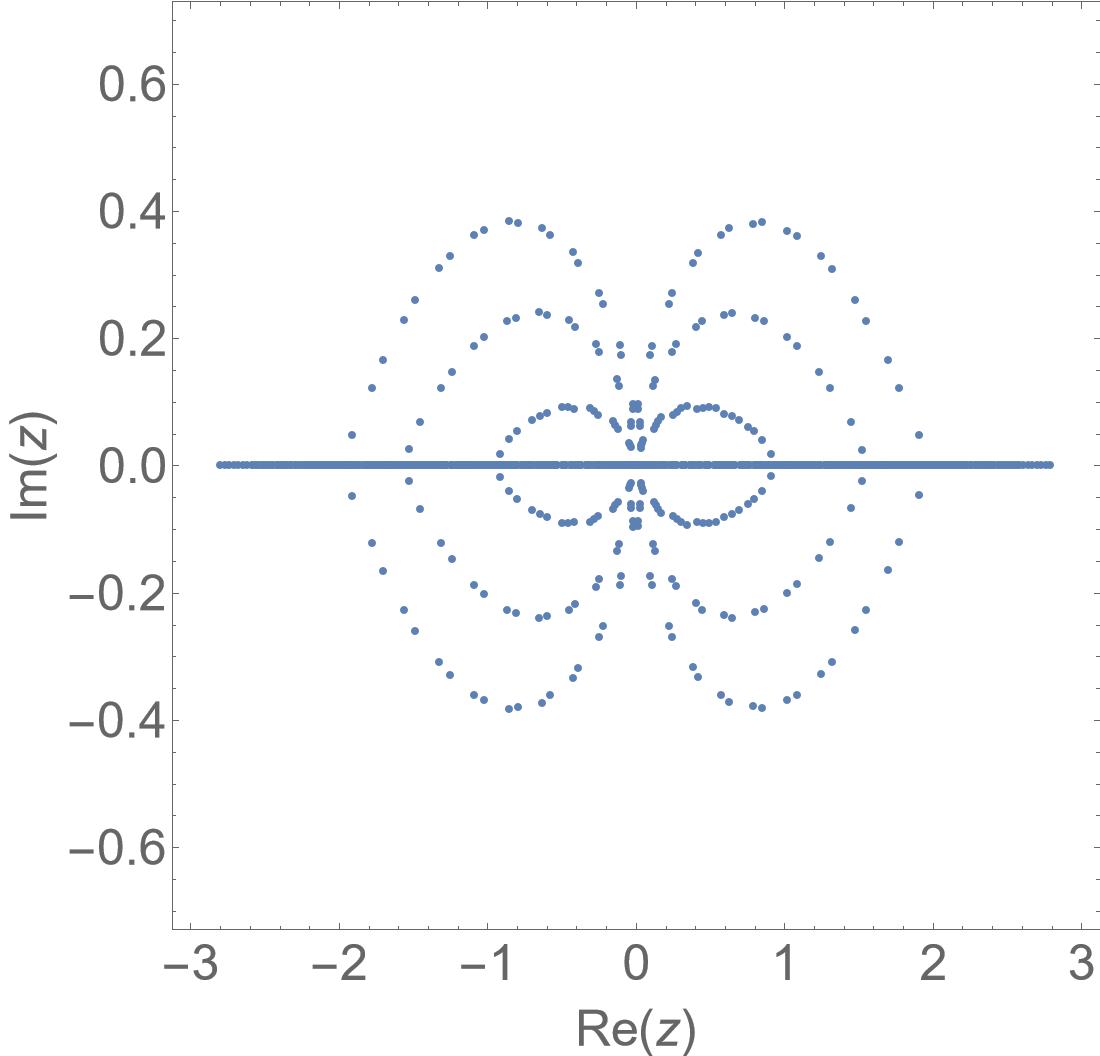}
        \caption{$\gamma=300$}
        \label{fS gamma=300}
    \end{subfigure}
    \hfill
    \begin{subfigure}{0.31\textwidth}
        \centering
        \includegraphics[width=\linewidth]{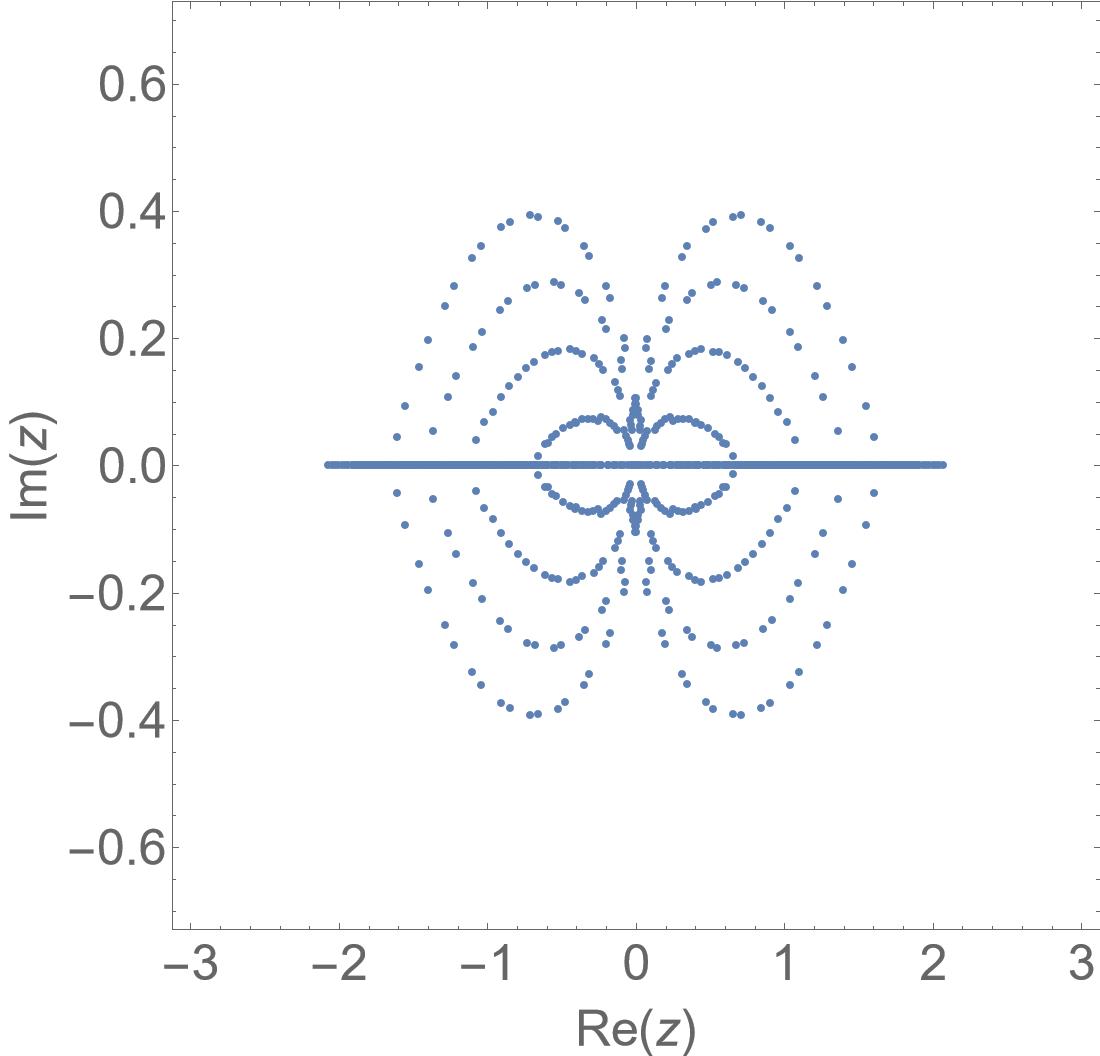}
        \caption{$\gamma=1000$}
        \label{fS gamma=1000}
    \end{subfigure}
    \hfill
    \begin{subfigure}{0.31\textwidth}
        \centering
        \includegraphics[width=\linewidth]{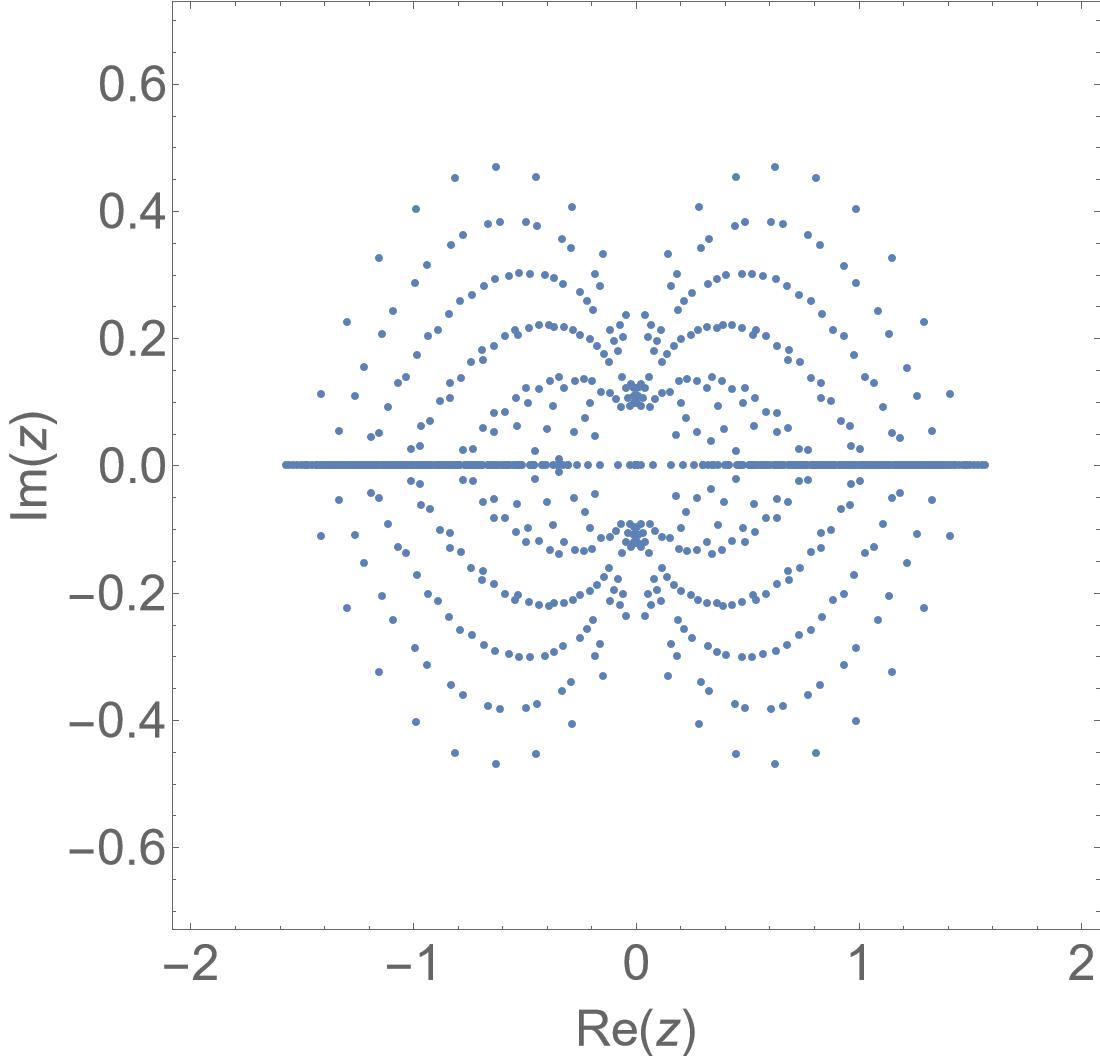}
        \caption{$\gamma=3000$}
        \label{fS gamma=3000}
    \end{subfigure}

    \caption{Root distributions with $\beta=1$ for $n=30$.}
    \label{fS beta=1 P=1}
\end{figure}

The root distributions for fixed $\beta$ and changed $\gamma$ are shown in Fig.\ref{fS beta=1 P=1}. From the distributions of the roots, we can clearly observe symmetry about the imaginary axis. For $\gamma=300$, the symmetric pattern is simple, consisting of three circular shapes. Increasing the value of $\gamma$, the number of circles increases and the pattern becomes more complicated.
\vskip.1in

\section{Singular anharmonic oscillator}
The singular anharmonic potential $\displaystyle V(x)=  \omega^2 x^2+\frac{2e}{x^4}+\frac{2d}{x^6}$ was studied in \cite{Zno1998,Kau1991}. Exact polynomial solutions in the even sector were derived in  \cite{Agb2012}. One of our main new results in this section is the proof that the gauged transformed Hamiltonian of this model has no polynomial solutions of odd degrees, in contrast to the (non-singular) sextic potential case in the previous section (for which both even and odd solutions exist for the gauged transformed Hamiltonian).

The Hamiltonian of the model reads
\begin{align}
   \label{A H(x)}
        H(x)=-\frac{d^2}{dx^2}+\omega^2 x^2 +\frac{2e}{x^4}+\frac{2d}{x^6},
\end{align}
where $\omega$, $e$, $d>0$ are real constant parameters. The time-independent Schr\"odinger equation is 
\begin{equation}\label{A Schrodinger equation}
    \frac{d^2 Y(x)}{dx^2}-\left(\omega^2 x^2 +\frac{2e}{x^4}+\frac{2d}{x^6} \right) Y(x)+{E}Y(x)=0.
\end{equation}
Let
\begin{align}\label{A Y0(x)}
     Y_0(x)=x^{3/2+e/\sqrt{2d}} \exp\left[-\frac{\omega}{2}x^2-\frac{\sqrt{2d}}{2} \frac{1}{x^2}\right],
\end{align}
which has no nodes when $\displaystyle \frac{3}{2}+\frac{e}{\sqrt{2d}}\leq 0$. It can be checked that a gauge transformation $Y_0(x)^{-1}HY_0(x)$ gets rid of the terms involving $\frac{1}{x^4}$ and $\frac{1}{x^6}$ in the transformed Hamiltonian. Note that $Y_0(x)$ is either even or odd, depending the value of $\displaystyle \frac{3}{2}+\frac{e}{\sqrt{2d}}$.
In what follows, we assume that $\displaystyle \frac{3}{2}+\frac{e}{\sqrt{2d}}\leq 0$ so that $Y_0(x)$ has no nodes.

We assume that $Y(x)$ has the factorizable form, $Y(x)=Y_0(x)\,y(x)$. We set 
\begin{align}\label{A Y(x)}
     Y(x)=x^{3/2+e/\sqrt{2d}} \exp\left[-\frac{\omega}{2}x^2-\frac{\sqrt{2d}}{2} \frac{1}{x^2}\right] y(x).
\end{align}
Under this substitution, the Schr\"odinger equation becomes
\begin{equation}\label{A Schrodinger y(x)}
    \begin{split}
        y''(x) & +\frac{2}{x} \left(-\omega x^2+\frac{3}{2}+\frac{e}{\sqrt{2d}}+\frac{\sqrt{2d}}{x^2}\right)y'(x)\\
        &+\left\{\left[{E}-2 \omega \left(2+\frac{e}{\sqrt{2d}}\right) \right]-\frac{1}{x^2}\left[2\omega\sqrt{2d}+\frac{1}{4}-\left(\frac{e}{\sqrt{2d}}+1\right)^2\right] \right\} y(x)=0,
    \end{split}
\end{equation}
To treat solutions in even and odd sectors in unified way, we set 
\begin{align}\label{A y(x)}
     y(x)=x^p \psi(x), \quad p=0,1.
\end{align}
Here if 
$(-x)^{3/2+e/\sqrt{2d}}=x^{3/2+e/\sqrt{2d}}$ under $x\rightarrow -x$ then  $p=0,1$ correspond to even and odd solutions, respectively; if $(-x)^{3/2+e/\sqrt{2d}}=-x^{3/2+e/\sqrt{2d}}$ under $x\rightarrow -x$ then  $p=0,1$ correspond to odd and even solutions, respectively.

Substituting (\ref{A y(x)}) into (\ref{A Schrodinger y(x)}), we have 
\begin{equation}
    \begin{aligned}\label{A psi(x)}
\psi^{\prime \prime}(x) &+ \frac{2}{x}\left(-\omega x^2+p+\frac{3}{2}+\frac{e}{\sqrt{2d}}+\frac{\sqrt{2d}}{x^2}\right) \psi^{\prime}(x)\\
&+ \frac{2p}{x^2}\left(-\omega x^2+p+\frac{3}{2}+\frac{e}{\sqrt{2d}}+\frac{\sqrt{2d}}{x^2}\right)\psi(x)\\
&+\left\{\left[{E}-2 \left(2+\frac{e}{\sqrt{2d}}\right) \omega\right]-\frac{1}{x^2}\left[\frac{1}{4}-\left(1+\frac{e}{\sqrt{2d} }\right)^2+2\omega \sqrt{2d}\right] \right\}\psi(x)=0.
    \end{aligned}
\end{equation}
Finally we make the change of variable, $z=x^2$, to get
\begin{equation}
    \begin{split}\label{A psi(z)}
    z^3 \psi^{\prime \prime}(z) & +\left[\sqrt{2d}z+\left(2+\frac{e}{\sqrt{2d}}+p\right) z^2-\omega z^3\right] \psi^{\prime}(z)\\
    &+ \frac{z}{4} \left[2p \left(\frac{3}{2}+\frac{ e}{\sqrt{2d}}\right)+\left(1+\frac{e}{\sqrt{2d} }\right)^2-\frac{1}{4}-2\omega\sqrt{2d}\right] \psi(z)\\
    &+\frac{1}{4} \left\{z^2\left[{E}-2\omega \left(p+2+\frac{e}{\sqrt{2d}}\right) \right]+2p\sqrt{2d}\right\}\psi(z)=0.
    \end{split}
\end{equation}

We seek exact polynomial solutions of this ODE of the form
\begin{equation}\label{A Polynomial psi(z)}
    \psi(z)=\prod_{i=1}^n(z-z_i),\quad \psi(z)\equiv 1~{\rm for}~n=0.
\end{equation}
By means of the Bethe ansatz method \cite{Zha2012}, 
applying (\ref{c2}), (\ref{general Bethe ansatz}) and (\ref{c1}), we find that $\psi(z)$ is solution of (\ref{A psi(z)}) with energies
\begin{align}
    \label{Enp}
     E_n=2\left(2+2n++2p+\frac{e}{\sqrt{2d}}\right) \omega,
\end{align}
provided that the model parameters and the roots $z_1,\dots, z_n$ satisfy the constraints
\begin{align}
   &\begin{aligned}\label{A c1}
        &2p\left( \frac{3}{2} +\frac{e}{\sqrt{2d}}\right)-2 \omega \sqrt{2d}-\frac{1}{4}+ \left(\frac{e}{\sqrt{2d}}+1\right)^2\\
       &\qquad\qquad\qquad =4 \omega \sum_{i=1}^n z_i -4n(n-1)-4n\left(p+2+\frac{e}{\sqrt{2d}}\right),
    \end{aligned}\\
    &\begin{aligned}\label{A c0}
        p \sqrt{2d}=2\omega \sum_{i=1}^n z_i^2-2 \left[2(n-1)+\left(p+2+\frac{e}{\sqrt{2d}}\right)\right] \sum_{i=1}^n z_i-2n\sqrt{2d},
    \end{aligned}
\end{align}
and the BAEs
\begin{equation}\label{A Bethe ansatz}
    \sum_{j \neq i}^n \frac{2 z_i^2}{z_i-z_j}+\omega z_i^2+(p+2+\frac{e}{\sqrt{2d}})z_i+\sqrt{2d}=0,\quad i=1,2,\ldots, n.
\end{equation}

It turns out that only solution for $p$ from  (\ref{A c1}), (\ref{A c0}) and (\ref{A Bethe ansatz}) is  $p=0$. This means that the ODE (\ref{A psi(z)}) does not have any polynomial solutions of odd degrees in $z$.
We have the following main new result in this section:
\vskip.1in
\noindent {\bf Proposition.} The value of $p$ in (\ref{A y(x)}) is identically equal to zero for closed-form solutions of the Schr\"odiner equation (\ref{A Schrodinger equation}). In other words, (\ref{A Schrodinger equation}) only has either even or odd (i.e. not both) exact solutions, depending on the property of  $x^{3/2+e/\sqrt{2d}}$ under $x\rightarrow -x$. 
\vskip.1in
\noindent{\em Proof.}
Summing over $i$ on both side of the BAEs (\ref{A Bethe ansatz}), we obtain
\begin{equation}\label{A sum Bethe ansatz}
    \sum_{\substack{i,j=1\\(i\neq j)}}^n \frac{2 z_i^2}{z_i-z_j}=\omega \sum_{i=1}^n z_i^2+\left(p+2+\frac{e}{\sqrt{2d}}\right)\sum_{i=1}^n z_i+\sqrt{2d}\sum_{i=1}^n 1=0,
\end{equation}
Now the first term on the left hand side is
\begin{equation}
    \begin{split}
        \sum_{\substack{i,j=1\\(i\neq j)}}^n \frac{2 z_i^2}{z_i-z_j} & =\sum_{\substack{i,j=1\\(i\neq j)}}^n \frac{z_i^2}{z_i-z_j}+\sum_{\substack{i,j=1\\(i\neq j)}}^n \frac{z_j^2}{z_j-z_i}= \sum_{\substack{i,j=1\\(i\neq j)}}^n \frac{z_i^2-z_j^2}{z_i-z_j}=\sum_{\substack{i,j=1\\(i\neq j)}}^n (z_i+z_j)\\
     &=\left( n-1 \right) \sum_{i=1}^n z_i+\left( n-1 \right) \sum_{j=1}^n z_j=2\left( n-1 \right) \sum_{i=1}^n z_i,
    \end{split}
\end{equation}     
So (\ref{A sum Bethe ansatz}) can be written as
\begin{equation}\label{A new sum Bethe ansatz}
   \omega \sum_{i=1}^n z_i^2=2\left( n-1 \right) \sum_{i=1}^n z_i+\left(p+2+\frac{e}{\sqrt{2d}}\right)\sum_{i=1}^n z_i+n\sqrt{2d}
\end{equation}
Substituting (\ref{A new sum Bethe ansatz}) into (\ref{A c0}) gives
\begin{equation}
    2p\sqrt{2d}=0 \implies p=0.
\end{equation}
This completes our proof of the Proposition. 

\vskip.1in
For $p=0$, the ODE (\ref{A psi(z)}) simplifies to
\begin{equation} \label{A new psi(z)}
    \begin{split}
     z^2 \psi^{\prime \prime}(z) &+ \left[ \sqrt{2d} +\left(2+\frac{e}{\sqrt{2d}}\right)z-\omega z^2 \right] \psi^{\prime}(z)\\
     &+\left\{ \frac{1}{4}z\left[{E}-2\left(2+\frac{e}{\sqrt{2d}}\right) \omega\right]-\frac{1}{4}\left[\frac{1}{4}-\left(1+\frac{e}{\sqrt{2d} }\right)^2+2 \omega\sqrt{2d}\right] \right\}\psi(z)=0.
    \end{split}   
\end{equation}
This ODE can be recast in the form
\begin{equation} \label{A doubly-confluent Heun}
    \begin{split}
     \psi^{\prime \prime}(z) &+ \left[-\frac{\sqrt{2d}}{z^2} +\frac{2+{e}/{\sqrt{2d}}}{z}-\omega  \right] \psi^{\prime}(z)\\
     &+ \frac{\left[{E}-2\left(2+{e}/{\sqrt{2d}}\right) \omega\right]z-\frac{1}{4}+\left(1+{e}/{\sqrt{2d} }\right)^2-2 \omega\sqrt{2d}}{4z^2}\psi(z)=0.
    \end{split}   
\end{equation}
It is a doubly-confluent Heun equation and has exact polynomial solutions (\ref{A Polynomial psi(z)}) with energies \cite{Agb2012}
\begin{align}\label{A En}
     E_n=2\left(2+2n+\frac{e}{\sqrt{2d}}\right) \omega, \qquad n=0,1, \dots,
\end{align}
provided that the model parameters satisfy the constraint
\begin{align}\label{A parameter constrain n}
    -2 \omega \sqrt{2d}-\frac{1}{4}+ \left(\frac{e}{\sqrt{2d}}+1\right)^2=4 \omega \sum_{i=1}^n z_i -4n\left(n+1+\frac{e}{\sqrt{2d}}\right).
\end{align} 
The roots $z_i$ are determined by the BAEs
\begin{equation}\label{A roots Bethe ansatz}
    \sum_{j\neq i}^n \frac{2}{z_i-z_j}+\frac{-\omega z_i^2+\left(2+\frac{e}{\sqrt{2d}}\right)z_i+\sqrt{2d}}{z_i^2}=0, \quad i=1,2,\ldots, n.
\end{equation}
The corresponding closed-form wave functions for the model are then given by
\begin{equation}\label{Yn}
    Y_n(x)=x^{3/2+e/\sqrt{2d}} \exp\left[-\frac{\omega}{2}x^2-\frac{\sqrt{2d}}{2} \frac{1}{x^2}\right] \left(\prod_{i=1}^n (x^2-z_i)\right).
\end{equation}

The ODE (\ref{A new psi(z)}) has a hidden $sl(2)$ algebra symmetry. To see this, we rewrite (\ref{A new psi(z)}) as 
\begin{eqnarray}
 &&\mathcal{H}\psi(z)=\varepsilon \psi(z), \qquad\varepsilon = \frac{1}{4}\left[\frac{1}{4}-\left(1+\frac{e}{\sqrt{2d} }\right)^2+2 \omega\sqrt{2d}\right],\nonumber\\
 &&\mathcal{H}=z^2 \frac{d^2}{dz^2} + \left[ \sqrt{2d} +\left(2+\frac{e}{\sqrt{2d}}\right)z-\omega z^2 \right] \frac{d}{dz}+\frac{z}{4}\left[{E}-2\left(2+\frac{e}{\sqrt{2d}}\right) \omega\right]. \label{calH A}
\end{eqnarray}
Applying theorem A2 (\ref{general algebraization}), the Hamiltonian $\mathcal{H}$ (\ref{calH A}) has a hidden $sl(2)$ algebra symmetry if
\begin{equation}
   \frac{1}{4}\left[{E}-2\left(2+\frac{e}{\sqrt{2d}}\right) \omega\right]\equiv c_1=-n[(n-1)a_3+b_2]\equiv n \omega,
\end{equation}
which gives the exact energies (\ref{A En}) above, i.e. $E=2\left(2+2n+\frac{e}{\sqrt{2d}}\right) \omega$. For such $E$ values, $\mathcal{H}$ depends on integer $n$ and can be expressed in terms of the $sl(2)$ generators (\ref{sl(2) operator}) as
\begin{equation}\label{A algebraic sl(2)}
    \mathcal{H}=J^0 J^0+\omega J^+ + \left(n+1+\frac{e}{\sqrt{2d}}\right) J^0+\sqrt{2d} J^- +n \left(\frac{1}{2}+\frac{n}{4}+\frac{e}{2\sqrt{2d}}\right).
\end{equation}
This provides a $sl(2)$ algebraization of the singular anharmonic oscillator. This algebraization had not been realized previously. The eigenvalues $\varepsilon$ of $\mathcal{H}$ with the polynomial eigenfunctions $\psi(z)=\prod_{i=1}^n(z-z_i)$ (\ref{A Polynomial psi(z)}) are given by 
\begin{equation}
    \varepsilon= n\left(n+1+\frac{e}{\sqrt{2d}}\right)- \omega \sum_{i=1}^n z_i,
\end{equation}
which is nothing but the constraint (\ref{A parameter constrain n}), and the roots $z_i$ are determined by the BAEs (\ref{A roots Bethe ansatz}).

We now present some explicit results for the first few solutions.

When $n=0$, we have $\psi(z)=1$. In this case the wave function is given by $Y_0(x)$ (\ref{A Y0(x)}) with energy $\displaystyle E_0=2(2+e/\sqrt{2d})\omega$ and the constraint for the model parameters $\displaystyle
2\omega\sqrt{2d}+{1}/{4}=\left({e}/{\sqrt{2d}}+1\right)^2$. $Y_0(x)$ given in (\ref{A Y0(x)}) represents the ground level solution of the system.

For $n=1$, the roots of (\ref{A roots Bethe ansatz}) are given by
\begin{align}\label{A z1}
    z_{1\pm}=\frac{1}{2\omega} \left[2+\frac{e}{\sqrt{2d}} \pm \sqrt{4\omega\sqrt{2d}+\left(2+\frac{e}{\sqrt{2d}}\right)^2}\right].
\end{align}
It follows from the constraint (\ref{A parameter constrain n}) that 
\begin{equation}\label{A parameter constrain  omega 1}
    \begin{split}
        \omega=\frac{1}{16d} \left(16 e+{35\sqrt{2d}}+\frac{4 e^2}{\sqrt{2d}} \pm 8\sqrt{{35d+12e\sqrt{2d}+3e^2}}\right).
    \end{split}
\end{equation}
The energies and the wave functions read 
\begin{align}
    &\begin{aligned}\label{A E1}
        E_1=2\left(4+\frac{e}{\sqrt{2d}}\right) \omega,
    \end{aligned}\\
    &\begin{aligned}\label{A Y1}
        Y_{1\pm}(x)=x^{3/2+e/\sqrt{2d}} \exp\left[-\frac{\omega}{2}x^2-\frac{\sqrt{2d}}{2} \frac{1}{x^2}\right] \left(\prod_{i=1}^n (x^2-z_{1\pm})\right).
    \end{aligned}
\end{align}
$Y_{1\pm}(x)$ given above are the first excited wave functions. Note that corresponding to one energy level $E_1$, i.e. one of the two $\omega$ expressions above, there are two wave functions $Y_{1\pm}(x)$, as expected from the hidden $sl(2)$ symmetry.

\subsection{Roots and allowed model parameters for $n=2, 3$}

\begin{figure}[ht]
    \centering
    \includegraphics[width=0.6\textwidth]{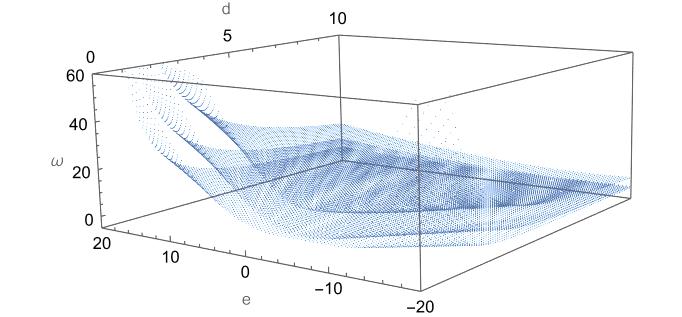}
    \caption{Relationship between model parameters $\omega, d, e$ and solutions of the BAEs for $n=2$.}
    \label{fA n2}
\end{figure}

When $n=2$, the energy is
\begin{equation}\label{A E2}
    E_2=2\left(6+\frac{e}{\sqrt{2d}}\right) \omega,
\end{equation}
We solve the roots $z_1, z_2$ of (\ref{A roots Bethe ansatz}) with $n=2$ and the constraint for the model parameters 
\begin{equation}\label{A parameter constrain 2}
    -2 \omega \sqrt{2d}-\frac{1}{4}+ \left(\frac{e}{\sqrt{2d}}+1\right)^2=4 \omega (z_1+z_2) -8\left(3+\frac{e}{\sqrt{2d}}\right)
\end{equation}
simultaneously via a numerical method. We obtain the the roots and the allowed model parameters, as shown in figure \ref{fA n2}. This figure shows the relationship between the parameters $e, \omega$ and $d$ for the BAEs to have solutions.
We set the range of $d$ from 0 to 10 and the $e$ from -20 to 20. From the constraint, the value of $\omega$ is determined by the roots of the BAEs. The hidden $sl(2)$ algebra symmetry of the model indicates that when $n=2$ the BAEs have 3 (=2+1) sets of solutions (i.e. 3 sets of roots) This is shown by the three layers of $\omega$ values in figure \ref{fA n2}.  These layers are symmetric about $e=0$ and thus the roots of the Bethe ansatz equations are also symmetric about $e=0$.

\begin{figure}[ht]
    \centering
    \includegraphics[width=0.6\textwidth]{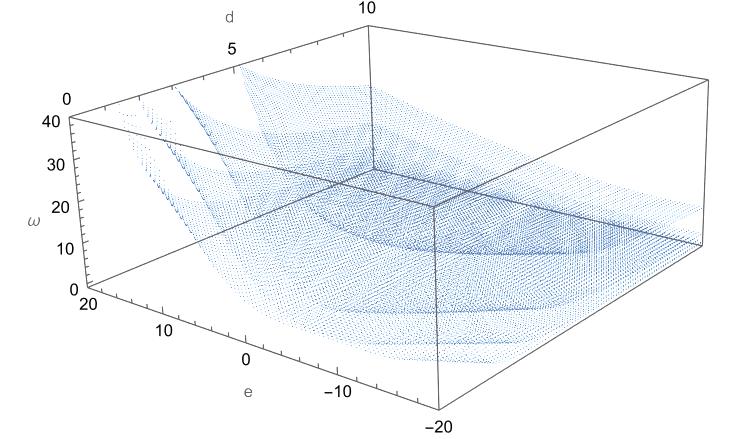}
    \caption{Relationship between model parameters $\omega, d, e$ and solutions of the BAEs for $n=3$.}
    \label{fA n3}
\end{figure}

When $n=3$, the energy is 
\begin{equation}\label{A E3}
    E_3=2\left(8+\frac{e}{\sqrt{2d}}\right) \omega,
\end{equation}
We numerically solve the roots $z_1, z_2, z_3$ of (\ref{A roots Bethe ansatz}) and the constraint for the model parameters
\begin{equation}\label{A parameter constrain 3}
    -2 \omega \sqrt{2d}-\frac{1}{4}+ \left(\frac{e}{\sqrt{2d}}+1\right)^2=4 \omega (z_1+z_2+z_3) -12\left(4+\frac{e}{\sqrt{2d}}\right).
\end{equation}
We obtain the relationship between the roots and the allowed parameters as shown in figure \ref{fA n3}. From the figure, there are 4 layers of $\omega$ values associated with each fixed $e$ and $d$. This means that there 4 different sets of roots of the BAEs for a fixed value of each parameter, in agreement with the prediction from the hidden $sl(2)$ algebraic structure. 

\subsection{Root distributions for higher level wave functions}
In this subsection, we will present the distributions and properties of the roots of the BAEs for fixed values of the model parameters satisfying the constraint.

\begin{figure}[ht]
    \centering
    \begin{subfigure}{0.24\textwidth}
        \centering
        \includegraphics[width=\linewidth]{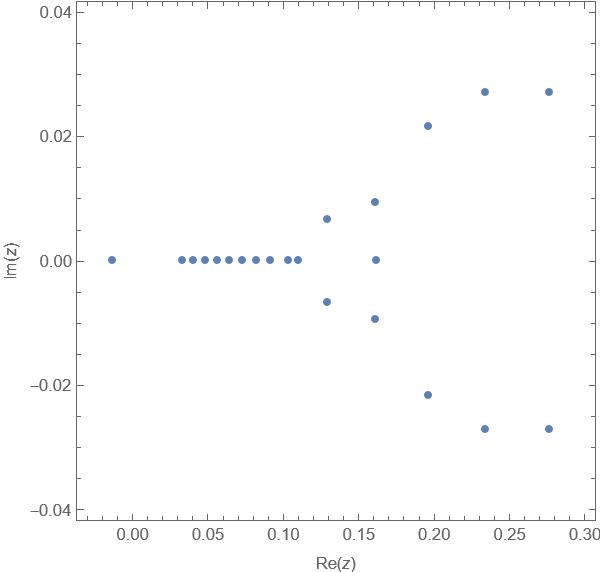}
        \caption{$\omega=272.8123$}
        \label{fA omega=272.8123 }
    \end{subfigure}
    \hfill
    \begin{subfigure}{0.24\textwidth}
        \centering
        \includegraphics[width=\linewidth]{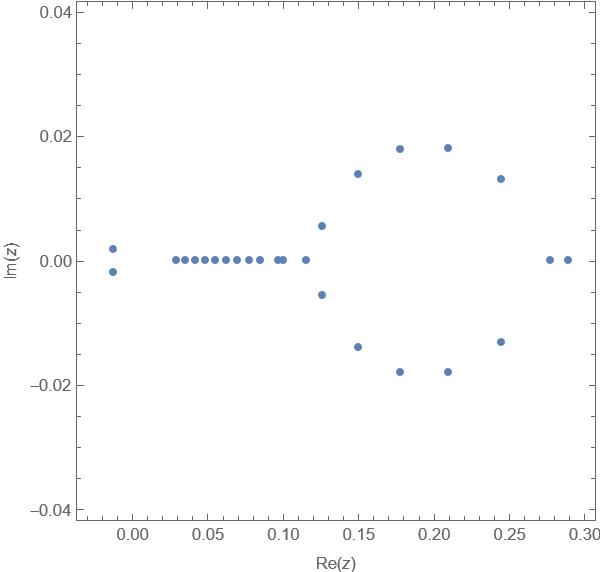}
        \caption{$\omega=358.2266$}
        \label{fA omega=358.2266}
    \end{subfigure}
    \hfill
    \begin{subfigure}{0.24\textwidth}
        \centering
        \includegraphics[width=\linewidth]{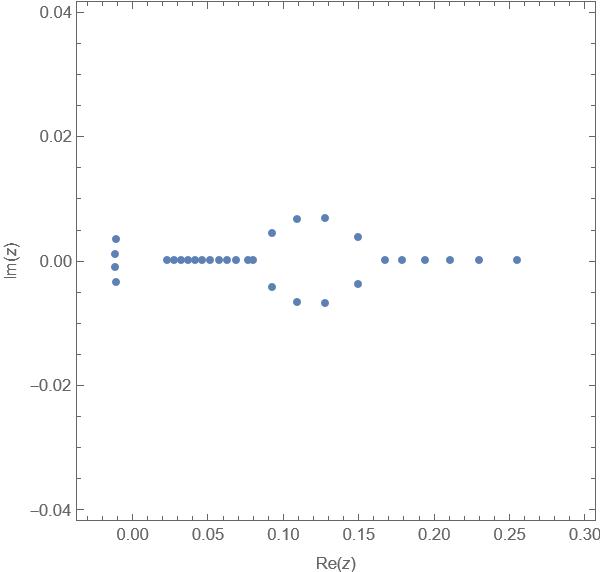}
        \caption{$\omega=581.0744$}
        \label{fA omega=581.0744}
    \end{subfigure}
    \hfill
    \begin{subfigure}{0.24\textwidth}
        \centering
        \includegraphics[width=\linewidth]{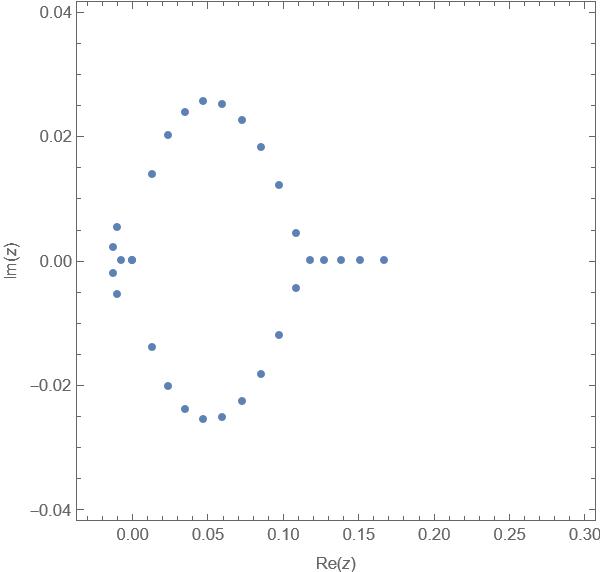}
        \caption{$\omega=883.7025$}
        \label{fA omega=883.7025}
    \end{subfigure}
        \hfill
    \begin{subfigure}{0.24\textwidth}
        \centering
        \includegraphics[width=\linewidth]{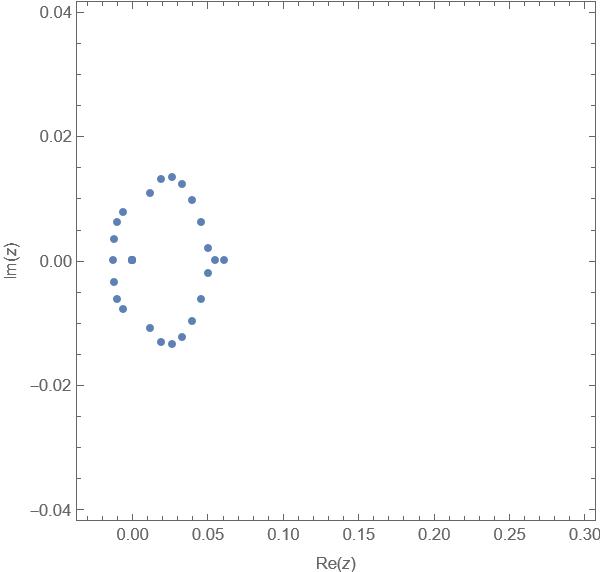}
        \caption{$\omega=2392.0027$}
        \label{fA omega=2392.0027}
    \end{subfigure}
        \hfill
    \begin{subfigure}{0.24\textwidth}
        \centering
        \includegraphics[width=\linewidth]{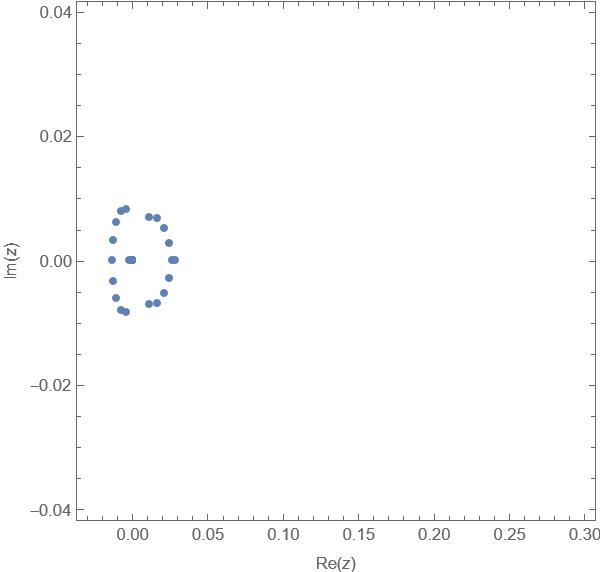}
        \caption{$\omega=5041.7783$}
        \label{fA omega=5041.7783}
    \end{subfigure}
        \hfill
    \begin{subfigure}{0.24\textwidth}
        \centering
        \includegraphics[width=\linewidth]{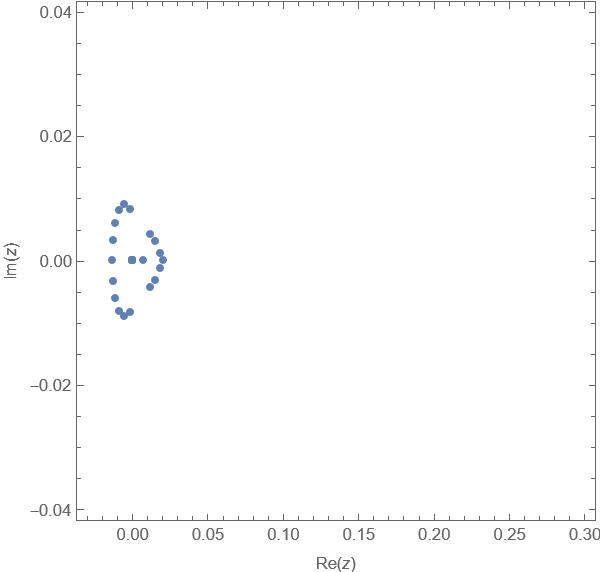}
        \caption{$\omega=6883.0782$}
        \label{fA omega=6883.0782}
    \end{subfigure}
        \hfill
    \begin{subfigure}{0.24\textwidth}
        \centering
        \includegraphics[width=\linewidth]{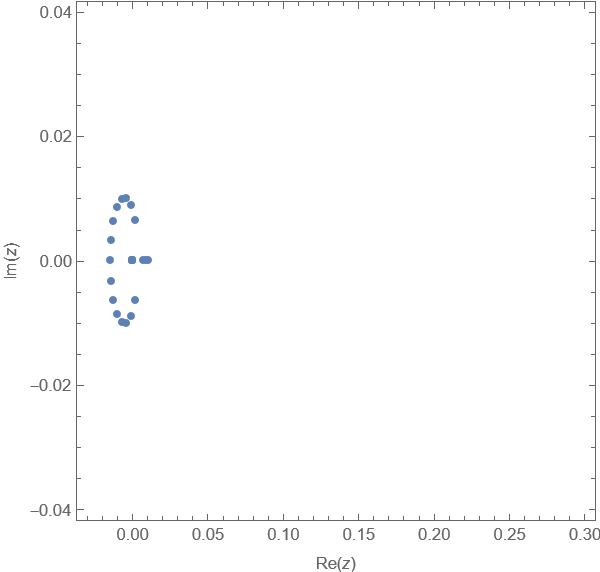}
        \caption{$\omega=11749.1839$}
        \label{fA omega=11749.1839}
    \end{subfigure}
    \caption{Involution of root distributions with different values of $\omega$.}
    \label{fA omega}
\end{figure}

We take $n=30$ and fix the values of the parameters $e, d$ to be $e=10, d=0.1$. From the $sl(2)$ algebraization of the model, there are 31 linear independent sets of roots for the BAEs, corresponding to 31 linear independent sets of coefficients $a_0, \ldots, a_{19}$ and 21 different values of the parameter $\omega$. Rearranging the values of $\omega$ from the smallest to the greatest, equivalently energy from the lowest to the highest, the distribution of roots of the BAEs changes with the value of the $\omega$, shown in Fig.\ref{fA omega}. The root distributions are shown in the complex plane. 
The caption shows the corresponding value of $\omega$ for each figure. As indicated by these figures, most roots appear in a certain area and become closer and closer with higher energies.  From (\ref{A En}), the energy depends on $\omega$ with fixed $e, d$. The roots distribute separately with lower energy, and closer with higher energy. The shape of roots distribution changes a lot when $\omega$ changes from 581.0744 to 883.702. The circle consisting of part of the roots enlarges and moves to the back, which may contain a phase changing process. The energy with $\omega=883.7025$ can be seen as a critical point.

\begin{figure}[ht]
    \centering
    \begin{subfigure}{0.31\textwidth}
        \centering
        \includegraphics[width=\linewidth]{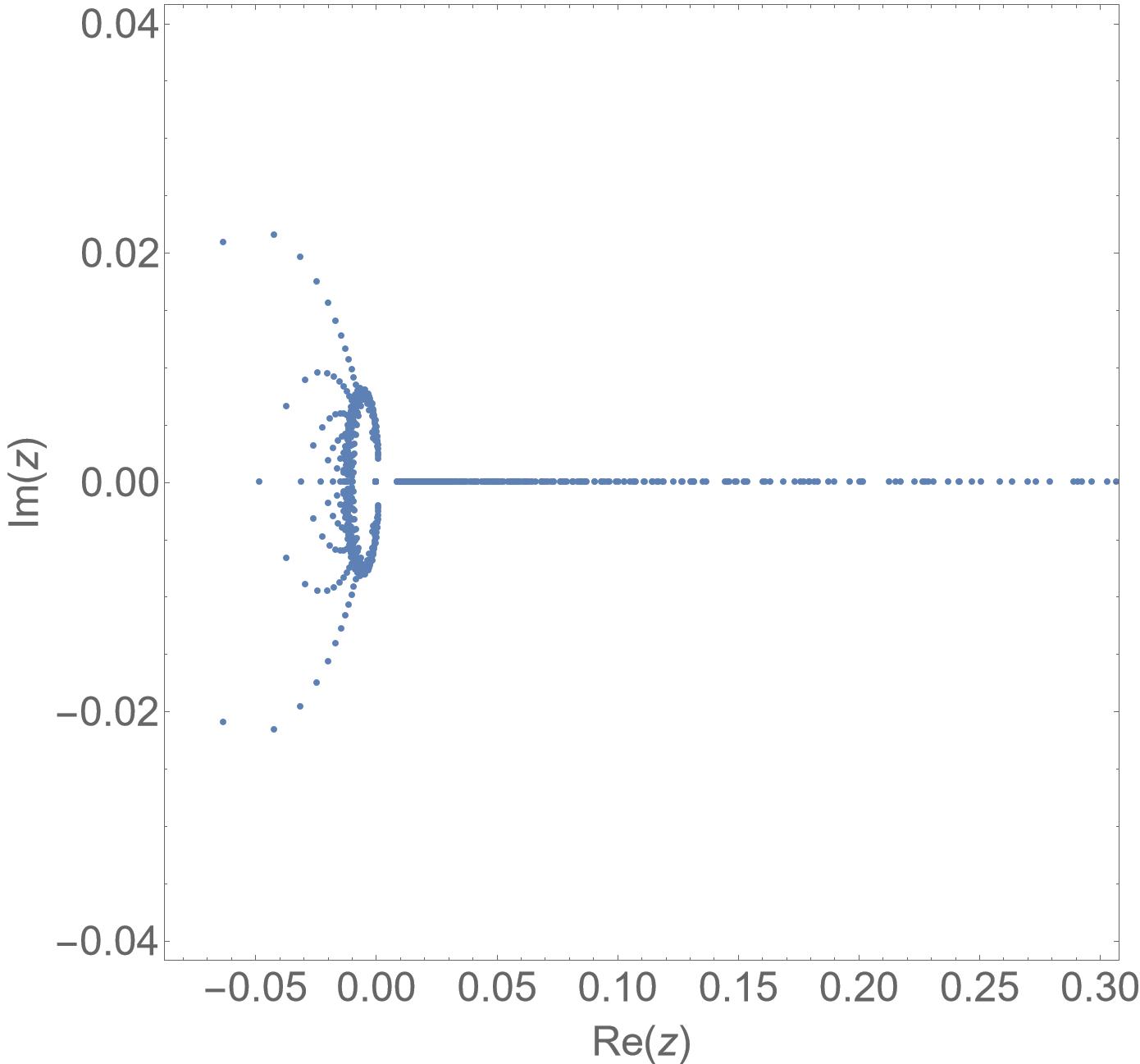}
        \caption{$e=0.1, d=0.1$}
        \label{fA e=0.1 d=0.1}
    \end{subfigure}
    \hfill
        \begin{subfigure}{0.31\textwidth}
        \centering
        \includegraphics[width=\linewidth]{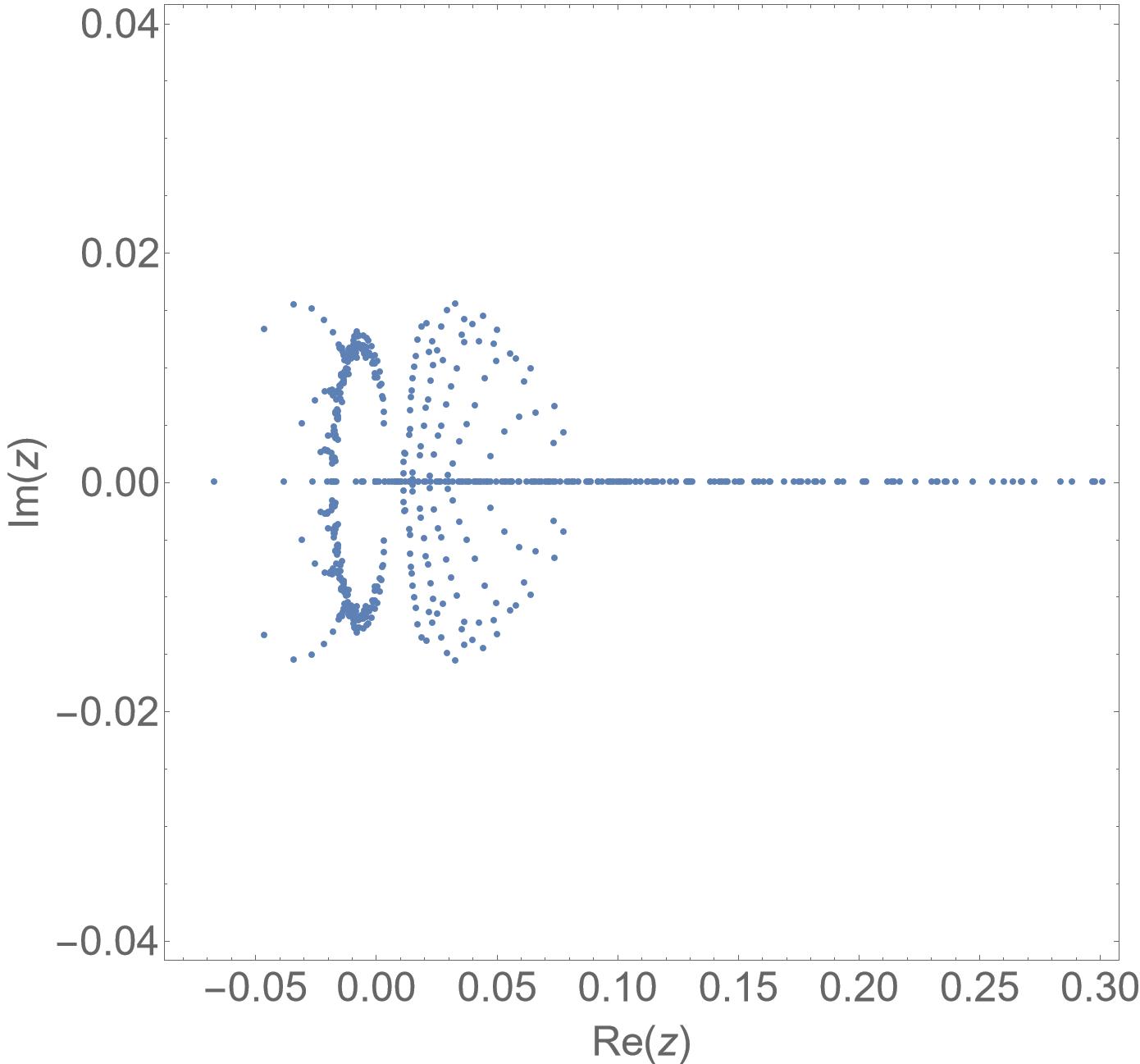}
        \caption{ $e=1, d=0.1$}
        \label{fA e=1 d=0.1}
    \end{subfigure}
    \hfill
        \begin{subfigure}{0.31\textwidth}
        \centering
        \includegraphics[width=\linewidth]{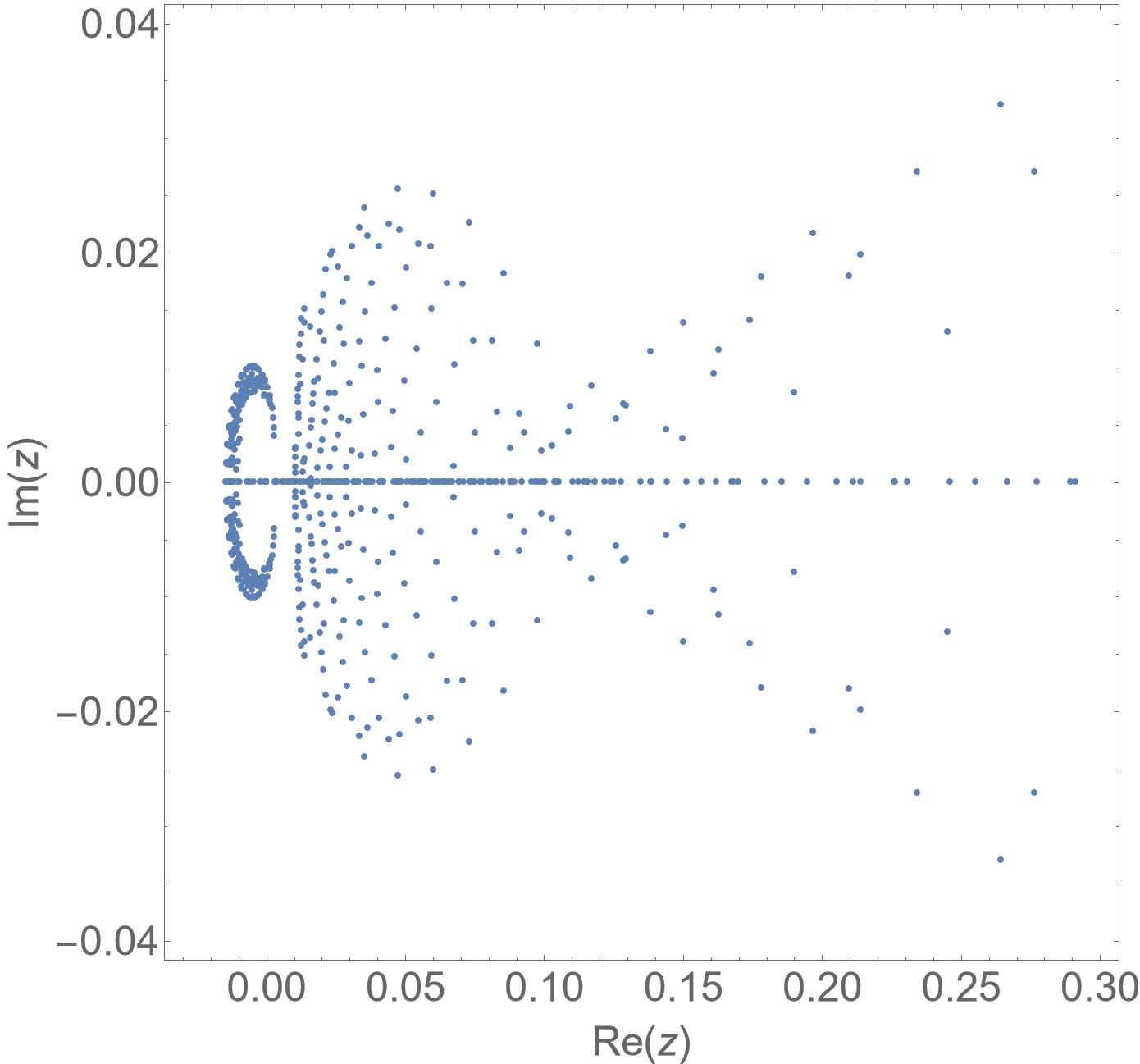}
        \caption{$e=10, d=0.1$}
        \label{fA e=10 d=0.1}
    \end{subfigure}
    \caption{Root distributions with fixed $d$ but increase of $e$  for $n=30$. }
    \label{fA d=0.1 n=30}
\end{figure}

All the roots corresponding to different $\omega$ values are the roots of the BAEs for $n=30$. Changing the value of $e$, we can find how all the roots of the BAEs change. Most of them are shown in Fig.\ref{fA d=0.1 n=30}, each of them corresponds to a different value of $e$ from 0.1 to 10. With $e$ increase, the curve on the left becomes smaller. Most roots move to the right side, forming many little curves and presenting a scattering shape. The root distributions are symmetric about the real axis.

\section{Generalized quantum isotonic oscillator}

Generalized quantum isotonic oscillators appeared in different contexts \cite{Car2008,Fel2009,Ses2010,hall2010}. In particular, they are connected with supersymmetric quantum mechanics \cite{Fel2009}. In this section we consider a quasi-exactly solvable extension with general potential of the form $\displaystyle V(x)=\omega^2 x^2 +2\beta(\beta-1) \frac{x^2-a^2}{(x^2+a^2)^2}$, where $\omega>0, \beta$ are real parameters. Exact solutions of this model were studied in \cite{Ses2010} on a case by case basis by means of the recursion relations. Moreover solutions in the even sector were derived in \cite{Agb2012}. We present a unified treatment of the model in even and odd sectors.

The Hamiltonian of the model reads 
\begin{align}
    \label{B H(x)}
        H(x)=-\frac{d^2}{dx^2}+\omega^2 x^2 +2\beta(\beta-1) \frac{x^2-a^2}{(x^2+a^2)^2}.
\end{align}
The time-independent Schr\"odinger equation is 
\begin{equation}\label{B Schrodinger equation}
    \frac{d^2 \Psi (x)}{d x^2}-\left[\omega^2 x^2 +2\beta(\beta-1) \frac{x^2-a^2}{(x^2+a^2)^2}\right] \Psi(x)+{E} \Psi(x)=0,
\end{equation}
Let 
\begin{equation}
   \Psi_0(x)=(x^2+a^2)^\beta\,\exp\left[-\frac{\omega x^2}{2}\right]. 
\end{equation}
We gauge-transform $H$ by means of $\Psi_0(x)$ to eliminate terms proportional to $\displaystyle \frac{1}{(x^2+a^2)^2}$ in the transformed Hamiltonian. This is equivalent to assuming that $\Psi(x)$ is factorized to the form $\Psi(x)=\Psi_0(x)\,y(x)$. So we set
\begin{align}\label{B Psi(x)}
        \Psi(x)=(x^2+a^2)^\beta \exp \left[-\frac{\omega x^2}{2}\right] y(x).
\end{align}
Substituting it into (\ref{B Schrodinger equation}), we obtain
\begin{align}\label{B Schrodinger y(x)}
    (a^2+x^2)y''(x)-2x\left[\omega (a^2+x^2)-2\beta\right] y'(x)+[2\beta^2+({E}-\omega)(x^2+a^2)-4\omega \beta x^2]y(x)=0.
\end{align}

To get odd and even solutions of the model in unified way, we set 
\begin{align}\label{B y(x)}
        y(x)=x^p \psi(x),\qquad p=0,1,
    \end{align}
where $p=0,1$ correspond to even and odd solutions, respectively.
Substituting (\ref{B y(x)}) into (\ref{B Schrodinger y(x)}) gives the ODE
\begin{equation}\label{B psi(x)}
    \begin{split}
        \left( a^2+x^2 \right) \psi''(x) & +\left[\frac{2p(x^2+a^2)}{x}+2\beta x-2\omega x (x^2+a^2)\right] \psi'(x)\\
        &+\left\{p\left[4\beta-2\omega (x^2+a^2)\right]+2\beta^2-4\omega\beta x^2+(E-\omega)(x^2+a^2)\right\}\psi(x)=0.
    \end{split}
\end{equation}
Making the change of variable, $\xi=x^2+a^2$, we have
\begin{equation}\label{B psi(xi)}
    \begin{split}
        4\xi(\xi-a^2)\psi''(\xi) & +\left[-4\omega \xi^2+2(1+2p+4\beta+2 \omega a^2)\xi-8\beta a^2\right] \psi'(\xi)\\
        &+\left\{[{E}-\omega(1+2p+4\beta)]\xi+2\beta(\beta+2p+2 \omega a^2)\right\}\psi(\xi)=0,
    \end{split}
\end{equation}
which can be recast in the form of the confluent Heun equation
\begin{equation}\label{B confluent Heun}
    \begin{split}
        \psi''(\xi) +\left[-\omega +\frac{2\beta}{\xi}+\frac{p+1/2}{\xi-a^2}\right] \psi'(\xi)
        +\frac{[{E}-\omega(1+2p+4\beta)]\xi+2\beta(\beta+2p+2 \omega a^2)}{4\xi(\xi-a^2)}\psi(\xi)=0.
    \end{split}
\end{equation} 

The above ODE is quasi-exactly solvable provided that the model parameters obey certain constraint and exact solutions are given by the degree $n$ polynomials
\begin{equation}\label{B polynomial psi}
    \psi(\xi)=\prod_{i=1}^n(\xi-\xi_i),\qquad \psi(\xi)\equiv 1~{\rm for}~n=0.
\end{equation}
Indeed, applying the Bethe ansatz method \cite{Zha2012}, (\ref{c2}), (\ref{general Bethe ansatz}) and (\ref{c1}), we get the energies and the constraint for the model parameters, 
\begin{align}
    &\begin{aligned}\label{B En}
        E^{(p)}_n=\omega(1+4n+2p+4\beta),
    \end{aligned}\\
       &\begin{aligned}\label{B parameter constrain n}
        \beta(\beta+2p+2\omega a^2) -2\omega\sum_{i=1}^n \xi_i+2n(n-1)+n(1+2p+4\beta-2\omega a^2)=0,
    \end{aligned}
\end{align}
where $n=0,1,2,\ldots $ and $\xi_1,\xi_2,\dots \xi_n$ satisfy the BAEs
\begin{equation}\label{B roots of Bethe ansatz}
    \sum_{j \neq i}^n \frac{2}{\xi_i-\xi_j}+\frac{-2\omega \xi_i^2+(1+2p+4\beta+2\omega a^2) \xi_i-4\beta a^2}{2 \xi_i^2-2a^2 \xi_i}=0,\quad i=1,2,\ldots,n.
\end{equation}
The corresponding closed-form expressions for the wave functions of (\ref{B Schrodinger equation}) in the even and odd sectors are given by
\begin{align}\label{B Psin}
        \Psi^{(p)}_n(x)=(x^2+a^2)^\beta \exp \left[-\frac{\omega x^2}{2}\right]\, x^p\, \left[\prod_{i=1}^n(x^2-\xi_i+a^2) \right].
\end{align}

In the following, we show that the ODE (\ref{B psi(xi)}) has a hidden $sl(2)$ algebra symmetry. To this end, we rewrite (\ref{B psi(xi)}) as 
\begin{equation}\label{calH B}
    \begin{split}
       &\mathcal{H}\psi(\xi)=\varepsilon\psi(\xi),\qquad
          \varepsilon=-2\beta(\beta+2p++2 \omega a^2),\\
        &\mathcal{H}=4\xi(\xi-a^2)\frac{d^2}{d\xi^2} +\left[-4\omega \xi^2+2(1+2p+4\beta+2 \omega a^2)\xi-8\beta a^2\right] \frac{d}{d\xi}
        +[{E}-\omega(1+2p+4\beta)]\xi.
    \end{split}
\end{equation} 
Applying theorem A2 (\ref{general algebraization}), the Hamiltonian $\mathcal{H}$ (\ref{calH B}) is a polynomial of the $sl(2)$ generators (\ref{sl(2) operator}) if
\begin{equation}
   {E}-\omega(1+2p+4\beta)\equiv c_1=-n[(n-1)a_3+b_2]\equiv 4n \omega, 
\end{equation}
which gives the energies (\ref{B En}) above, i.e. $E=(1+4n+2p+4\beta)\omega$.  For such $E$ values, $\mathcal{H}$ (\ref{calH B}) depends on integer $n$ and is given in terms of the $sl(2)$ generators (\ref{sl(2) operator}) by
\begin{equation}\label{B algebraic sl(2)}
    \begin{split}
    \mathcal{H}=&4 J^0 J^0-4a^2 J^0 J^- +4\omega J^+ + (4n+4p+8\beta+4a^2 \omega-2) J^0\\
    &-2a^2(n+4\beta) J^- +n(n+2p+4\beta+2a^2\omega-1). 
    \end{split}
\end{equation}
This provide an unified $sl(2)$ algebraization of the generalized quantum isotonic oscillator in the even and odd sectors. This algebraization is new. The eigenvalues $\varepsilon$ of $\mathcal{H}$ with polynomial eigenfunctions (\ref{B polynomial psi}) are given by
\begin{equation}
    \varepsilon=4n(n-1)+2n(1+2p+4\beta+2\omega a^2)-4\omega\sum_{i=1}^n \xi_i,
\end{equation}
which coincides with the constraint (\ref{B parameter constrain n}), and the roots $\xi_i$ are determined by the BAEs (\ref{B roots of Bethe ansatz}).

We now give the explicit results for the first few solutions.

When $n=0$, we have $\psi(\xi)=1$. The constraint in this case is $\beta=-2p-2\omega a^2$. The corresponding energies and wave functions in even and odd sectors are
\begin{equation}
    E_0^{(p)}=(4\beta+2p+1)\omega,\qquad \Psi_0^{(p)}(x)=\frac{x^p}{(x^2+a^2)^{2p+2\omega a^2}}\exp \left[-\frac{\omega x^2}{2}\right].
\end{equation}
$\Psi_0^{(0)}(x)$ is even and is the ground level wave function, and $\Psi_0^{(1)}(x)$ is odd and represents the first excited wave function.

For $n=1$, the Bethe ansatz equation (\ref{B roots of Bethe ansatz}) has solutions 
\begin{equation}\label{B xi1 p}
 \xi^{(p)}_{1\pm}=\frac{1}{4\omega}\left[2p+1+4\beta+2\omega a^2\pm\sqrt{(2p+1+4\beta)^2+4(2p+1-4\beta+\omega a^2)\omega a^2}\right].
\end{equation}
Substituting (\ref{B xi1 p}) into (\ref{B parameter constrain n}), we have the constraint for the model parameters
\begin{equation}\label{B parameter constrain 1 p}
\begin{split}
  \omega a^2={-\frac{2\beta^2+(5+4p)(\beta+1) \pm \sqrt{25\beta^2+(32p+46)\beta+(25+32p)}}{4(1+\beta)}}.
\end{split}
\end{equation}
The corresponding energies and wave functions in even and odd sectors are 
\begin{align}
    &\begin{aligned}\label{B E1 p}
     E_1^{(p)}=\omega(2p+5+4\beta),
    \end{aligned}\\
   &\begin{aligned}\label{B Psi 1 p}
       \Psi^{(p)}_{1\pm}(x)= (x^2+a^2)^\beta\exp \left[-\frac{\omega x^2}{2}\right]\,x^p\, (x^2-\xi^{(p)}_{1\pm}+a^2).
   \end{aligned}
\end{align}
$\Psi^{(0)}_{1\pm}(x)$ and $\Psi^{(1)}_{1\pm}(x)$ represent the first even and odd excited wave functions, respectively. We remark that associated with one energy level $E_1^{(0)}$ (resp. $E_1^{(1)}$) there are two even wave functions $\Psi^{(0)}_{1\pm}(x)$ (resp. two odd wave functions $\Psi^{(1)}_{1\pm}(x)$). This is in agreement with the prediction of the hidden $sl(2)$ symmetry.

\subsection{Roots and allowed model parameters for $n=2, 3$}

\begin{figure}[ht]
    \centering
    \begin{subfigure}{0.48\textwidth}
        \centering
        \includegraphics[width=\linewidth]{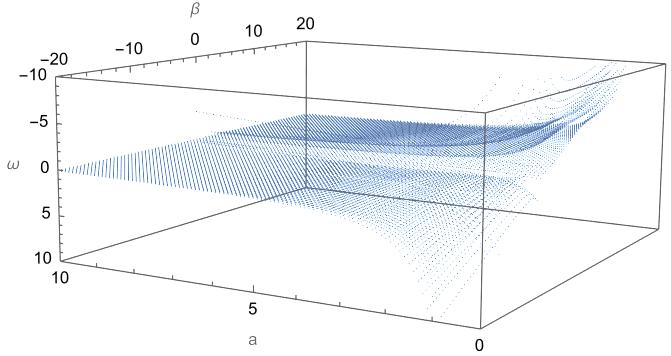}
        \caption{Relationship in even space.}
        \label{fB n2 p0}
    \end{subfigure}
    \hfill
    \begin{subfigure}{0.48\textwidth}
        \centering
        \includegraphics[width=\linewidth]{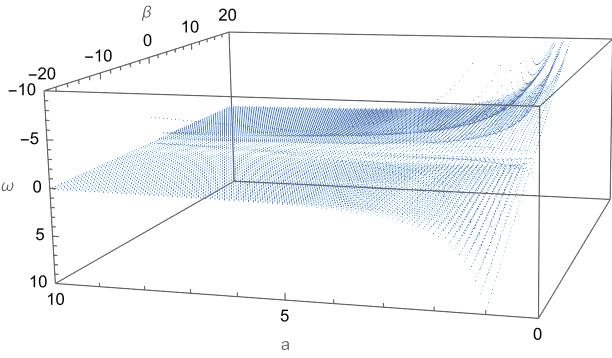}
        \caption{Relationship in odd space.}
        \label{fB n2 p1}
    \end{subfigure}
    \caption{Relationship between $\omega, a, \beta$ and solutions of the BAEs for $n=2$.}
    \label{fB n2}
\end{figure}

For $n=2$, the energies (in the even and odd sectors) and the parameter constraint are
\begin{align}
&\begin{aligned}\label{B E2}
        E_2^{(p)}=\omega(9+2p+4\beta),
    \end{aligned}\\
&\begin{aligned}
 \label{B parameter constrain 2}
        \beta^2+(8+2p)\beta+2a^2\omega(\beta-2)-2\omega(\xi_1+\xi_2)+4+2(1+2p)=0  
\end{aligned}     
\end{align} 
We solve the roots $\xi_1^{(p)}, \xi_2^{(p)}$ of \ref{B roots of Bethe ansatz}  and the parameter constraint (\ref{B parameter constrain 2})
simultaneously via numerical techniques. We obtain the roots and the values of the allowed model parameters, as shown Fig.\ref{fB n2}. The figure shows the relationship between the parameters $a, \beta$ and $\omega$ for which the BAEs have solutions. Fig.\ref{fB n2 p0} corresponds to even ($p=0$) solutions  and Fig.\ref{fB n2 p1} to odd ($p=1$) solutions. The ranges of $a$ and $\beta$ are from 0 to 10 and from -20 to 20, respectively. 
As expected from the hidden $sl(2)$ algebraic structure of the model, three different sets of values for the parameter $\omega$ and the roots of the BAEs are obtained, corresponding to the 3 layers in the figure. We remark that the value of $\beta$ near -2, only 1 layer can be shown in a figure. This is because the $\omega$ values corresponding to two sets of roots of the BAEs become complex numbers, leading to complex energies. So these roots are discarded,  leading to only one layer in the figure for the value of $\beta$ near -2. 
The work precision for numerical calculation is 20, however, we reserve 4 decimal places in this paper for easy reading. From the tables, there are $n+1$ number of roots for each parameter, satisfying the algebraic structure. For value of $\beta$ near -2, two of three independent roots of BAEs will lead to $\omega$ be a complex number. Also, as seen from Fig.\ref{fB n2 p0} and Fig.\ref{fB n2 p1}, parameter space figures for the even and odd sectors are similar. 

\begin{figure}[ht]
    \centering

    \begin{subfigure}{0.48\textwidth}
        \centering
        \includegraphics[width=\linewidth]{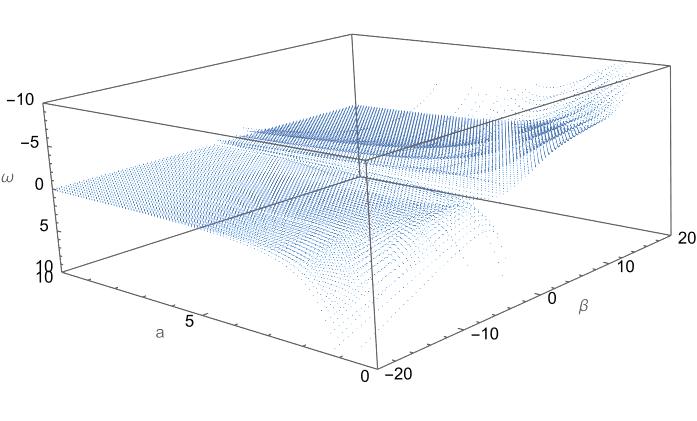}
        \caption{Relationship in even space.}
        \label{fB n3 p0}
    \end{subfigure}
    \hfill
    \begin{subfigure}{0.48\textwidth}
        \centering
        \includegraphics[width=\linewidth]{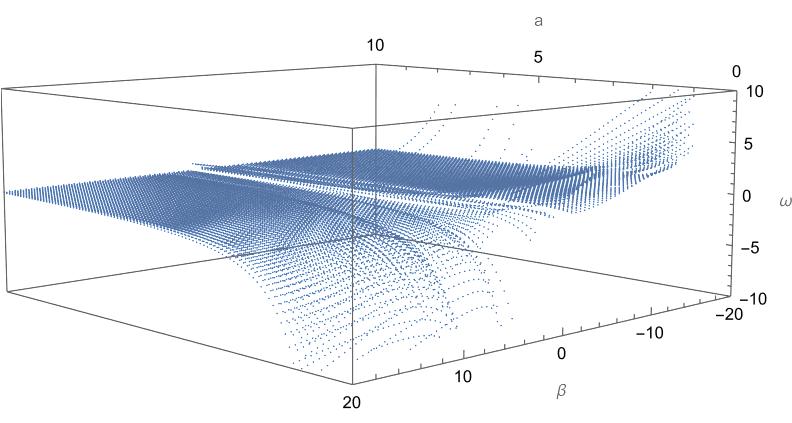}
        \caption{Relationship in odd space.}
        \label{fB n3 p1}
    \end{subfigure}
    \caption{Relationship between $\omega, a, \beta$ and solutions of the BAEs for $n=3$.}
    \label{fB n3}
\end{figure}

For $n=3$, the energies and the constraint for the model parameters are
\begin{align}
    &\begin{aligned}\label{B E3}
        E_3^{(p)}=\omega (13+2p+4 \beta),
    \end{aligned}\\
    &\begin{aligned}\label{B parameter constrain 3}
        \beta^2+(12+2p)\beta+2\omega a^2(\beta-3)-2\omega(\xi_1+\xi_2+\xi_3)+12+3(1+2p)=0,
    \end{aligned}
\end{align}
Similar to the $n=2$ case, we numerically solve the BAEs (\ref{B roots of Bethe ansatz}) and the constraint (\ref{B parameter constrain 3}) simultaneously. Fig.\ref{fB n3} shows the relationship between parameters $a, \beta$ and $\omega$ for which the BAEs have solutions. Fig.\ref{fB n2 p0} is for the even solutions and Fig.\ref{fB n2 p1} is for the odd solutions. The method of plotting the figures is similar to that of plotting Fig.\ref{fB n2}. When the value of $\beta$ near to -2, the parameter $\omega$ become complex for 3 different sets of roots of the BAEs, 
implying that there is only one layer when $\beta$ is near to -2. For other values, there are 4 layers in the figure, in agreement with the prediction from the hidden $sl(2)$ algebraic structure. 

\subsection{Root distributions for higher level wave functions}
\begin{figure}[ht]
    \centering

    \begin{subfigure}{0.24\textwidth}
        \centering
        \includegraphics[width=\linewidth]{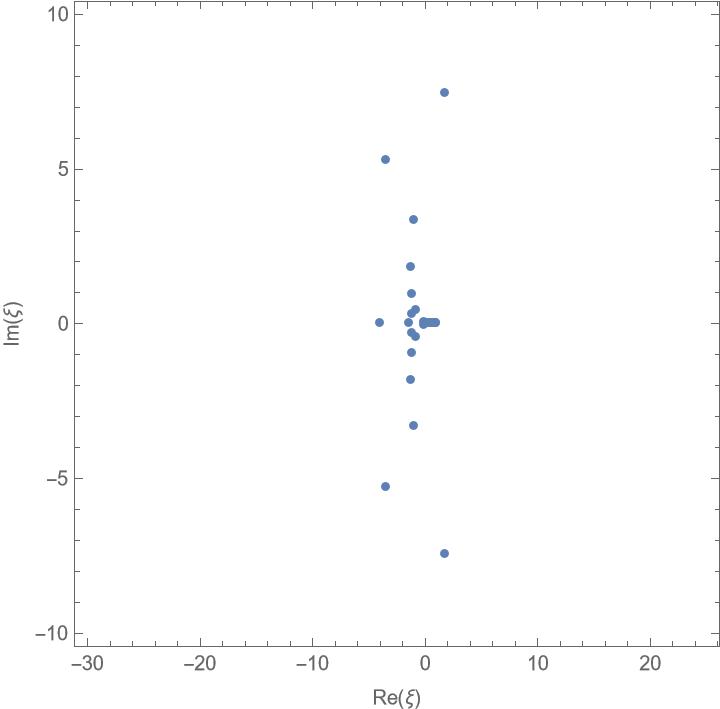}
        \caption{$\omega=-7.1370$}
        \label{fB omega=-7.1370 p1}
    \end{subfigure}
    \hfill
    \begin{subfigure}{0.24\textwidth}
        \centering
        \includegraphics[width=\linewidth]{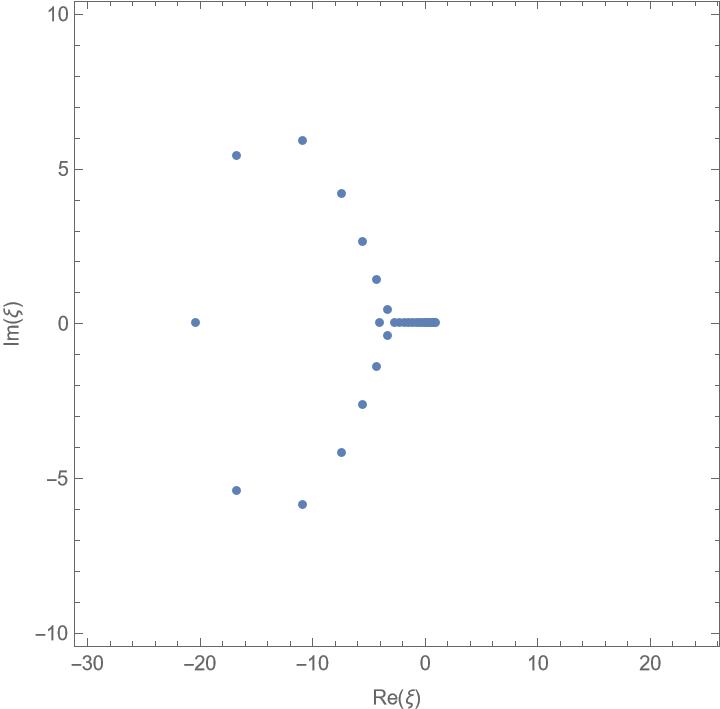}
        \caption{$\omega=-5.8792$}
        \label{fB omega=-5.8792 p1}
    \end{subfigure}
    \hfill
    \begin{subfigure}{0.24\textwidth}
        \centering
        \includegraphics[width=\linewidth]{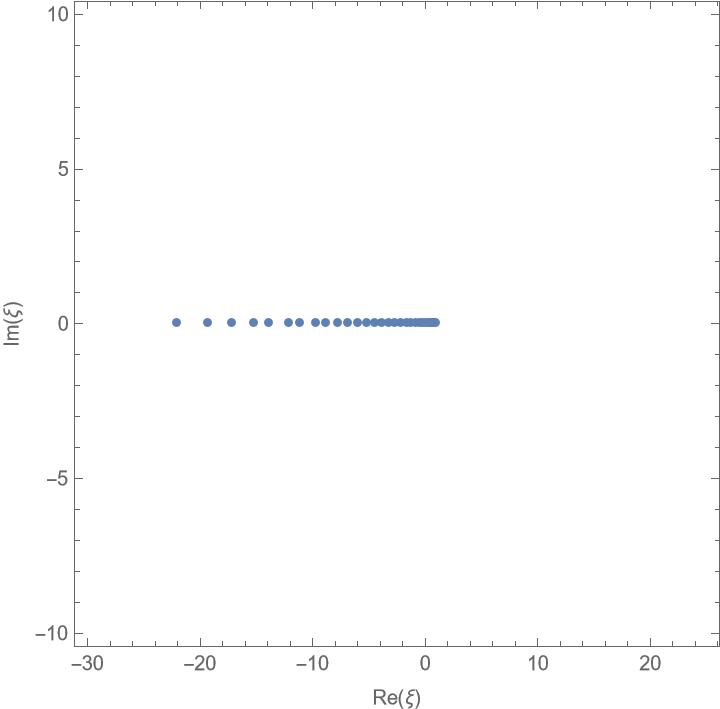}
        \caption{$\omega=-4.5587$}
        \label{fB omega=-4.5587 p1}
    \end{subfigure}
        \hfill
    \begin{subfigure}{0.24\textwidth}
        \centering
        \includegraphics[width=\linewidth]{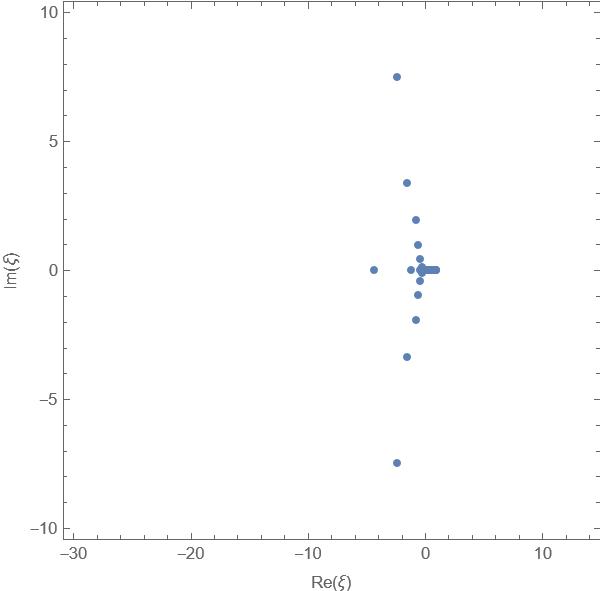}
        \caption{$\omega=-3.4680$}
        \label{fB omega=-3.4680 p1}
    \end{subfigure}
        \hfill
    \begin{subfigure}{0.24\textwidth}
        \centering
        \includegraphics[width=\linewidth]{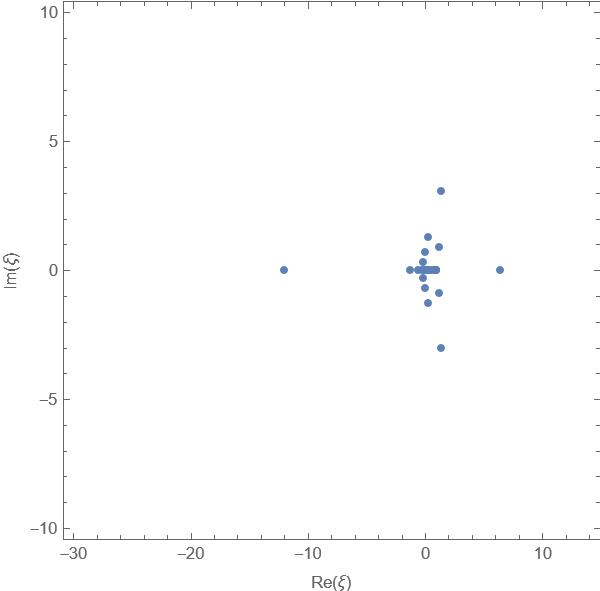}
        \caption{$\omega=-3.0494$}
        \label{fB omega=-3.0494 p1}
    \end{subfigure}
        \hfill
    \begin{subfigure}{0.24\textwidth}
        \centering
        \includegraphics[width=\linewidth]{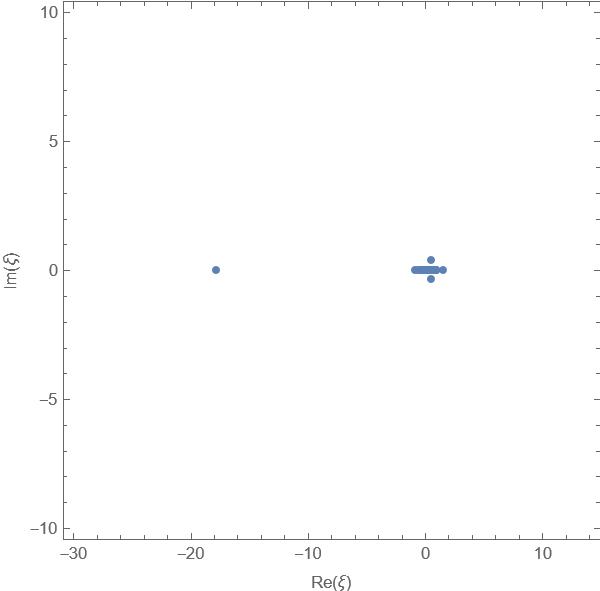}
        \caption{$\omega=-2.5398$}
        \label{fB omega=-2.5398 p1}
    \end{subfigure}
    \hfill
    \begin{subfigure}{0.24\textwidth}
        \centering
        \includegraphics[width=\linewidth]{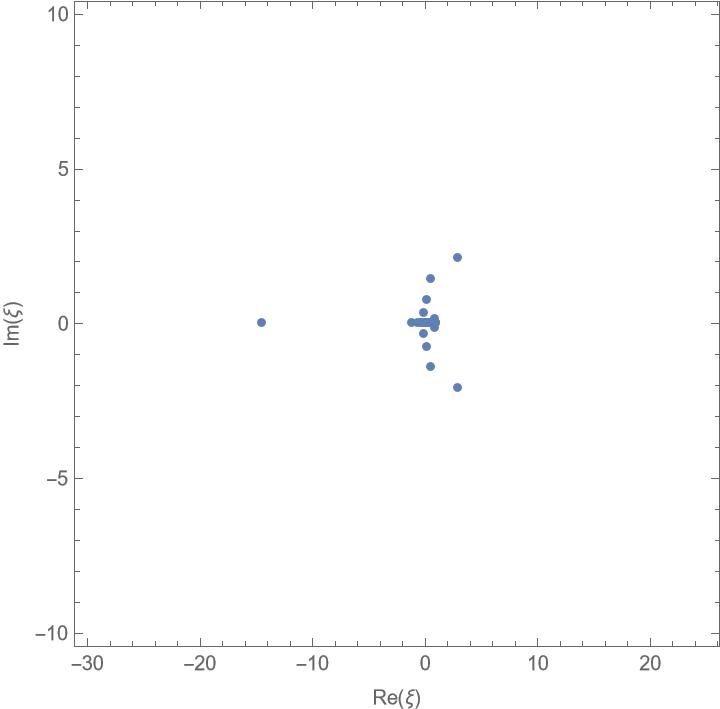}
        \caption{$\omega=1.1753 \times 10^{-18}$}
        \label{fB omega=1.1753 p1}
    \end{subfigure}
        \hfill
    \begin{subfigure}{0.24\textwidth}
        \centering
        \includegraphics[width=\linewidth]{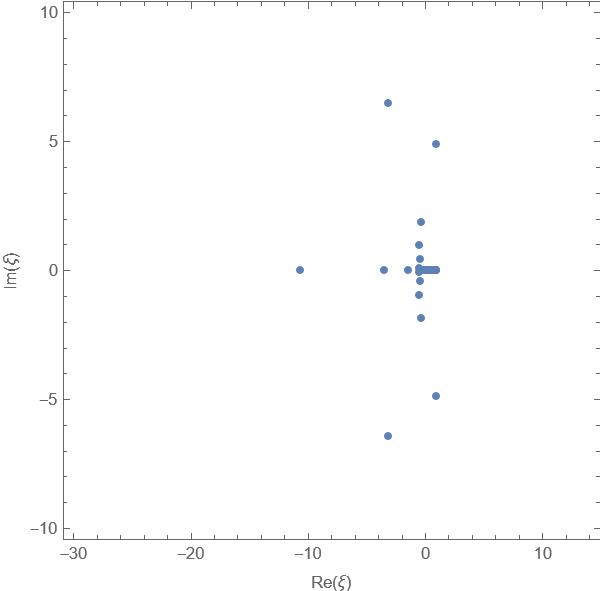}
        \caption{$\omega=9.4780 \times 10^{-10}$}
        \label{fB omega=9.4780 p1}
    \end{subfigure}
    \caption{Change root distributions with $\omega$ for odd solutions.}
    \label{fB omega p=1}
\end{figure}

We use the procedure similar to that in subsection 2.2. We take $n=30$ and  
solve the corresponding bilinear matrix equation for the coefficients and the parameter $\omega$. As examples, the root distributions for $p=1$ with varying values of $\omega$ and fixed values of $\alpha=1, \beta=-0.025$ are shown in Fig.\ref{fB omega p=1}.  
The roots distribution is very different with different values of $\omega$. It might contain many critical points and the phase changing always happens. For $\omega<0$, the roots distribute discretely and become closer, and they will separate again with $\omega>0$. 

\begin{figure}[ht]
    \centering

    \begin{subfigure}{0.31\textwidth}
        \centering
        \includegraphics[width=\linewidth]{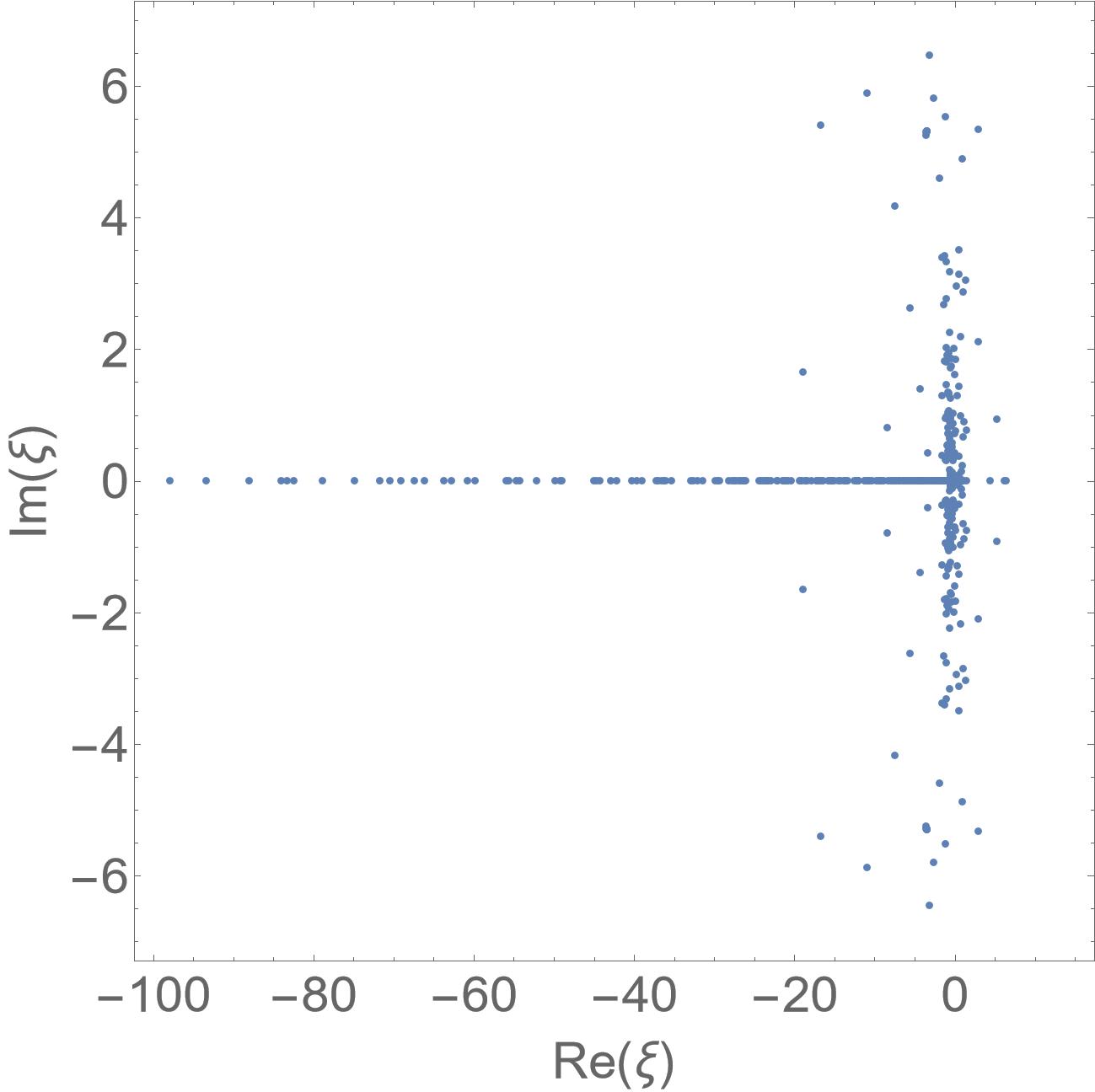}
        \caption{$a=1, \beta=-0.025$}
        \label{fB a=1 beta=-0.025}
    \end{subfigure}
    \hfill
     \begin{subfigure}{0.31\textwidth}
        \centering
        \includegraphics[width=\linewidth]{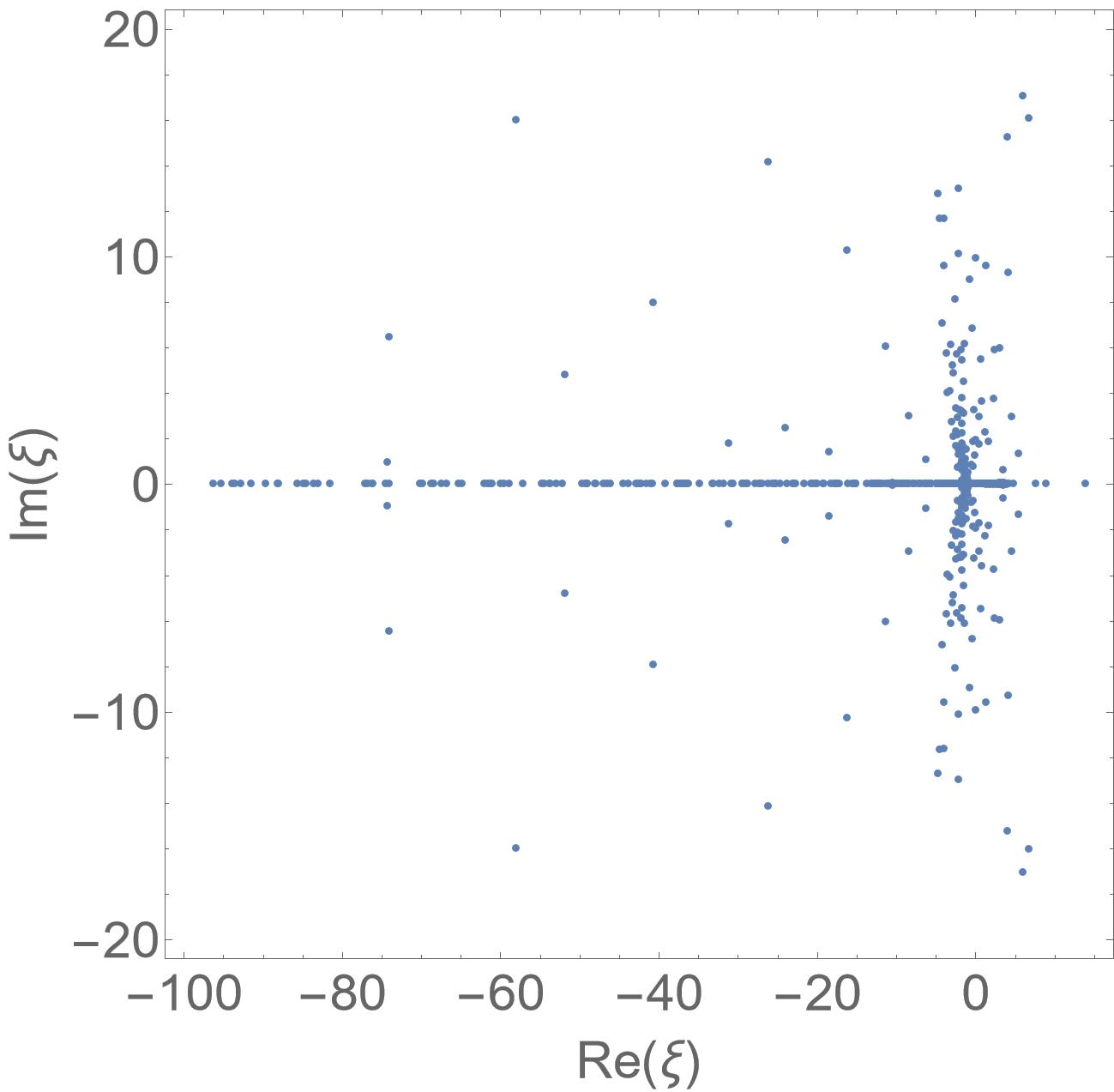}
        \caption{$a=2,\beta=-0.025$}
        \label{fB a=1 beta=-0.025 }
    \end{subfigure}
    \hfill
     \begin{subfigure}{0.31\textwidth}
        \centering
        \includegraphics[width=\linewidth]{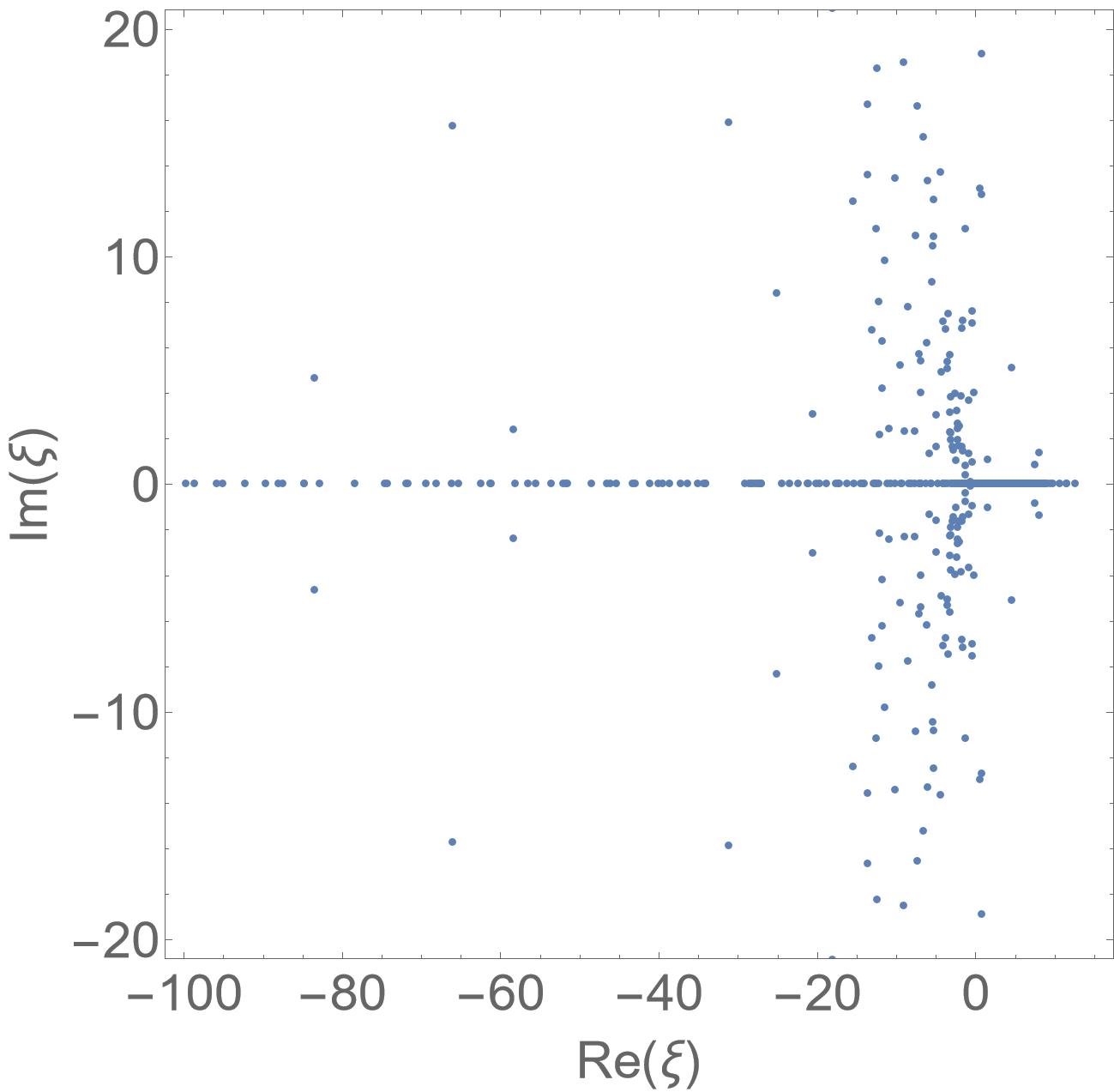}
        \caption{$a=3,\beta=-0.025$}
        \label{fB a=3 beta=-0.025}
    \end{subfigure}
    \hfill
    \caption{Change of the root distributions with the model parameters for $n=30$.}
    \label{fB beta=-0.025 n=30}
\end{figure}

Fig.\ref{fB beta=-0.025 n=30} are the distributions of the roots of the BAEs for $n=30$. The roots are in the real axis for $\xi<0$, and they distribute at the imaginary axis for $\xi \geq 0$. Increasing the value of $\alpha$, The roots at the imaginary axis begin to move from the right side to the left. They are symmetric about the real axis.

\newpage
\section{Non-polynomially modified oscillator}
The non-polynomially modified oscillator with potential of the form  $\displaystyle V(x)=\omega^2 x^2+2 \lambda \frac{x^2}{1+\delta x^2}$ was introduced in \cite{risken1967,mitra1978,bessis1980}.  Solutions to this model were examined in \cite{whitehead1982,marcilhacy1985} on a case by case basis. The even sector of the model was studied in \cite{Agb2012} from the point of view of quasi-exact solvability. Here we provide derivation of generic closed-form solutions for constraints of the model parameters, the corresponding energies and wavefunctions in the even and odd sectors in a unified way.

The Hamiltonian of the model reads
\begin{align}
    \label{D H(x)}
        H(x)=-\frac{d^2}{dx^2}+\omega^2 x^2+2 \lambda \frac{x^2}{1+\delta x^2}
\end{align}
The corresponding time-independent
Schr\"odinger equation is 
\begin{equation}\label{D Schrodinger equation}
    \frac{d^2 \phi(x)}{dx^2}-\left[\omega^2 x^2+2 \lambda \frac{x^2}{1+\delta x^2}\right] \phi(x)+{E} \phi(x)=0.
\end{equation}
We carry out a similarity transformation $\phi_0^{-1}(x)H\phi_0(x)$, where
\begin{equation}
    \phi_0(x)=(1+\delta x^2)\exp\left[-\frac{\omega x^2}{2}\right].
\end{equation}
Equivalently, we set $\phi_(x)=\phi_0(x)\,y(x)$, i.e. we make the following transformation
\begin{align}\label{D f(x)}
        \phi(x)=(1+\delta x^2)\exp\left[-\frac{\omega x^2}{2}\right] y(x),
\end{align}
we have
\begin{align}\label{D Schrodinger y(x)}
    (1+\delta x^2)y''(x) & -2x\left[\omega (1+\delta x^2)-2\delta\right]y'(x)\nonumber\\
    & +\left[(E-\omega)(1+\delta x^2)+2\delta-2\lambda x^2-4\delta \omega x^2\right]y(x)=0.
\end{align}

To treat odd and even solutions in unified way, we set 
\begin{align}\label{D y(x)}
        y(x)=x^p \psi(x),\qquad p=0,1,
    \end{align}
where $p=0,1$ correspond to even and odd solutions, respectively.    
Substituting (\ref{D y(x)}) into (\ref{D Schrodinger y(x)}) gives 
\begin{equation}
    \begin{split}\label{D ODE x}
    (1+\delta x^2)\psi''(x) & +\frac{2}{x}\left[(p-\omega x^2)(1+\delta x^2)+2\delta x^2\right]\psi'(x)\\
    &+\left[(E-(2p+1)\omega)(1+\delta x^2)+2(2p+1)\delta-\left(4 \omega+\frac{2\lambda}{\delta}\right)\delta x^2\right]\psi(x)=0.
    \end{split}
\end{equation}
Making the variable change, $\xi=1+\delta\, x^2$, we have
\begin{align}\label{D ODE xi}
    \xi(\xi-1) \psi''(\xi) & +\left[-\frac{\omega}{\delta} \xi^2+\xi \left(\frac{5}{2}+p+\frac{\omega}{\delta}\right) -2 \right] \psi'(\xi)\nonumber\\
    & +\left[ \frac{\xi}{4\delta} \left({E}-\frac{2\lambda}{ \delta}-(5+2p)\omega\right) +\frac{1}{2\delta}\left((2p+1)\delta+2\omega+\frac{\lambda}{\delta}\right) \right]\psi(\xi)=0,
\end{align}
which can be recast in the form of the confluent Heun equation
\begin{align}\label{D confluent Heun}
     \psi''(\xi) & + \left[-\frac{\omega}{\delta} +\frac{2}{\xi}+\frac{p+1/2}{\xi-1}\right] \psi'(\xi)\nonumber\\
     &+ \frac{\left({E}-\frac{2\lambda}{ \delta}-(5+2p)\omega\right) \xi+2(2p+1)\delta+4\omega+\frac{2\lambda}{\delta}}{4\delta\,\xi(\xi-1)}\psi(\xi)=0.
\end{align}

The above ODE is quasi-exactly solvable and exact solutions are given by polynomials of degree $n$ of the form
\begin{equation}\label{D polynomial psi(xi)}
    \psi(\xi)=\prod_{i=1}^n(\xi-\xi_i),\qquad \psi(\xi)\equiv 1~{\rm for}~n=0.
\end{equation}
Indeed, applying (\ref{c2}), (\ref{general Bethe ansatz}) and (\ref{c1}), we can show that (\ref{D polynomial psi(xi)}) is solution to (\ref{D ODE xi}) with energies 
\begin{align}
    \label{D En}
        E^{(p)}_n=\frac{2 \lambda}{\delta}+(5+4n+2p)\omega,\qquad n=0,1,2,\ldots, 
\end{align}
provided that the model parameters and the roots $z_i$ obey the constraint 
\begin{align}\label{D parameter constrain n}
      \frac{1}{2\delta}\left((2p+1)\delta+2\omega+\frac{\lambda}{\delta}\right)=\frac{\omega}{\delta}\sum_{i=1}^n \xi_i-n\left(n+p+\frac{3}{2}+\frac{\omega}{\delta}\right)
\end{align}
and the BAEs
\begin{equation} \label{D roots of the Bethe ansatz equation}
    \sum_{j\neq i}^n \frac{2}{\xi_i-\xi_j}+\frac{-{\omega} \xi_i^2+\left(\omega+(p+{5}/{2}){\delta}\right) \xi_i-2\delta}{\delta\, \xi_i (\xi_i-1)}=0,\quad i=1,2,\ldots,n.
\end{equation}
The corresponding wave functions in even and odd sectors are
\begin{align}\label{D fn(x)}
        \phi^{(p)}_n(x)=(1+\delta x^2)\exp \left[-\frac{\omega x^2}{2}\right] \,x^p\,\left[\prod_{i=1}^n (\delta x^2-\xi_i+1) \right].
    \end{align}

The ODE (\ref{D ODE xi}) which describes the even and odd sectors of the system in unified way has a hidden $sl(2)$ algebra symmetry. To demonstrate this, we rewrite (\ref{D ODE xi}) as
\begin{align}\label{calH D}
    &\mathcal{H}\psi(\xi)=\varepsilon\psi(\xi),\qquad \varepsilon=-\frac{1}{2\delta}\left((2p+1)\delta+2\omega+\frac{\lambda}{\delta}\right),\nonumber\\
    &\mathcal{H}=\xi(\xi-1) \frac{d^2}{d\xi^2} +\left[-\frac{\omega}{\delta} \xi^2+\xi \left(\frac{5}{2}+p+\frac{\omega}{\delta}\right) -2 \right] \frac{d}{d\xi}+ \frac{\xi}{4\delta} \left({E}-\frac{2\lambda}{ \delta}-(5+2p)\omega\right).
\end{align}
Applying theorem A2 (\ref{general algebraization}), the Hamiltonian (\ref{calH D}) has a hidden $sl(2)$ algebra symmetry if
\begin{equation}
    \frac{1}{4\delta} \left({E}-\frac{2\lambda}{ \delta}-(5+2p)\omega\right)\equiv c_1=-n[(n-1)a_3+b_2]\equiv \frac{\omega}{\delta}n,
\end{equation}
which gives the energies (\ref{D En}) above, i.e. $E=\frac{2 \lambda}{\delta}+(5+4n+2p)\omega$. For such $E$ values, (\ref{calH D}) depends on integer $n$ and 
is an element in the enveloping algebra of $sl(2)$ in terms of the differential operators (\ref{sl(2) operator}),
\begin{equation}
    \mathcal{H}=J^0 J^0-J^0 J^-+\frac{\omega}{\delta} J^+ +\left(\frac{3}{2}+n+p+\frac{\omega}{\delta}\right) J^0- \left(\frac{n}{2}+2\right) J^- +\frac{n}{4 \delta} \left[2\omega+(3+n+2p)\delta \right],
\end{equation}
This provides an unified $sl(2)$ algebraization of the non-polynomially modified oscillator in the even and odd sectors. This algebraization is one of the main new results in this section. The eigenvalues $\varepsilon$ of $\mathcal{H}$ with polynomial eigenfunctions (\ref{D polynomial psi(xi)}) are given by
\begin{equation}
    \varepsilon=n\left(n+p+\frac{3}{2}+\frac{\omega}{\delta}\right)-\frac{\omega}{\delta}\sum_{i=1}^n\xi_i,
\end{equation}
which is nothing but the constraint (\ref{D parameter constrain n}), and the roots $\xi_i$ are determined by the BAEs (\ref{D roots of the Bethe ansatz equation}).

We now give explicit expressions for the solutions of the first few energy levels.

When $n=0$, we have $\psi(\xi)=1$. The constraint for the model parameters in this case is $\displaystyle \frac{\lambda}{\delta}+(2p+1)\delta+2\omega=0$. The corresponding energies and wave functions in the even and odd sectors are given by
\begin{equation}
    E_0^{(p)}=\frac{2\lambda}{\delta}+(2p+5)\omega,\qquad \phi^{(p)}_0(x)=x^p\,(1+\delta x^2)\exp \left[-\frac{\omega x^2}{2}\right].
\end{equation}
$\phi^{(0)}_0(x)$ is even and is the ground level solution, while  $\phi^{(1)}_0(x)$ is odd and represents the first excited solution.

For $n=1$, the Bethe ansatz equation (\ref{D roots of the Bethe ansatz equation}) has the distinct roots $\xi_1$
\begin{equation}\label{D n1 p}
     \xi^{(p)}_{1\pm}=\frac{1}{4\omega}\left[2\omega+(2p+5)\delta \pm \sqrt{4\omega^2+4\omega(2p-3) \delta+(4p^2+20p+25)\delta^2}\right].
\end{equation}
Substituting this into the constraint (\ref{D parameter constrain n}), we obtain
\begin{equation}\label{D parameter constrain 1 p}
    \lambda=-\frac{1}{2}\delta \left((6p+7)\delta+6\omega \pm \sqrt{4\omega^2+4\omega(2p-3) \delta+(4p^2+20p+25)\delta^2}\right).
\end{equation}
The corresponding energies and wave functions in the even and odd sectors are 
\begin{align}
    &\begin{aligned}\label{D E1 p}
        E^{(p)}_1=\frac{2\lambda}{\delta}+(2p+9)\omega,
    \end{aligned}\\
    &\begin{aligned}\label{D f1 p0}
        &\phi^{(p)}_{1\pm}(x)=x^p\,(1+\delta x^2)\exp \left[-\frac{\omega x^2}{2}\right] (\delta x^2-\xi^{(p)}_{1\pm}+1).
    \end{aligned}
\end{align}
$\phi^{(0)}_{1\pm}(x)$ and $\phi^{(1)}_{1\pm}(x)$ give the even and odd next level solutions, respectively. Note that associated with one energy $E_1^{(0)}$ (resp. $E_1^{(1)}$) there are two even wave functions $\phi^{(0)}_{1\pm}(x)$ (resp. two odd wave functions $\phi^{(1)}_{1\pm}(x)$), as expected from the hidden $sl(2)$ symmetry of the model.

\subsection{Roots and allowed model parameters for $n=2,3$}

\begin{figure}[ht]
    \centering

    \begin{subfigure}{0.48\textwidth}
        \centering
        \includegraphics[width=\linewidth]{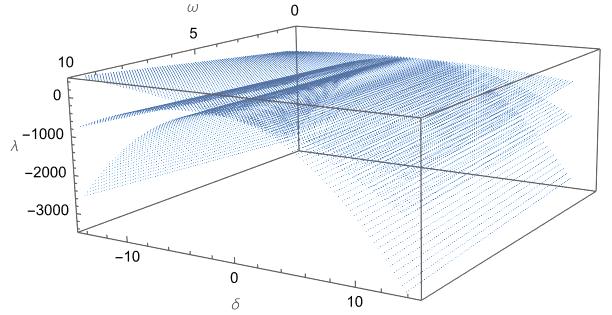}
        \caption{Relationship for $n=2$ in even space.}
        \label{fD n2 p0}
    \end{subfigure}
    \hfill
    \begin{subfigure}{0.48\textwidth}
        \centering
        \includegraphics[width=\linewidth]{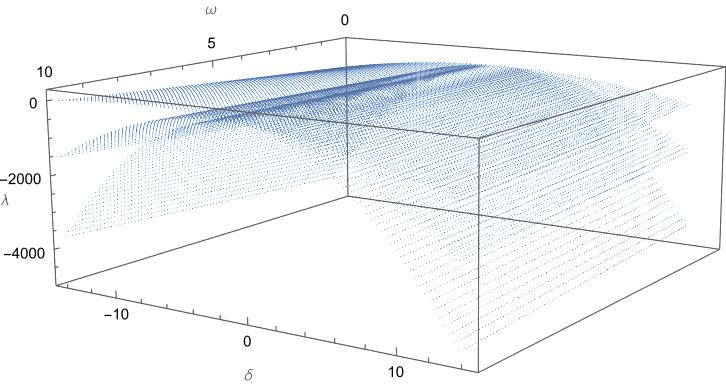}
        \caption{Relationship for $n=2$ in odd space.}
        \label{fD n2 p1}
    \end{subfigure}
    \caption{Relationship between $\omega, a, \beta$ and solutions of the BAEs for $n=3$.}
    \label{fD n2}
\end{figure}

For $n=2$,  the corresponding energies and parameter constraints are 
\begin{align}
    &\begin{aligned}\label{D E2}
         E_2^{(p)}=\frac{2 \lambda}{\delta}+(13+2p)\omega,
    \end{aligned}\\
    &\begin{aligned}\label{D parameter constrain 2}
        3\delta (5+2p)+\frac{\lambda}{\delta}+2\omega[3-(\xi_1+\xi_2)]=0,
    \end{aligned}
\end{align}
Similar to the treatment in the previous sections,
we search for the roots of (\ref{D roots of the Bethe ansatz equation}) via numerical methods. Solving the BAEs and the constraint equation simultaneously, we obtain the relationship between the parameters and the roots $\xi_1, \xi_2$ as shown in Fig.\ref{fD n2}. 
The root distributions are symmetry about $\delta=0$. 
From (\ref{D parameter constrain 2}), the constraint equation depends on the roots of the BAEs. Therefore, the number of the layers in the figure is the same as the number of the sets of roots of the BAEs. For $n=2$, there are 3 layers in the figures, which agrees with the number predicted from the hidden $sl(2)$ algebraic structure. 

\begin{figure}[ht]
    \centering

    \begin{subfigure}{0.48\textwidth}
        \centering
        \includegraphics[width=\linewidth]{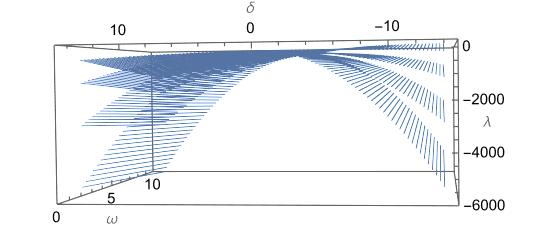}
        \caption{Relationship for $n=3$ in even space.}
        \label{fD n3 p0}
    \end{subfigure}
    \hfill
    \begin{subfigure}{0.48\textwidth}
        \centering
        \includegraphics[width=\linewidth]{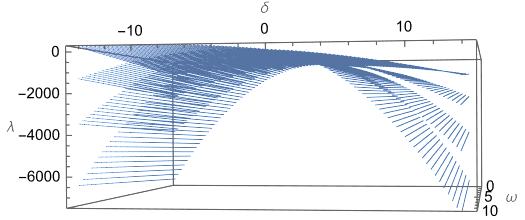}
        \caption{Relationship for $n=3$ in odd space.}
        \label{fD n3 p1}
    \end{subfigure}

    \caption{Relationship between $\omega, a, \beta$ and solutions of the BAEs for $n=3$.}
    \label{fD n3}
\end{figure}

For $n=3$, the energies and the parameter constraints are
\begin{align}
    &\begin{aligned}\label{D E3}
        E_3^{(p)}=\frac{2 \lambda}{\delta}+(17+2p)\omega,
    \end{aligned}\\
    &\begin{aligned}\label{D parameter constrain 3}
        4\delta(7+2p)+\frac{\lambda}{\delta}+2\omega[4-(\xi_1+\xi_2+\xi_3)]=0,
    \end{aligned}
\end{align}
The corresponding root distributions of the BAEs for $n=3$ are shown in Fig.\ref{fD n3}. 
The layers in the parameter spaces are symmetric about $\delta=0$. There are 4 layers in each figure, indicating 4 different sets of roots of the BAEs for $n=3$. This is as expected from the hidden $sl(2)$ algebraic structure of the model. 

\subsection{Root distributions for higher level wave functions}

\begin{figure}[ht]
    \centering

     \begin{subfigure}{0.24\textwidth}
        \centering
        \includegraphics[width=\linewidth]{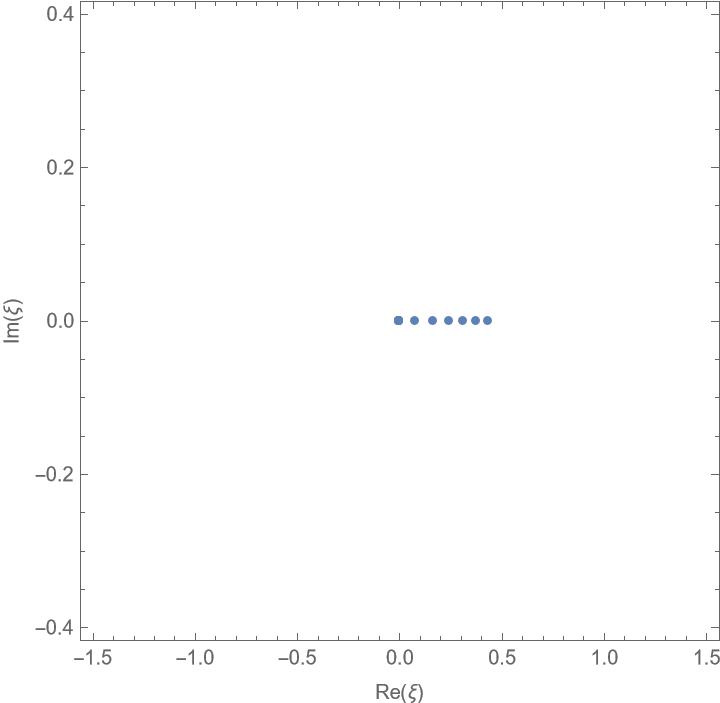}
        \caption{$\lambda=-0.0049$}
        \label{fD lambda=-0.0049 p1}
    \end{subfigure}
    \hfill
    \begin{subfigure}{0.24\textwidth}
        \centering
        \includegraphics[width=\linewidth]{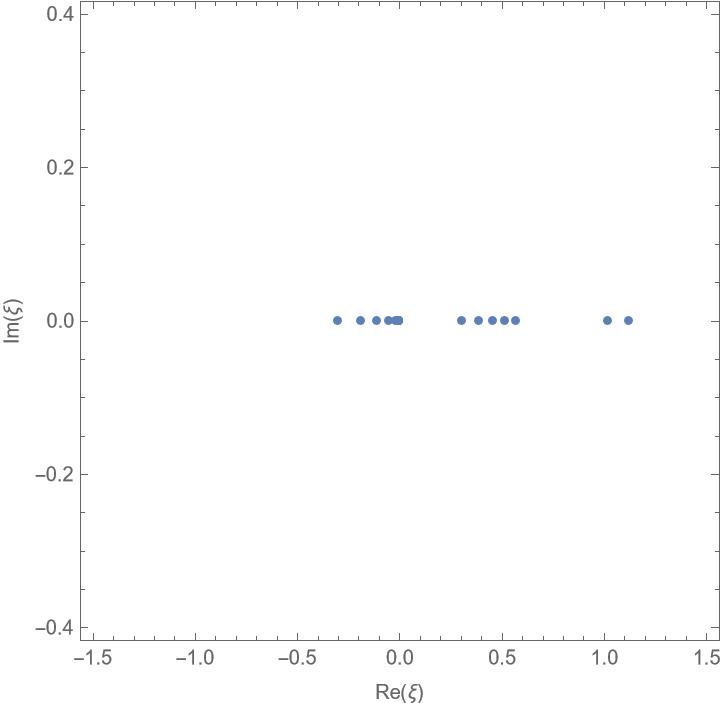}
        \caption{$\lambda=0.0451$}
        \label{fD lambda=0.0451 p1}
    \end{subfigure}
    \hfill
    \begin{subfigure}{0.24\textwidth}
        \centering
        \includegraphics[width=\linewidth]{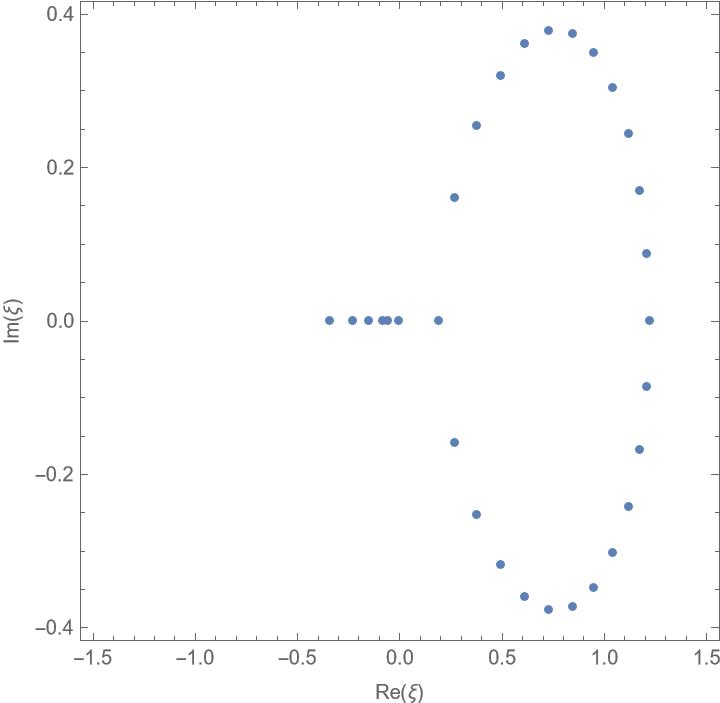}
        \caption{$\lambda=0.0590$}
        \label{fD lambda=0.0590 p1}
    \end{subfigure}
    \hfill
    \begin{subfigure}{0.24\textwidth}
        \centering
        \includegraphics[width=\linewidth]{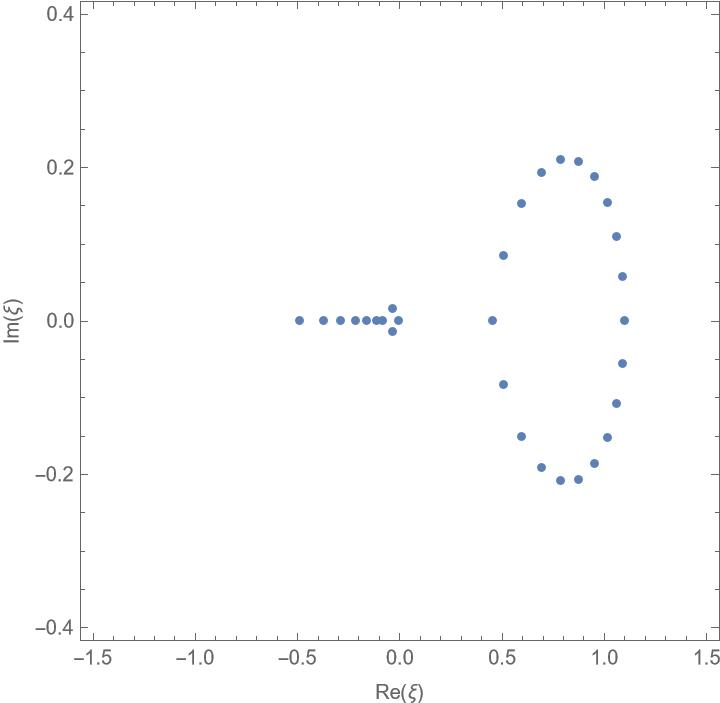}
        \caption{$\lambda=0.1242$}
        \label{fD lambda0.1242 p1}
    \end{subfigure}
    \begin{subfigure}{0.24\textwidth}
        \centering
        \includegraphics[width=\linewidth]{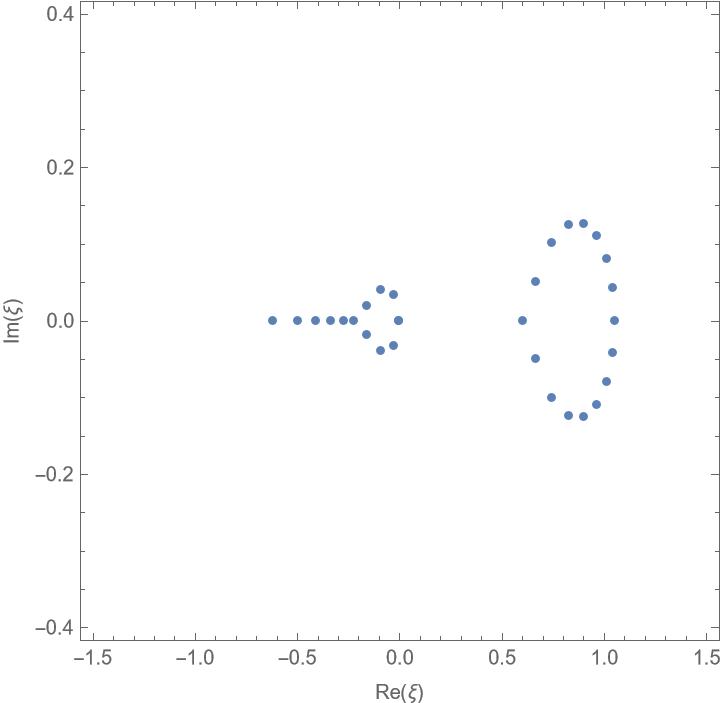}
        \caption{$\lambda=0.2024$}
        \label{fD lambda=0.2024 p1}
    \end{subfigure}
    \hfill
    \begin{subfigure}{0.24\textwidth}
        \centering
        \includegraphics[width=\linewidth]{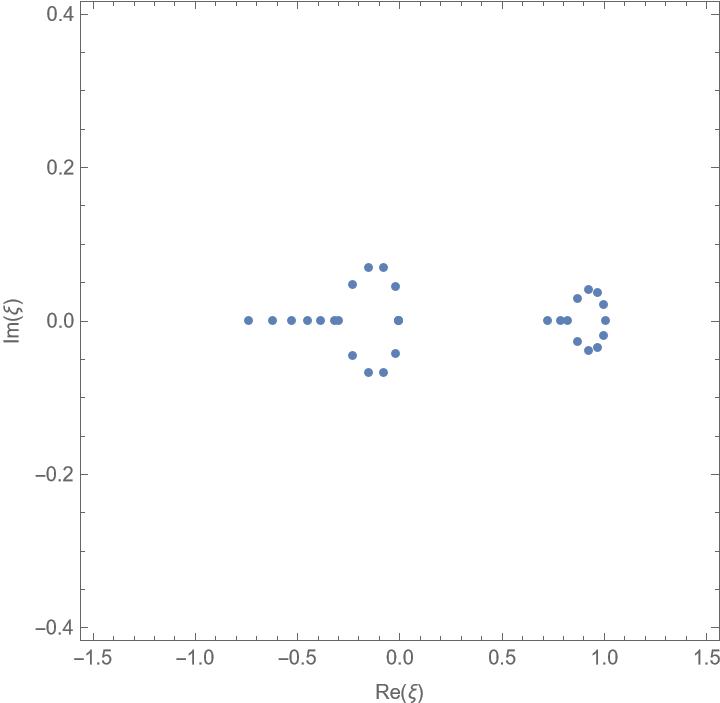}
        \caption{$\lambda=0.2914$}
        \label{fD lambda=0.2914 p1}
    \end{subfigure}
    \hfill
    \begin{subfigure}{0.24\textwidth}
        \centering
        \includegraphics[width=\linewidth]{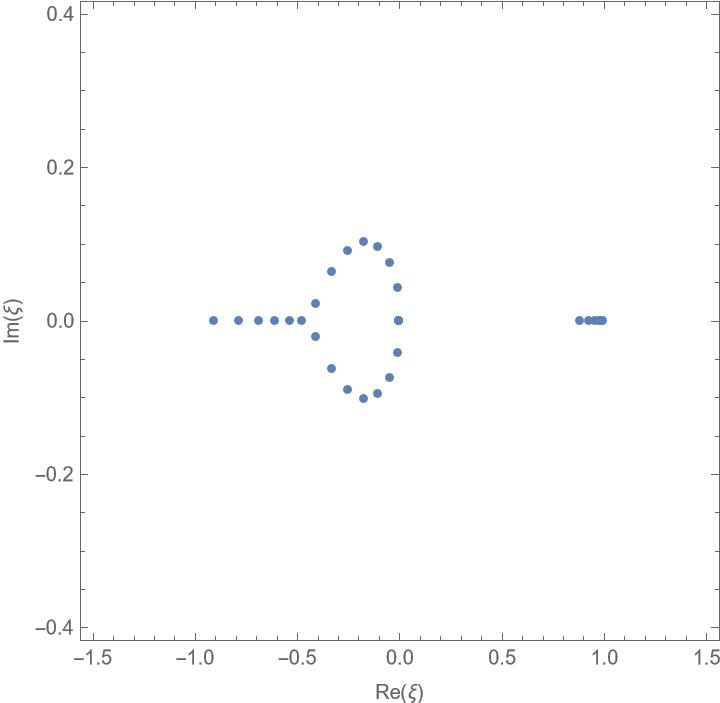}
        \caption{$\lambda=0.4425$}
        \label{fD lambda=0.4425 p1}
    \end{subfigure}
    \hfill
    \begin{subfigure}{0.24\textwidth}
        \centering
        \includegraphics[width=\linewidth]{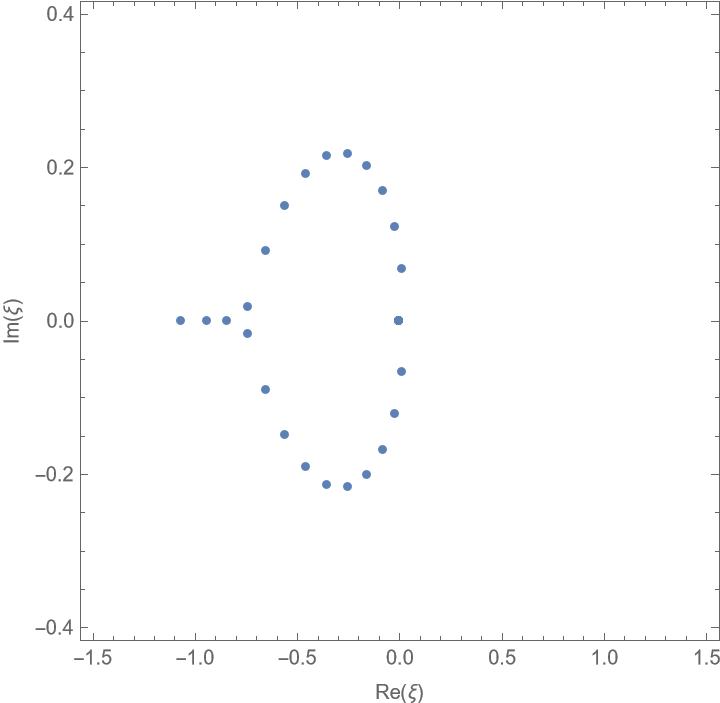}
        \caption{$\lambda=0.6125$}
        \label{fD lambda=0.6125 p1}
    \end{subfigure}
   
   \caption{Change of root distributions with different values of $\lambda$ for odd solutions.}
    \label{fD lambda p=1}
\end{figure}

We follow procedure similar to the previous sections. For the current case, we fix the values of the parameters $\omega=-1, \delta=0.01$. The root distributions corresponding to different $\lambda$ is shown in Fig.\ref{fD lambda p=1} for the odd sector. Like the models in the previous sections, root distributions in the even sector are similar. The roots distribute only in the real axis for $\lambda<0$, the energy is negative, in Fig.\ref{fD lambda=-0.0049 p1}. The shape of the roots changes from a line to a circle when $\lambda$ changes from 0.0451 to 0.0590, which indicates a phase change at that value. After that, the line and circle begin to separate and the left line starts to become a circle with the increase of $\lambda$. The left circle expands and the right circle shrinks. The right circle will disappear with the greatest value of $\lambda$, also the largest energy. 

\begin{figure}[ht]
    \centering

    \begin{subfigure}{0.31\textwidth}
        \centering
        \includegraphics[width=\linewidth]{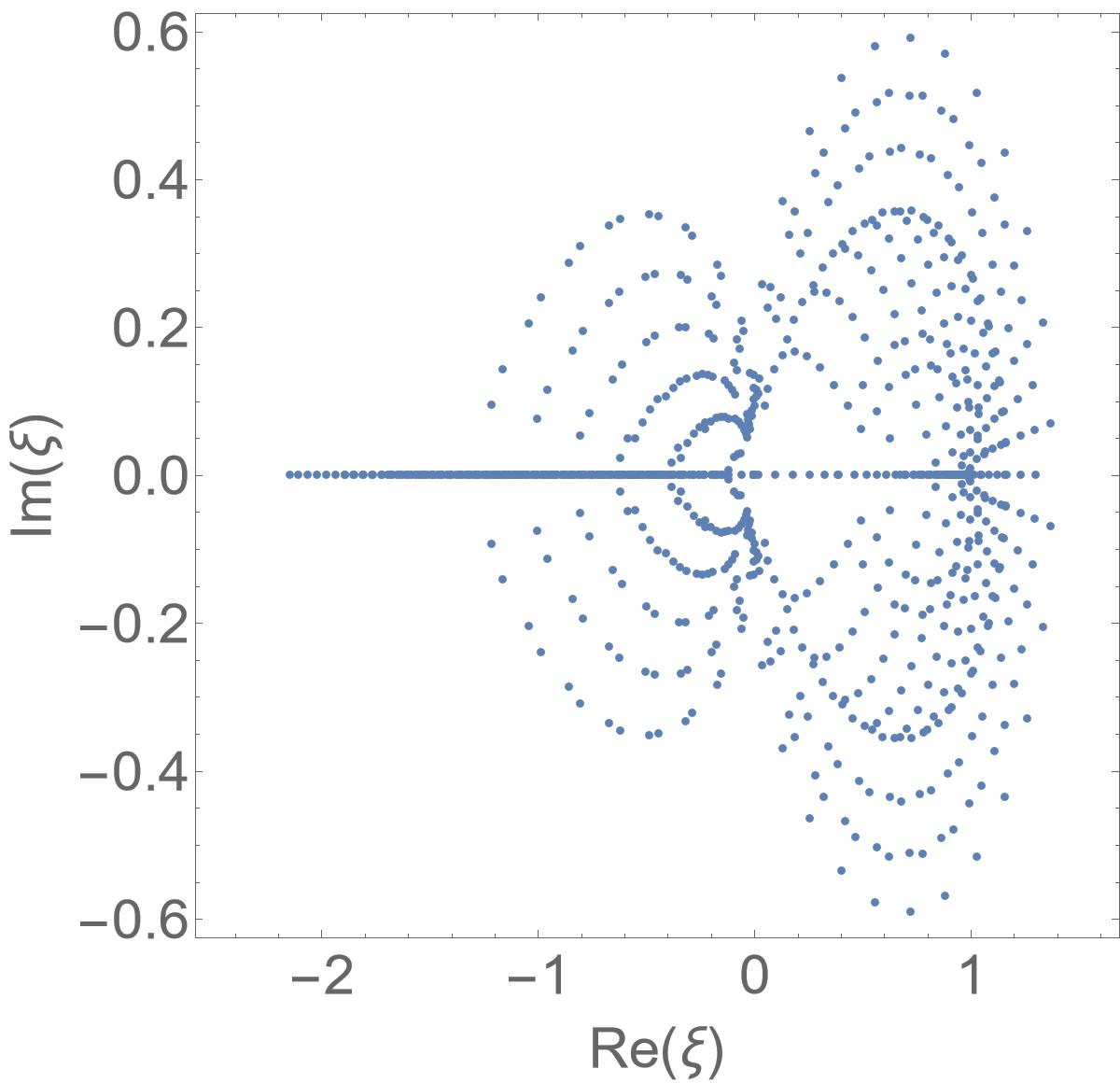}
        \caption{$\omega=-0.5$}
        \label{fD omega=-0.5}
    \end{subfigure}
    \hfill
    \begin{subfigure}{0.31\textwidth}
        \centering
        \includegraphics[width=\linewidth]{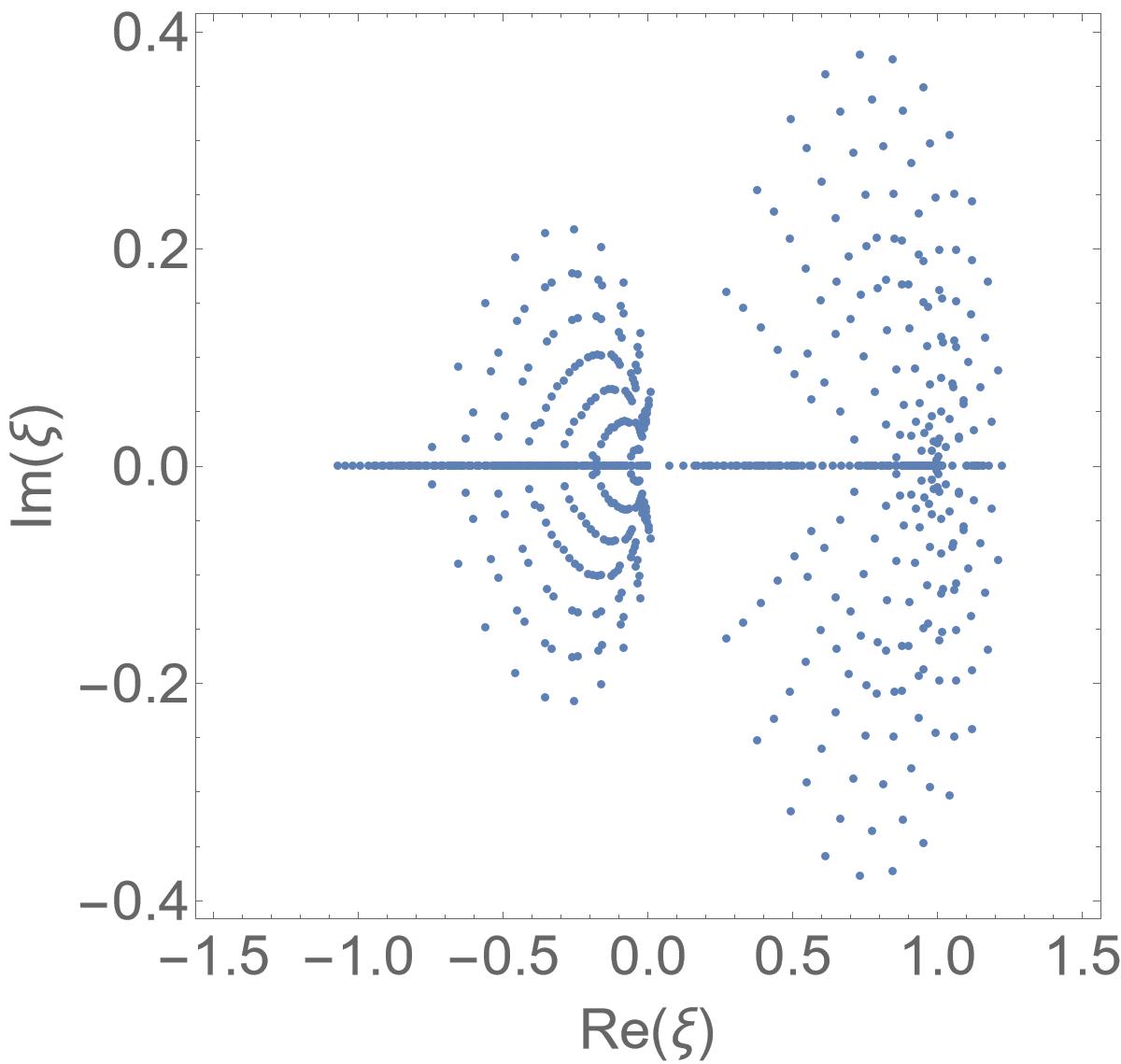}
        \caption{$\omega=-1$}
        \label{fD omega=-1}
    \end{subfigure}
    \hfill
    \begin{subfigure}{0.31\textwidth}
        \centering
        \includegraphics[width=\linewidth]{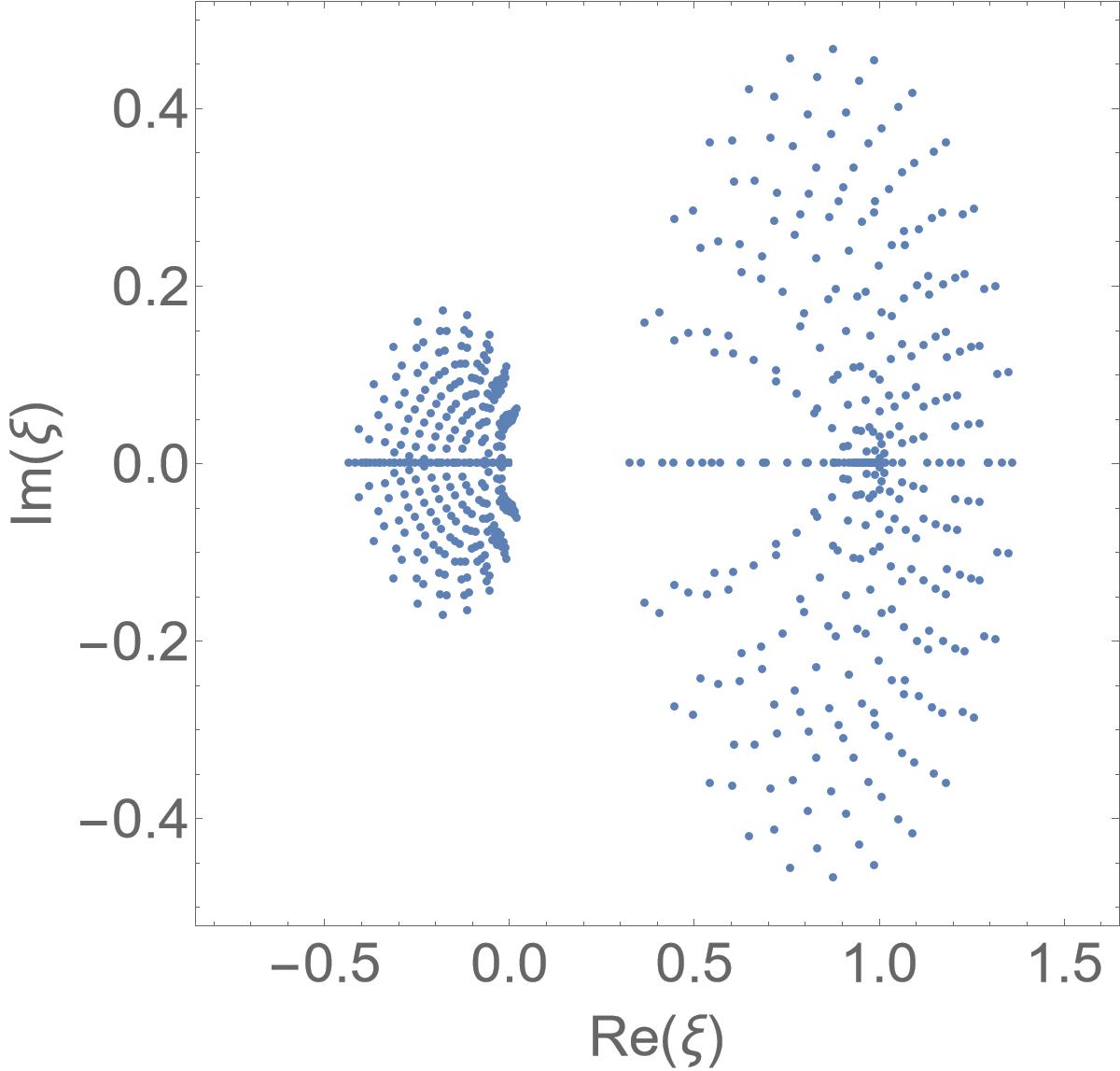}
        \caption{$\omega=-2.5$}
        \label{fD omega=-2.5}
    \end{subfigure}
    
    \caption{The changing of roots distribution with different values of $\omega$ for $n=30$.}
    \label{fD delta=0.01 n=30}
\end{figure}

The combination of roots for all the allowed values of $\lambda$ for $\omega=1, \delta=2$ is Fig.\ref{fD delta=0.01 n=30}. We keep the value of $\delta=0.01$ and change the value of $\omega$ from -0.5 to -2.5, which is shown in Fig.\ref{fD delta=0.01 n=30}. There are two circle and a line, consisting of the shape of the roots. For $\omega=-0.5$, two circles and the line connect with each other. Then, decreasing the value of $\omega$, two circles begin to separate for $\omega=-1$ and they become two parts when $\omega=-2.5$. The left circle also shrinks with the decreasing of $\omega$. 

\newpage
\section{Schr\"odinger equation from kink stability analysis of $\phi^6$ field theory}

The $\phi^6$-type field theory 1+1 dimensions is characterized by the Largrangian
\begin{equation}\label{kink L}
    \mathcal{L}=\frac{1}{2} \partial_{\lambda} \phi \partial^{\lambda} \phi-\frac{\mu^2}{8 g^2 (1+\epsilon^2)}(g^2 \phi^2+\epsilon^2)(1-g^2 \phi^2)^2,
\end{equation}
where the $\mu$ has the dimension of mass and $\epsilon^2$ is the dimensionless constant. 
It is well-known that the field equation of this theory has kink solutions. By performing the stability analysis around the kink solution, the authors in \cite{Chr1975} arrived at the Schr\"odinger equation,
\begin{equation}\label{kink Schrodinger}
H\psi(x)=\left\{-\frac{d^2}{dx^2}+V(x)\right\}\psi(x)=E\psi(x),
\end{equation}
where 
the potential $V(x)$ is given by
\begin{equation}\label{kink V(x)}
V(x)=\mu^2\frac{8\sinh^4\frac{\mu x}{2}-\left({20}{\nu}-4\right)\sinh^2\frac{\mu x}{2}
    +2\left(\nu+1\right)\left(\nu-2\right)}
    {8\left(1+\nu+\sinh^2\frac{\mu x}{2}\right)^2}.  
\end{equation}
Here we have introduced the notation
$$\nu=1/\epsilon^2$$.

Exact solutions to the Schr\"odinger system were studied by brute force and on a case by case basis in \cite{Chr1975,Jat1989}. In \cite{Zha2012}, Bethe ansatz method was applied to derive general formulas for energies, constraint of the model parameter $\nu$ and wave functions. However, explicit expressions were obtained only for two lower level solutions. In this section, we improve the procedure presented in \cite{Zha2012}. This enables us to present explicit expressions for several higher level solutions, which constitutes one of the main new results in this section.

We let 
\begin{equation}
    \psi_0(x)=\frac{1}{(1+\nu+\sinh^2 \frac{\mu x}{2})^{\frac{3}{2}}}
\end{equation}
and perform a similarity transformation $\psi^{-1}_0(x)H\psi_0(x)$ to get rid of the term involving $1(1+\nu+\sinh^2 \frac{\mu x}{2})^{2}$ in the transformed Hamiltonian. This is equivalent to the following transformation of the wave function has the form
\begin{equation}
    \psi(x)=\frac{y(x)}{(1+\nu+\sinh^2 \frac{\mu x}{2})^{\frac{3}{2}}}.
\end{equation}
This transformation gives the ODE for $y(x)$ \cite{Zha2012},
\begin{equation}
 -y''+3\mu\frac{\sinh\frac{\mu x}{2}\cosh\frac{\mu x}{2}}{1+\nu+\sinh^2\frac{\mu x}{2}}y'
   +\frac{\frac{3}{2}\mu^2\left(1+\nu\right)}{1+\nu+\sinh^2\frac{\mu x}{2}}y
    =\left(E+\frac{5}{4}\mu^2\right)y. 
\end{equation}
Moreover, making a change variable $z=\cosh\frac{\mu x}{2}$ gives rise to the ODE \cite{Zha2012} 
\begin{equation}\label{kink ODE}
    \begin{split}
        \left[ z^4 +\left(\nu-1 \right)z^2 -\nu\right] y''(z) & -\left[5 z^3-\left(\nu+6\right)z \right]y'(z)\\
        &+\left[\left(\frac{4E}{\mu^2}+5\right)z^2 +\frac{4\nu E}{ \mu^2}-\nu-6\right]y(z)=0.
    \end{split}
\end{equation}

Setting 
\begin{equation}
 y(z)=z^p\, \phi(z), \qquad p=0,1,    
\end{equation}
the above ODE for $y(z)$ becomes 
\begin{equation}\label{kink ODE phi}
    \begin{split}
       (\nu+z^2)(z^2-1) \phi''(z) &+\left[\frac{2p(\nu+z^2)(z^2-1)}{z}+z(6+\nu-5z^2)\right]\phi'(z)\\
       &+\left[z^2 \left(5+\frac{4E}{\mu^2}\right) +\nu \left(\frac{4E}{\mu^2}-1\right) +p(6+\nu-5z^2)-6\right]\phi(z)=0,
    \end{split}
\end{equation}
Changing the variable $\xi=z^2+\nu$, we obtain
\begin{equation}\label{kink ODE phi(xi)}
    \begin{split}
        \left[\xi^3-(1+2\nu)\xi^2+\nu(\nu+1)\xi\right] \phi''(\xi)& +\left[(p-2)\xi^2+ \left( \frac{5}{2}-\nu(p-5)-p\right)\xi-3\nu(\nu+1) \right]\pi'(\xi)\\
        &+\left\{\left[\frac{E}{\mu^2}-\frac{5}{4}(p-1)\right]\xi+\frac{3}{2}(1+\nu)(p-1)\right\}\phi(\xi)=0.
    \end{split}
\end{equation}
This equation can be recast in the form
\begin{equation}\label{kink Heun}
    \begin{split}
         \phi''(\xi) +\left[-\frac{3}{\xi}+\frac{p+1/2}{\xi-\nu}+\frac{1/2}{\xi-\nu-1}\right]\pi'(\xi)
        +\frac{\left[\frac{E}{\mu^2}-\frac{5}{4}(p-1)\right]\xi+\frac{3}{2}(1+\nu)(p-1)}{\xi(\xi-\nu)(\xi-\nu-1)}\phi(\xi)=0.
    \end{split}
\end{equation}
This is the general Heun equation
\begin{equation}\label{kink general Heun}
    \begin{split}
         \phi''(\xi) +\left[\frac{\gamma}{\xi}+\frac{\delta}{\xi-\nu}+\frac{\sigma}{\xi-\nu-1}\right]\pi'(\xi)
        +\frac{\alpha\beta\xi-q}{\xi(\xi-\nu)(\xi-\nu-1)}\phi(\xi)=0
    \end{split}
\end{equation}
with $\gamma=-3,\,\beta=p+\frac{1}{2},\,\sigma=\frac{1}{2},\, q=\frac{3}{2}(1-p)(\nu+1)$ and 
\begin{eqnarray}
    \alpha=\frac{p-3}{2}+\sqrt{\frac{(p+1)^2}{4}-\left(\frac{E}{\mu^2}+\frac{1-p}{4}\right)},\quad
    \beta=\frac{p-3}{2}+\sqrt{\frac{(p+1)^2}{4}-\left(\frac{E}{\mu^2}-\frac{1-p}{4}\right)}.
\end{eqnarray}
It is easily checked that $\alpha+\beta+1 = \gamma+\delta+\sigma$, as required.

The above ODE is quasi-exactly solvable and its exact solutions are given by polynomials of degree $n$ of the form
\begin{equation}
    \phi(\xi)=\prod_{i=1}^n(\xi-\xi_i),\qquad \phi(\xi)\equiv 1~{\rm for}~ n=0.
\end{equation}
Indeed, applying the Bethe ansatz method \cite{Zha2012}, (\ref{c2}), (\ref{general Bethe ansatz}) and (\ref{c1}), we obtain the energies
\begin{align}\label{kink En}
        E^{(p)}_n=-\frac{\mu^2}{4}(2n+p-1)(2n+p-5),\qquad n=0,1,2,\dots,
    \end{align}
and the constraint for the model parameter $\nu$,
\begin{equation}
    \begin{split}\label{kink parameter constrain n}
       (2n+p-4)\sum_{i=1}^n \xi_i-n\left(n-\frac{7}{2}\right)(2\nu+1)-\left[p\left(n-\frac{3}{2}\right)+\frac{3}{2}\right](\nu+1)=0.
    \end{split}
\end{equation}
Here the roots $\xi_1,\ldots,\xi_n$ are determined by the BAEs,
\begin{equation}\label{kink roots of Bethe ansatz}
    \sum_{j \neq i}^n \frac{2}{\xi_i-\xi_j}+\frac{(p-2)\xi_i^2+ \left( \frac{5}{2}-\nu(p-5)-p\right)\xi_i-3\nu(\nu+1)}{\xi_i(\xi_i-\nu)(\xi_i-\nu-1)}=0,\quad i=1,2,\ldots,n.
\end{equation}
The corresponding wave functions are
\begin{align}\label{kink psi n}
        \psi^{(p)}_n(z)=\frac{\cosh^p\frac{\mu x}{2}}{\left(1+\nu+\sinh^2 \frac{\mu x}{2} \right)^{\frac{3}{2}}}\, \prod_{i=1}^n\left(\cosh^2\frac{\mu x}{2}-\xi_i+\nu\right).
\end{align}

In what follows, we show that the ODE (\ref{kink ODE phi(xi)}) possess a hidden $sl(2)$ algebra symmetry. We rewrite (\ref{kink ODE phi(xi)}) as
\begin{equation}\label{calH kink}
    \begin{split}
    \mathcal{H}&\phi(\xi)=\varepsilon\phi(\xi),\qquad \varepsilon=+\frac{3}{2}(1+\nu)(1-p),\\
        \mathcal{H}=&\xi(\xi-\nu)(\xi-\nu-1) \frac{d^2}{d\xi^2} \\
        &+\left[(p-2)\xi^2+ \left( \frac{5}{2}-\nu(p-5)-p\right)\xi-3\nu(\nu+1) \right]\frac{d}{d\xi}
        +\left[\frac{E}{\mu^2}-\frac{5}{4}(p-1)\right]\xi.
    \end{split}
\end{equation}
Applying theorem A2 (\ref{general algebraization}), the Hamiltonian $\mathcal{H}$ (\ref{calH kink}) has a hidden $sl(2)$ algebra symmetry if
\begin{equation}
    \frac{E}{\mu^2}-\frac{5}{4}(p-1)\equiv c_1=-n[(n-1)a_3+b_2]\equiv -n[(n-1)+p-2],
\end{equation}
which gives the exact energies (\ref{kink En}), i.e. $E=-\frac{\mu^2}{4}(2n+p-1)(2n+p-5)$. For such $E$ values, $\mathcal{H}$ (\ref{calH kink}) depends on integer $n$ and can be expressed as
\begin{equation}\label{kink algebraic sl(2)}
    \begin{split}
        \mathcal{H}=&-J^+ J^0-(1+2\nu) J^0 J^0 +(\nu+\nu^2) J^0 J^- -\left(\frac{3n}{2}+p-3\right)J^+ +\frac{1}{2}\nu(1+\nu)(n-6) J^- \\
        &+\frac{1}{2}\left[(7-2n-p)(1+2\nu)-p\right]J^0  -\frac{1}{4}n[(1+2\nu)(p+n-7)+p], 
    \end{split}
\end{equation}
in terms of the $sl(2)$ differential operators (\ref{sl(2) operator}). This provides a $sl(2)$ algebraization for the Schr\"odinger system from the kink stability analysis of $\phi^6$ field theory. This result is in contrast to \cite{Jat1989} which claims that no $sl(2)$ algebraization is possible for this system. The algebraization given above represents one of the new results in this section. 

When $n=0$, we have $\phi(\xi)=1$. For $p=0$, the constraint for the model parameter $\nu$ is $\nu=-1$. The corresponding energy and wave function are
\begin{equation}
    E_0^{(0)}=-\frac{5\mu^2}{4},\qquad 
    \psi_0^{(0)}(x)=\frac{1}{\sinh^3 \frac{\mu x}{2} },
\end{equation}
which is the solution for the Schr\"odinger system (\ref{kink Schrodinger}) with the constraint $\nu=-1$. For $p=1$, there is no constraint for model parameter $\nu$ and the energy and wave function are given by
\begin{equation}
    E_0^{(1)}=0,\qquad
    \psi_0^{(1)}(x)=\frac{\cosh\frac{\mu x}{2}}{\left(1+\nu+\sinh^2 \frac{\mu x}{2} \right)^{\frac{3}{2}}}.
\end{equation}
$\psi_0^{(1)}(x)$ is the solution of the Schr\"odinger system (\ref{kink Schrodinger}) for arbitrary model parameter $\nu$. This solution was also obtained in \cite{Chr1975,Zha2012}. 

As a by product, we remark that when $\nu=0$,  $\psi_0^{(1)}(x)$ becomes 
\begin{equation}
    \psi_0^{(1)}(x)=\frac{1}{1+\sinh^2 \frac{\mu x}{2}}.
\end{equation}
This gives the ground level solution (with $E_0=0$) of the Schr\"odiner equation with the hyperbolic potential well of the form
\begin{equation}
    V(x)=-\mu^2\,\frac{1-\sinh^2\frac{\mu x}{2}-2\sinh^4\frac{\mu x}{2}}
    {2\cosh^4\frac{\mu x}{2}}.
\end{equation}

For $n=1$, we obtain two sets of solutions of (\ref{kink roots of Bethe ansatz}) together with the associated values of the model parameter $\nu$,
\begin{align} \label{kink xi1 p}
    &\xi^{(p)}_{11}=\frac{5}{2(p-2)}, \qquad \nu^{(p)}_{11}=-1,\\
    &\xi^{(p)}_{12}=\frac{8-5p}{2}, \qquad \nu^{(p)}_{12}=2(1-p).
\end{align}
The corresponding energies and wave functions are
\begin{align}
    &\begin{aligned} \label{kink E1 p0}
        E^{(p)}_1=-\frac{(p+1)(p-3)}{4}\mu^2,
    \end{aligned}\\
    &\begin{aligned}\label{kink psi 1 p}
        &\psi^{(p)}_{11}(z)=\frac{\cosh^p\frac{\mu x}{2}}{\left(1+\nu^{(p)}_{11}+\sinh^2 \frac{\mu x}{2} \right)^{\frac{3}{2}}}\, \left(\cosh^2\frac{\mu x}{2}-\xi^{(p)}_{11}+\nu^{(p)}_{11}\right),\\
        &\psi^{(p)}_{12}(z)=\frac{\cosh^p\frac{\mu x}{2}}{\left(1+\nu_{12}^{(p)}+\sinh^2 \frac{\mu x}{2} \right)^{\frac{3}{2}}}\, \left(\cosh^2\frac{\mu x}{2}-\xi^{(p)}_{12}+\nu^{(p)}_{12}\right).
    \end{aligned}
\end{align}
The solutions $\psi_{12}^{(0)}(x)$ and $\psi_{12}^{(1)}(x)$ given above
were obtained in \cite{Chr1975}. Here we re-derive them by means of the Bethe ansatz method. The other two solutions $\psi_{11}^{(p)}(x)$ in (\ref{kink psi 1 p}) are new.

When $n=2$, we have energies 
\begin{equation}\label{kink E2 p}
    E_2^{(p)}=-\frac{(p+3)(p-1)}{4}\mu^2.
\end{equation}
For $n=2, p=0$, the energy is $E_2^{(0)}=\frac{3\mu^2}{4}$, same as the energy for $n=1, p=0$. In this case we find that the Bethe ansatz equations (\ref{kink roots of Bethe ansatz}) has two roots
\begin{align}\label{kink xi2 p0}
   \xi^{(0)}_1=\frac{8 + 12\sqrt{2}}{21}, \qquad \xi^{(0)}_2=\frac{8 - 12\sqrt{2}}{21},
\end{align}
and the constraint (\ref{kink parameter constrain n}) yields
 $   \nu=-\frac{1}{3}. $
The corresponding wave function is given by 
\begin{align}
    \label{kink psi 2 p0}
        \psi^{(0)}_2(z)=\frac{\left[\cosh^2\frac{\mu x}{2}-\frac{5+4\sqrt{2}}{7}\right]\left[\cosh^2\frac{\mu x}{2}-\frac{5-4\sqrt{2}}{7}\right]}{\left(\frac{2}{3}+\sinh^2 \frac{\mu x}{2}\right)^{\frac{3}{2}}}.
\end{align}
This solution of the Schr\"odinger equation (\ref{kink Schrodinger}) has not been found previously and is new. 

For $n=2, p=1$, the energy is $E_2^{(1)}=0$, same as the energy for $n=0, p=1$.   As can be seen from the (\ref{kink ODE phi(xi)}), when $E=0, p=1$, the ODE (\ref{kink ODE phi(xi)}) degenerates to the one for $n=0, p=1$, for which the Bethe ansatz equations (\ref{kink roots of Bethe ansatz}) become irrelevant. So solution for $n=2, p=1$ is the same as that for $n=0, p=1$, i.e. same energy level and same wave function. This is a very interesting phenomena of this model. As far as we know, this kind of situation has never appeared previously in the literature.

\subsection{Root distributions for higher level wave functions}

\begin{figure}[ht]
    \centering
    \begin{subfigure}{0.24\textwidth}
        \centering
        \includegraphics[width=\linewidth]{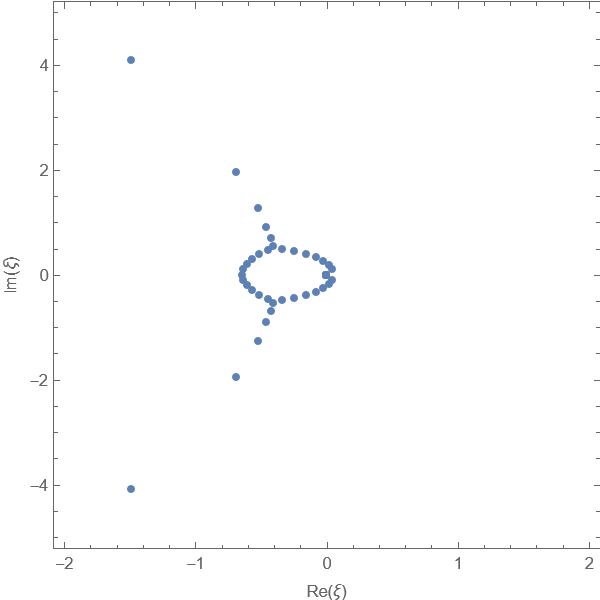}
        \caption{$\nu=-0.99$}
        \label{fD nu=-0.99 p1}
    \end{subfigure}
    \hfill
    \begin{subfigure}{0.24\textwidth}
        \centering
        \includegraphics[width=\linewidth]{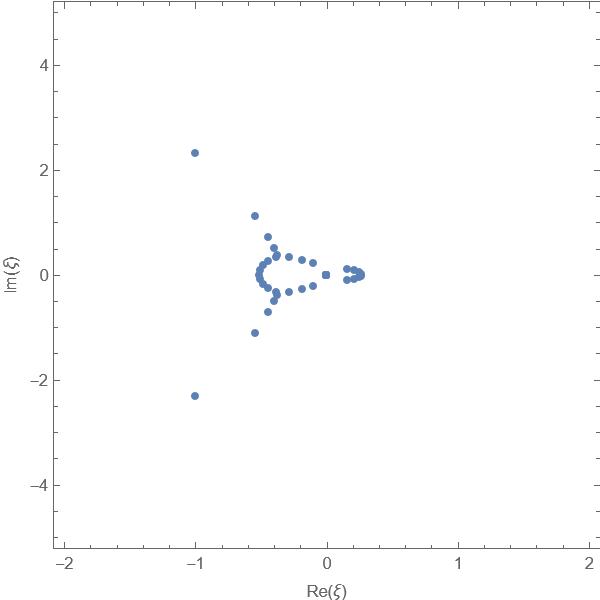}
        \caption{$\nu=-0.78$}
        \label{fD nu=-0.78 p1}
    \end{subfigure}
    \hfill
    \begin{subfigure}{0.24\textwidth}
        \centering
        \includegraphics[width=\linewidth]{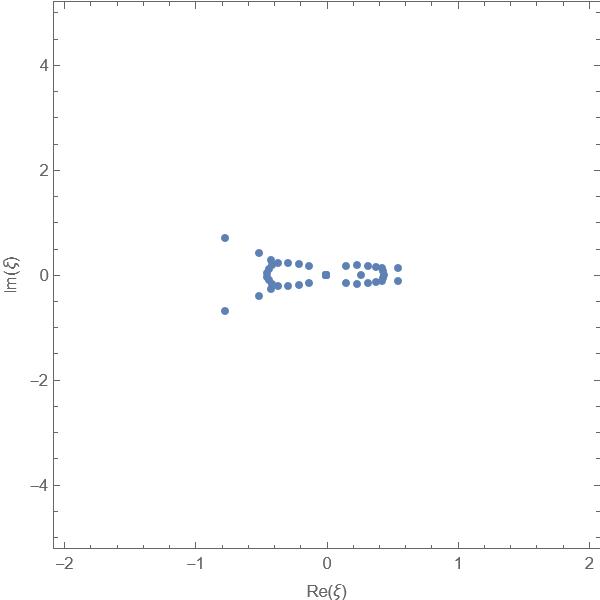}
        \caption{$\nu=-0.58$}
        \label{fD nu=-0.58 p1}
    \end{subfigure}
    \hfill
    \begin{subfigure}{0.24\textwidth}
        \centering
        \includegraphics[width=\linewidth]{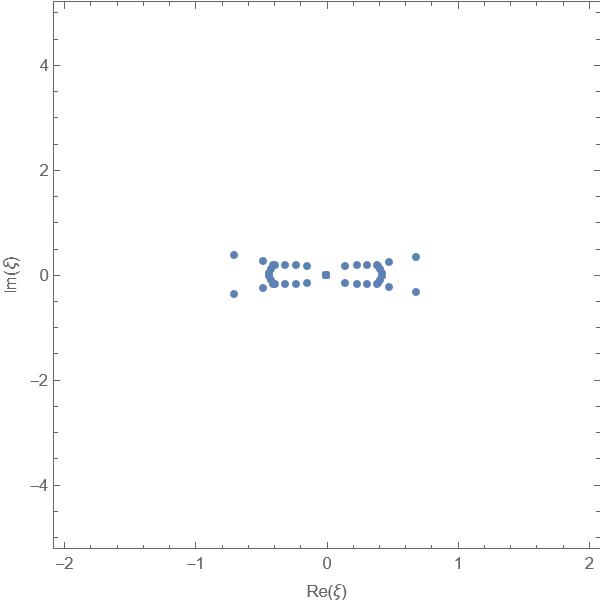}
        \caption{$\nu=-0.51$}
        \label{fD nu=-0.51 p1}
    \end{subfigure}
    \hfill
    \begin{subfigure}{0.24\textwidth}
        \centering
        \includegraphics[width=\linewidth]{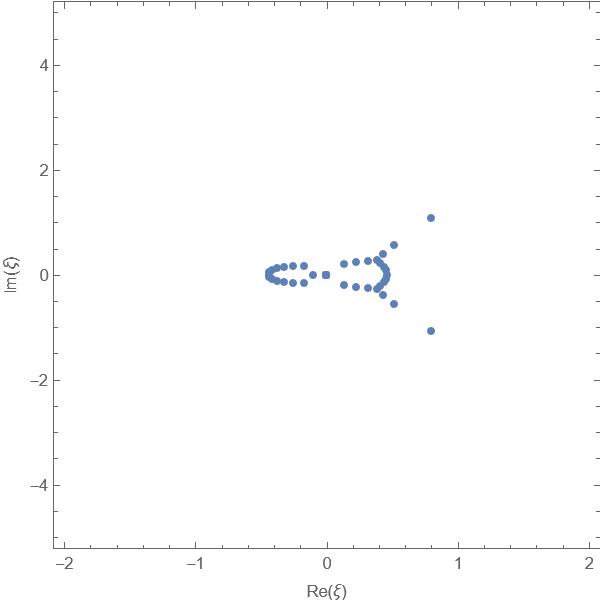}
        \caption{$\nu=-0.38$}
        \label{fD nu=-0.38 p1}
    \end{subfigure}
    \hfill
    \begin{subfigure}{0.24\textwidth}
        \centering
        \includegraphics[width=\linewidth]{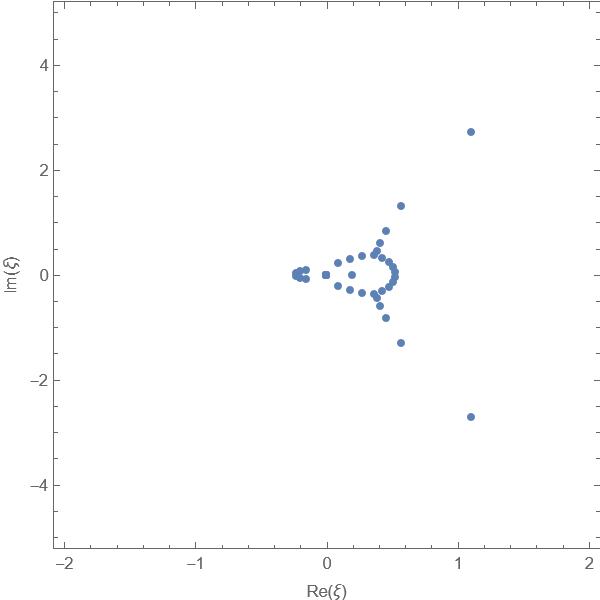}
        \caption{$\nu=-0.18$}
        \label{fD nu=-0.18 p0}
    \end{subfigure}
    \hfill
    \begin{subfigure}{0.24\textwidth}
        \centering
        \includegraphics[width=\linewidth]{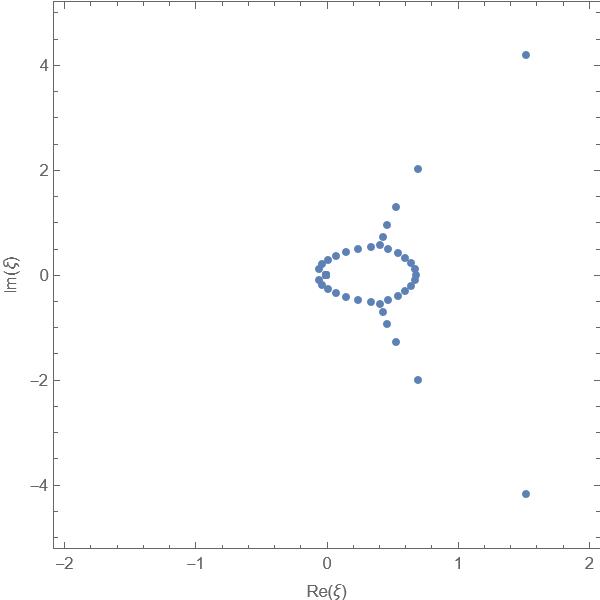}
        \caption{$\nu=-0.01$}
        \label{fD nu=-0.01 p1}
    \end{subfigure}  
    \caption{Change of root distributions with the change of $\nu$ for $p=1$ and $n=50$.}
    \label{fk nu p1}
\end{figure}

This model is different from the models in other sections: the BAEs only depend on the model parameter $\nu$ and the parameter $\mu$ is a free parameter. Thus, root distributions of the BAEs would only change with $\nu$. We take $n=50$ and calculate the associated root distributions numerically. They are shown in 
Fig.\ref{fk nu p1} for 
$p=1$. 
As seen from the figures, with the increase of $\nu$ values, the roots move from left to right. We remark that the root distributions for $p=0$ are similar to the ones for $p=1$ shown in the figure. 

\newpage

\begin{figure}[ht]
    \centering
    \includegraphics[width=0.6\textwidth]{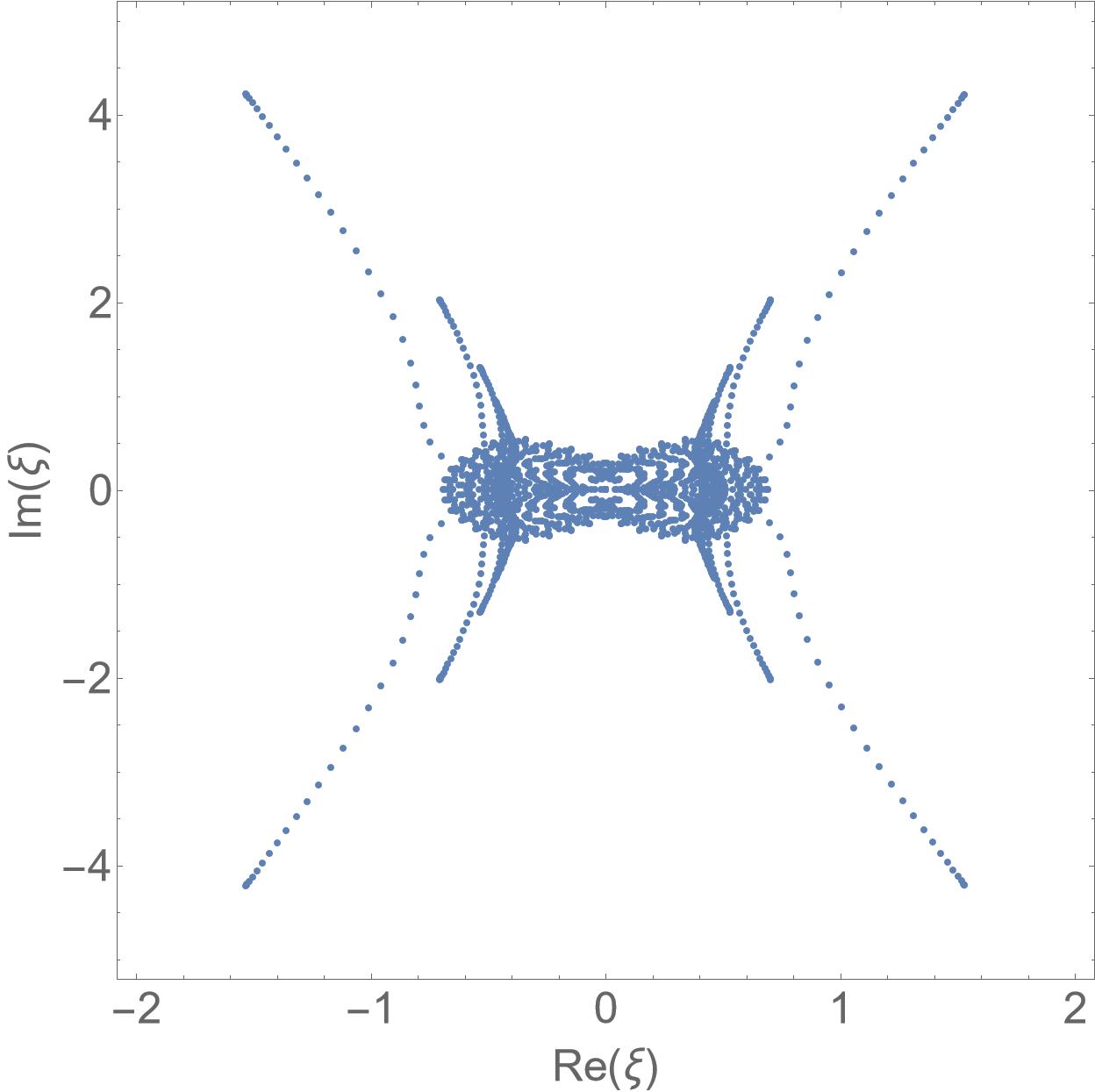}
    \caption{Root distribution for $n=50$}\label{fk p0}
\end{figure}

The distribution of all roots of the BAEs for $n=50$ is shown in Fig.\ref{fk p0}. The distribution has a strong symmetry: it is not only symmetric about the real axis, but also symmetric about the imaginary axis.
 
\section{Conclusions}
We have provided a systematic and unified treatment for quasi-exact solutions for an interesting class of models: the sextic anharmonic oscillator, the singular anharmonic oscillator, the generalized quantum isotonic oscillator, the non-polynomially modified oscillator, and the Schr\"odinger system from the kink stability analysis of the $\phi^6$ field theory.  The Bethe ansatz method enables us to obtain generic closed-form expressions for the constraints of the model parameters, the corresponding energies and wave functions in the odd and even sectors in a unified way. We have also provided unified $sl(2)$ algebraizations for the even and odd sectors of the models. 

We have given explicit analytic formulas for the ground states and first excited states of the models. For higher level cases we have performed detailed analysis of the BAEs via numerical methods and plotted the relationships between the model parameters and solutions of the BAEs. 
We are able to present numerical results for levels up to $n=30$ or $n=50$. The root distributions for given choice of allowed model parameters in the complex plane exhibit very interesting features: Their shapes change drastically across different parameter regions, indicating a phenomenon analogous to phase transitions observed in the pairing integrable models \cite{marquette2012}. 
Zeroes of polynomials associated with wavefunctions of quasi-exactly solvable systems have been largely unexplored and our plots for the zeroes of the polynomial solutions should provide insight for further study in this direction.

\section*{Acknowledgement}
IM was supported by by Australian Research Council Future Fellowship FT180100099, and YZZ was supported by Australian Research Council Discovery Project DP190101529.

\appendix

\section {Bethe ansatz method and algebraization}

Consider the 2nd order linear ODE,
\begin{equation}\label{general ODE}
    \left[X(z) \frac{d^2}{dz^2}+Y(z) \frac{d}{dz}+Z(z)\right] S(z)=0,
\end{equation}
where $X(z), Y(z)$ and $Z(z)$ are polynomials of degrees 4,3 and 2 respectively,
\begin{align}\label{X Y Z}
    X(z)=\sum_{k=0}^4 a_k z^k, \quad Y(z)=\sum_{k=0}^3 b_k z^k,\quad Z(z)=\sum_{k=0}^2 c_k z^k.
\end{align}
Let
\begin{equation}\label{sl(2) operator}
    J^+=-z^2 \frac{d}{dz}+nz,\quad J^0=z\frac{d}{dz}-\frac{n}{2},\quad J^-=\frac{d}{dz} 
\end{equation}
be the 1st-order differential operator in single variable $z$. These differential operators satisfy the $\mathfrak{sl}(2)$ commutation relations for any value of the parameter $n$.
\begin{equation}\label{sl(2) commutation relations}
  \left[J^0,J^{\pm}\right]=\pm J^{\pm},\quad \left[J^+,J^-\right]=2 J^0,
\end{equation}
If $n$ is a non-negative integer, $n=0,1,2, \ldots,$ then (\ref{sl(2) operator}) provide a $(n + 1)$-dimensional irreducible representation  $P_{n+1}(z) =< 1, z z^2,\ldots, z^n >$ of the $sl(2)$ algebra. From this it is evident that any differential operator which is a polynomial of the $sl(2)$ generators (\ref{sl(2) operator}) with non-negative $n$ will have the space $P_{n+1}(z)$ as its invariant subspace. This is the main idea behind quasi-exact solvability of a differential operator with a $sl(2)$ algebraization (i.e. a hidden $sl(2)$ algebraic structure) \cite{Tur1988,Tur1994}. 

The following two results were respectively proved in \cite{Zha2012,zhang2016}. 
\vskip.1in
\noindent {\bf Theorem A1.} Let $n$ be a non-negative integer. Given a pair of polynomials $X(z)$ and $Y(z)$, then the values of the coefficients $c_2, c_1, c_0$ of the polynomial $Z(z)$ such that the ODE (\ref{general ODE}) has polynomial solutions of degree $n$
\begin{equation}\label{S(z)}
    S(z)=\prod_{i=1}^n (z-z_i),\qquad S(z)\equiv 1~{\rm for}~ n=0
\end{equation}
with distinct roots $z_1,z_2,\ldots, z_n$ are given by
\begin{align}
    &\begin{aligned}\label{c2}
        c_2=-n(n-1) a_4-n b_3,
    \end{aligned}\\
    &\begin{aligned}\label{c1}
        c_1=-[2(n-1) a_4+b_3]\sum_{i=1}^n z_i-n(n-1) a_3-n b_2,
    \end{aligned}\\
    &\begin{aligned}\label{c0}
        c_0=&-[2(n-1)a_4+b_3] \sum_{i=1}^n z_i^2-2a_4 \sum_{i<j}^n z_i z_j\\
        &-[2(n-1)a_3+b_2]\sum_{i=1}^n z_i-n(n-1) a_2-n b_1.
    \end{aligned}
\end{align}
Here the roots $z_1,z_2,\ldots, z_n$ are determined by the BAEs
\begin{equation}\label{general Bethe ansatz}
    \sum_{j\neq i}^n \frac{2}{z_i-z_j}+\frac{b_3 z_i^3+b_2 z_i^2+b_1 z_i+b_0}{a_4 z_i^4+a_3 z_i^3+a_2 z_i^2+a_1 z_i+a_0}=0, \quad i=1,2,\dots, n.
\end{equation}
The above equations (\ref{c2})-(\ref{general Bethe ansatz}) give all polynomials $Z(z)$ such that the ODE (\ref{general ODE}) has polynomial solutions of the form (\ref{S(z)}).

\vskip.1in
\noindent {\bf Theorem A2.} Let $n$ be positive integer. The differential operator
\begin{equation}
    {\cal H}=X(z) \frac{d^2}{dz^2}+Y(z) \frac{d}{dz}+Z(z),
\end{equation}
where $X(z), Y(z), Z(z)$ are given in (\ref{X Y Z}), allows for a $sl(2)$ algebraization if and only if 
\begin{equation}\label{general algebraization}
    b_3=-2(n-1)a_4,\qquad c_2=n(n-1)a_4,\qquad c_1=-n[(n-1)a_3+b_2].
\end{equation}

\bibliographystyle{unsrt} 
\bibliography{name.bib}

\begin{thebibliography}{10}

\bibitem{Tur1988}
A.V. Turbiner.
\newblock Quasi-exactly-solvable problems andsl(2) algebra.
\newblock {\em Communications in Mathematical Physics}, 118(3):467--474, sep
  1988.

\bibitem{Tur1994}
A.V. Turbiner.
\newblock Hidden algebra of the n-body calogero problem.
\newblock {\em Physics Letters B}, 320(3-4):281--286, jan 1994.

\bibitem{Gon1993}
A.~Gonz{\'{a}}lez-L{\'{o}}pez, N.~Kamran, and P.J. Olver.
\newblock Normalizability of one-dimensional quasi-exactly solvable
  schrödinger operators.
\newblock {\em Communications in Mathematical Physics}, 153(1):117--146, apr
  1993.

\bibitem{Ush1994}
A.G. Ushveridze.
\newblock {\em Quasi-exactly solvable models in quantum mechanics}.
\newblock IOP Publishing, jul 1994.

\bibitem{turbiner2016}
A.V. Turbiner.
\newblock One-dimensional quasi-exactly solvable schr{\"o}dinger equations.
\newblock {\em Physics Reports}, 642:1--71, 2016.

\bibitem{Ben1996}
C.M. Bender and G.V. Dunne.
\newblock Quasi-exactly solvable systems and orthogonal polynomials.
\newblock {\em Journal of Mathematical Physics}, 37(1):6--11, jan 1996.

\bibitem{Sas2009}
R.~Sasaki.
\newblock Bethe ansatz solutions to quasi exactly solvable difference
  equations.
\newblock {\em Symmetry, Integrability and Geometry: Methods and Applications},
  nov 2009.

\bibitem{Zha2012}
Y.-Z. Zhang.
\newblock Exact polynomial solutions of second order differential equations and
  their applications.
\newblock {\em Journal of Physics A: Mathematical and Theoretical},
  45(6):065206, 2012.

\bibitem{Agb2012}
D.~Agboola and Y.-Z. Zhang.
\newblock Unified derivation of exact solutions for a class of quasi-exactly
  solvable models.
\newblock {\em Journal of Mathematical Physics}, 53(4):042101, 2012.

\bibitem{agboola2014}
D.~Agboola, J.~Links, I.~Marquette, and Y.-Z. Zhang.
\newblock New quasi-exactly solvable class of generalized isotonic oscillators.
\newblock {\em Journal of Physics A: Mathematical and Theoretical},
  47(39):395305, 2014.

\bibitem{quesne2018}
C.~Quesne.
\newblock Quasi-exactly solvable polynomial extensions of the quantum harmonic
  oscillator.
\newblock In {\em Journal of Physics: Conference Series}, volume 1071, page
  012016. IOP Publishing, 2018.

\bibitem{Que2018}
C.~Quesne.
\newblock Quasi-exactly solvable schr\"odinger equations, symmetric polynomials
  and functional bethe ansatz method.
\newblock {\em Acta Polytechnica}, 58(2):118, apr 2018.

\bibitem{quesne2017}
C.~Quesne.
\newblock Extended nikiforov-uvarov method, roots of polynomial solutions, and
  functional bethe ansatz method.
\newblock {\em arXiv preprint arXiv:1704.01406}, 2017.

\bibitem{Kra1997}
A.~Krajewska, A.~Ushveridze, and Z.~Walczak.
\newblock Bender{\textendash}dunne orthogonal polynomials general theory.
\newblock {\em Modern Physics Letters A}, 12(16):1131--1144, may 1997.

\bibitem{Atre2003}
R.~Atre and P.K. Panigrahi.
\newblock Quasi-exactly solvable hamiltonians: a new approach and an
  approximation scheme.
\newblock {\em Physics Letters A}, 317(1-2):46--53, 2003.

\bibitem{marquette2012}
I.~Marquette and J.~Links.
\newblock Generalized heine--stieltjes and van vleck polynomials associated
  with two-level, integrable bcs models.
\newblock {\em Journal of Statistical Mechanics: Theory and Experiment},
  2012(08):P08019, 2012.

\bibitem{Zno1998}
M.~Znojil.
\newblock Quantum exotic: a repulsive and bottomless confining potential.
\newblock {\em Journal of Physics A: Mathematical and General},
  31(14):3349--3355, apr 1998.

\bibitem{Kau1991}
R.S. Kaushal.
\newblock Quantum mechanics of noncentral harmonic and anharmonic potentials in
  two-dimensions.
\newblock {\em Annals of Physics}, 206(1):90--105, feb 1991.

\bibitem{Car2008}
J.F. Cari{\~{n}}ena, A.M. Perelomov, M.F. Ra{\~{n}}ada, and M.~Santander.
\newblock A quantum exactly solvable nonlinear oscillator related to the
  isotonic oscillator.
\newblock {\em Journal of Physics A: Mathematical and Theoretical},
  41(8):085301, feb 2008.

\bibitem{Fel2009}
J.M. Fellows and R.A. Smith.
\newblock Factorization solution of a family of quantum nonlinear oscillators.
\newblock {\em Journal of Physics A: Mathematical and Theoretical},
  42(33):335303, jul 2009.

\bibitem{Ses2010}
J.~Sesma.
\newblock The generalized quantum isotonic oscillator.
\newblock {\em Journal of Physics A: Mathematical and Theoretical},
  43(18):185303, apr 2010.

\bibitem{hall2010}
R.~L. Hall, N.~Saad, and {\"O}.~Ye{\c{s}}ilta{\c{s}}.
\newblock Generalized quantum isotonic nonlinear oscillator in d dimensions.
\newblock {\em Journal of Physics A: Mathematical and Theoretical},
  43(46):465304, 2010.

\bibitem{risken1967}
H.~Risken and H.D. Vollmer.
\newblock The influence of higher order contributions to the correlation
  function of the intensity fluctuation in a laser near threshold.
\newblock {\em Zeitschrift f{\"u}r Physik}, 201(3):323--330, 1967.

\bibitem{mitra1978}
A.K. Mitra.
\newblock On the interaction of the type $\lambda x^2/(1+ gx^2)$.
\newblock {\em Journal of Mathematical Physics}, 19(10):2018--2022, 1978.

\bibitem{bessis1980}
N.~Bessis, G.~Bessis, and G.~Hadinger.
\newblock The perturbed ladder operator method: closed form expressions of
  perturbed wavefunctions and matrix elements.
\newblock {\em Journal of Physics A: Mathematical and General}, 13(5):1651,
  1980.

\bibitem{whitehead1982}
R.R. Whitehead, A.~Watt, G.P. Flessas, and M.A. Nagarajan.
\newblock Exact solutions of the schrodinger equation (-$d/dx^2+ x^2+ \lambda
  x^2/(1+ gx^2)) \psi (x)= e\psi (x)$.
\newblock {\em Journal of Physics A: Mathematical and General}, 15(4):1217,
  1982.

\bibitem{marcilhacy1985}
G.~Marcilhacy and R.~Pons.
\newblock The schrodinger equation for the interaction potential $x^2+ \lambda
  x^2/(1+ gx^2)$ and the first heun confluent equation.
\newblock {\em Journal of Physics A: Mathematical and General}, 18(13):2441,
  1985.

\bibitem{Chr1975}
N.H. Christ and T.D. Lee.
\newblock Quantum expansion of soliton solutions.
\newblock {\em Physical Review D}, 12(6):1606--1627, sep 1975.

\bibitem{Jat1989}
D.P. Jatkar, C.N. Kumar, and A.~Khare.
\newblock A quasi-exactly solvable problem without sl(2) symmetry.
\newblock {\em Physics Letters A}, 142(4-5):200--202, dec 1989.

\bibitem{zhang2016}
Y.-Z. Zhang.
\newblock Hidden sl(2)-algebraic structure in rabi model and its 2-photon and
  two-mode generalizations.
\newblock {\em Annals of Physics}, 375:460--470, 2016.

\end{thebibliography}

\end{document}